\documentclass[amsa]{ipart}

\RequirePackage{hyperref}
\usepackage{graphicx}
\usepackage{hyperref}
\usepackage{cleveref}
\startlocaldefs
\endlocaldefs

\pubyear{2021}
\volume{6}
\issue{2}
\firstpage{1}
\lastpage{1}
\arxiv{2107.02323}

\begin{document}

\begin{frontmatter}

%%  "Title of the Paper"
\title[Light Cone Estimates]{On the non-blow up of energy critical nonlinear massless scalar fields in `3+1' dimensional globally hyperbolic spacetimes: Light cone estimates\protect\thanksref{T1}}
\thankstext{T1}{Light cone estimates}

\begin{aug}
%%  \author{\fnms{John} \snm{Smith}\thanksref{t2}\ead[label=e1]{smith@foo.com}\ead[label=e2,url]{www.foo.com}}
%%  \thankstext{t2}{The author is supported by ...}
%%  \address{line 1\\ line 2\\ \printead{e1}\\\printead{e2}}

\author{\fnms{Puskar} \snm{Mondal}\ead[label=e1]{puskar@cmsa.fas.harvard.edu}}
\address{Center of Mathematical Sciences and Applications,\\ Department of Mathematics, Harvard University,\\
Cambridge, MA, USA\\\printead{e1}}
%\author{\fnms{???} \snm{???}\thanksref{t2}\ead[label=e2]{???}}
%\address{???\\\printead{e2}}
%\and
%\author{\fnms{???} \snm{???}%
%        \ead[label=e3]{???}%
%        \ead[label=u1,url]{???}}
%\address{???\\\printead{e3}\\\printead{u1}}
%\thankstext{t2}{The author is supported by ...}
\end{aug}
%%  History:
%\received{\sday{3} \smonth{1} \syear{2019}}

\begin{abstract}
Here we prove a global existence theorem for the solutions of the semi-linear wave equation with critical non-linearity admitting a positive definite Hamiltonian. 
Formulating a parametrix for the wave equation in a globally hyperbolic curved spacetime, we derive an apriori pointwise bound for the solution of the nonlinear wave equation in terms of the initial energy, from which the global existence follows in a straightforward way. This is accomplished by two steps. First, based on Moncrief's light cone formulation we derive an expression for the scalar field in terms of integrals over
the past light cone from an arbitrary spacetime point to an `initial', Cauchy
hypersurface and additional integrals over the intersection of this cone with
the initial hypersurface. Secondly, we obtain apriori estimates for the energy associated with three quasi-local approximate time-like conformal Killing and one approximate Killing vector fields. Utilizing these naturally defined energies associated with the physical stress-energy tensor together with the integral equation, we show that the spacetime $L^{\infty}$ norm of the scalar field remains bounded in terms of the initial data and continues to be so as long as the spacetime remains singularity/Cauchy-horizon free. 
\end{abstract}

%\begin{keyword}[class=AMS]
%\kwd[Primary ]{}
%\kwd{Light cone estimates}
%\kwd[; secondary ]{}
%\end{keyword}

%%  Upper case for every keyword
\begin{keyword}
\kwd{Light Cone Estimates}
\kwd{Global Existence}
\kwd{Energy Estimates}
\kwd{Nonlinear Waves}
\kwd{Approximate Killing Vector Fields}
\end{keyword}

%\tableofcontents

\end{frontmatter}

%%  The body
\section{Introduction}
One of the most important open problems of twenty first century physics is the proof of Penrose's \textit{Cosmic Censorship} conjecture \cite{penrose1999question}. Present in its two forms, this conjecture essentially 
hints towards the validity of classical determinism. If one simply goes back to special relativity, then the underlying spacetime, the Minkowski space, does not have any singularity. Naturally one would expect that physically reasonable fields if evolving on the Minkowski spacetime background do not develop singularity at finite time. In other words physically acceptable classical fields evolving from regular Cauchy data in Minkowski spacetime should extend uniquely and continuously to globally defined, singularity free solutions of the associated field equations on the entire spacetime. This would imply that classical determinism holds in the realm of special relativity. Global well-posedness of several classical fields on the Minkowski background have been proven to hold true. If we, for now, focus on the physical $3+1$ spacetimes, this includes several linear and non-linear scalar fields admitting positive definite energy (sub-critical and critical but not super-critically nonlinear) \cite{grillakis1990regularity, struwe1988globally, jorgens1961anfangswertproblem, rauch1981u5}, Yang-Mills or Yang-Mills-Higgs fields \cite{eardley1982global, eardley1982global2, klainerman1995finite} etc. On the other hand there are explicit examples of classical fields that exhibit a finite time blow up property on the $3+1$ dimensional Minkowski background. These include focusing energy critical and sub-critical nonlinear wave fields \cite{krieger2009slow, donninger2014exotic}, wave maps (nonlinear sigma models in the physics terminology) from spacetime to curved target manifolds \cite{donninger2012stable, bizon2015generic, krieger2008renormalization1}, perfect fluids \cite{christodoulou2007formation}.

A natural question then would be whether such a result holds true in a globally hyperbolic curved spacetime. If a breakdown of the global existence were to occur then that would certainly be pathological in a sense that the violation of classical determinism would hint that the field under consideration is not physically adequate. In addition, one would ultimately want to study the evolution of the spacetime geometry while coupled to additional fields in order to address the `Cosmic Censorship' question. However, if the fields themselves blow up in finite time in a background spacetime, one would certainly not reasonably expect that this blow up feature would be suppressed by coupling to gravity. This is due to the fact that pure gravity in the absence of any additional source fields may itself blow up in finite time through curvature concentration. Therefore it is fundamentally important to investigate the temporal behaviour of classical fields in globally hyperbolic curved spacetimes. Classical Yang-Mills fields \cite{chrusciel1997global, ghanem2016global}, linear and non-linear sub-critical Klein-Gordon fields admitting positive definite energy [in prep. with V. Moncrief] are known to exhibit the global existence property on a curved background. Motivated by such results, in this article we study the temporal behaviour of the solutions of the wave equation in $3+1$ dimensions with critical nonlinearity
\begin{align}
\label{eq:main}
\hat{g}^{\mu\nu}\nabla_{\mu}\nabla_{\nu}\varphi-\alpha\varphi^{5}=0 
\end{align}
while properly formulated as a Cauchy problem (i.e., with the prescribed data of the field and associated conjugate momentum on an initial spacelike hypersurface). Here $\hat{g}^{-1}:\hat{g}^{\mu\nu}\partial_{\mu}\otimes \partial_{\nu}$ is the inverse of the spacetime metric, $\nabla$ is the metric compatible connection, and $\alpha>0$ is the coupling constant. This non-linearity implies that the corresponding Hamiltonian of this system (which controls the $\dot{H}^{1}$ Sobolev norm of $\varphi$) is invariant under suitable scaling. More precisely, if $x^{\mu}\mapsto \lambda x^{\mu}$ and $\varphi(x)\mapsto \frac{1}{\sqrt{\lambda}}\varphi(\lambda^{-1}x)$, then the wave equation (\ref{eq:main}) remains invariant. Furthermore, the associated Hamiltonian $\mathcal{E}\approx \int_{\Sigma}(g^{ij}\partial_{i}\varphi\partial_{j}\varphi+m^{2}+\alpha\varphi^{6})N\mu_{g}$, where $\Sigma$ is a $3$ dimensional space-like hypersurface, $m$ is the momentum conjugate to $\varphi$, $N$ is the Lapse function, and $g$ is the Riemannian metric induced on $\Sigma$ by $\hat{g}$ ($\mu_{g}$ is the induced volume form), remains invariant under this scaling too. Roughly this means that the energy dispersion by the derivative term and energy concentration by the nonlinearity are similar and therefore the terminology `critical'. In this borderline case, one dominates the other slightly leading to global existence or finite time blow up. In the current case, we shall see that a non-concentration type lemma holds i.e., if one shrinks the hypersurface $\Sigma$ to smaller and smaller sets, the integral $\int_{\Sigma}\alpha\varphi^{6}N\mu_{g}$ does not blow up. In other words, the non-linearity is unable to focus energy. If one slightly increases the strength of the non-linearity (i.e, if the equation is taken to be $\hat{g}^{\mu\nu}\nabla_{\mu}\nabla_{\nu}\varphi-\alpha\varphi^{5+\epsilon}=0,~\epsilon>0$ instead), whether a global existence or blow up occurs in a general case is still an open problem.     

Jorgens \cite{jorgens1961anfangswertproblem} proved the global existence property for semi-linear wave equations on $3+1-$ dimensional Minkowski space for sub-critical nonlinearities of the wave fields. Later classical work by Grillakis \cite{grillakis1990regularity} established the global well-posedness of the critically nonlinear wave equation [\ref{eq:main}] on the Minkowski spacetime background, that is for $\hat{g}^{\mu\nu}=\eta^{\mu\nu}$, the Minkowski metric. In both of these cases, one requires the positive definiteness of the associated energy functional and such property is referred to as defocusing. \cite{grillakis1990regularity} utilized the natural energy associated with the timelike Killing field $\partial_{t}$ and conformal Killing fields $D:=x^{\mu}\partial_{\mu}$, $K:=x_{\nu}x^{\nu}\partial_{t}+2tx^{\mu}\partial_{\mu}$ of the Minkowski spacetime. In addition, \cite{grillakis1990regularity} also utilized an additional vector field $R:=\frac{x^{\mu}}{t}\partial_{\mu}$ in conjunction with the aforementioned Killing and conformal Killing fields and an integral equation for the solution to derive a-priori estimates which helped to control the spacetime $L^{\infty}$ norm of the solution. Prior to \cite{grillakis1990regularity}, of course, global well-posedness results were known assuming a certain smallness condition on the data (in a suitable function space). 

This paper was motivated in part by the desire to adapt the Grillakis \cite{grillakis1990regularity}
argument to globally hyperbolic curved spacetimes. The global wellposedness result on curved spacetime is not obvious (it is in fact not obvious on flat spacetimes as well). The difficulty lies in constructing an integral expression for the solution since the so called Huygens principle does not hold on a general curved spacetime. In other words, the value of the wavefield at a point $p$ does not only depend on its integral over the mantle of the past light cone emanating from $p$ and the light cone's 2-dimensional intersection with the initial Cauchy hypersurface but also on its integral over the interior of the past light cone, the tail contribution. In flat spacetime, since the Huygens principle is known to hold true for linear waves, this tail contribution vanishes. Secondly, in a general curved spacetime, the existence of Killing and conformal vector fields is an undesirable restriction. Note that the existence of a time-like Killing field $\partial_{t}$ guaranteed a positive definite conserved energy for Minkowski space (simply by Noether's theorem). One may impose such symmetry on the spacetimes under study. But such an assumption proves to be too restrictive. For example, assuming a time-like Killing field would imply that the spacetime is stationary, which is an extremely strong condition to impose. We would therefore want to focus on the most generic spacetimes. If one simply defines an energy from the associated stress-energy tensor by fully contracting it with a time-like vector field $n$, then clearly this energy is not conserved. The obstruction to the conservative nature of this energy is precisely due to the non-vanishing strain tensor of $n$. However, if one assumes certain regularity on the background spacetime metric, then the energy which is no longer conserved, is nevertheless bounded by the initial energy. Therefore one may still obtain required estimates through the use of these appropriately defined approximate Killing/conformal Killing fields by paying a price of regularity of the background spacetime metric. Secondly, using Moncrief's light cone formulation, we derive an expression for $\varphi$ at $p$ in terms of its integral over the full past light cone and its intersection with the initial Cauchy hypersurface. This contains Huygens violating tail terms. However, solving an associated transport equation and using an integration by parts argument, this tail contribution may be converted to an integral over the 3 dimensional mantle of the past light cone and its 2 dimensional intersection with the initial hypersurface. These two main ingredients along with the use of a few additional inequalities yield a spacetime $L^{\infty}$ estimate of the wave field, from which the global existence follows in a straightforward way.  

We note that there are a few studies \cite{lai2012global, ibrahim2003solutions} in the literature which deal with the semi-linear wave equation with variable coefficients (only spatially varying or both spacetime varying) and critical nonlinearity on the Minkowski background. However, such an equation can be formulated as a critically nonlinear wave equation on a manifold equipped with a Lorentzian metric. Their method may be applied to prove a global existence theorem for critically nonlinear wave equations on curved spacetimes. Simply choosing a Gaussian normal coordinate yields an equivalence modulo additional innocuous first order terms. Choice of Gaussian normal slicing is well known to be pathological in a sense that the mean extrinsic curvature of the slice becomes singular in finite time (at least when the initial slice has a positive mean curvature). However, such singularity is merely a coordinate singularity and can be handled with a little additional work (see for example \cite{bernal2005smoothness}). Further, these studies have implemented energy estimates associated with suitable vector field multipliers in conjunction with Strichartz estimates. In our our study, we obtain a spacetime point-wise bound of the wave field by means of the aforementioned integral equation and the energy estimates associated with the three approximate Killing/conformal Killing vector fields. The later of the aforementioned studies \cite{ibrahim2003solutions}, where the coefficients are spatially (in a suitable sense) varying, is based on Klainerman's commuting vector field approach together with Strichartz estimates.  The former study \cite{lai2012global} is quite interesting and based on the technique developed by Klainerman \cite{klainerman2001commuting} and Cristodolou $\&$ Klainerman \cite{christodoulou1993global}, where the multiplier vector fields are constructed by means of an optical function generated by solving an Eikonal equation on a globally hyperbolic spacetime. The level sets of this optical function describe the null cones and the gradient vector field of the optical function is a null geodesic generator. Therefore, one needs to control the null geometry of the spacetime on which the wave field is evolving. However, assuming that the spacetime is globally hyperbolic, the null geometry is well behaved. While our method does not rely on solving for an optical function, we nevertheless require the additional integral equation. Their method can be extended in the gravity problem where the analysis to control the null geometry is heavy (since the geometry becomes unknown of the equations involved; see \cite{klainerman2010breakdown, klainerman2005rough} for example). Our method on the other hand is relatively simpler but requires closure of an extremely delicate bootstrap argument (in prep.). Therefore, these two different methods are, in a sense, complimentary to each other. There are of course numerous additional studies on the global behaviour of critically non-linear wave fields on Minkowski spacetimes (e.g., \cite{shatah1993regularity, shatah1994well}) as well as for wave equations  with variable coefficients (e.g., \cite{smith1998parametrix}).

The outline of the paper is as follows. We start with the derivation of the appropriate expression for the massless scalar field $\varphi$ at an arbitrary spacetime point $p$ in its geodesic normal neighbourhood. Next, we construct the suitable approximate Killing/conformal Killing vector fields (in an appropriate sense of course). Then, we derive necessary estimates using the energies associated with these vector fields. Utilizing a few additional inequalities together with the integral expression and the energy estimates, we then prove a spacetime point-wise bound on the solution $\varphi$ in terms of the initial energy. This $L^{\infty}$ estimate finishes the proof of the global wellposedness.

\section{Notations and facts}
The `$3+1$' dimensional spacetime manifold is denoted by $M$. Assuming global hyperbolicity while studying the Cauchy problem, one is led to spacetimes of the form of a product manifold $\mathbb{R}\times \Sigma$, where $\Sigma$ is diffeomorphic to a Cauchy hypersurface. This leads to the interpretation of an evolving physical universe $\Sigma$ embedded in the spacetime. We will designate by $\Sigma_{t}$ the constant $t$ hypersurfaces. Let $n$ denote the unit time-like future directed normal to $\Sigma_{t}$. The tangent space $T_{p}M$ at a point $p$ may be split as follows 
\begin{align}
T_{p}M=T_{\pi(p)}\Sigma_{t}\oplus \mathbb{R},
\end{align}
where $\pi$ is the natural projection $\pi: \Sigma_{t}\times \mathbb{R}\to \Sigma_{t}$. Therefore, the vector field $\frac{\partial}{\partial t}$ may be written as follows 
\begin{align}
\label{eq:shift}
\frac{\partial}{\partial t}=Nn+Y,
\end{align}
where $N$ is the lapse function and $Y$ is the shift vector field parallel to $\Sigma_{t}$ i.e., $Y$ is a section of the tangent bundle $T\Sigma_{t}$. Under this decomposition, the spacetime metric $\hat{g}$ takes following form in a local coordinate basis $\{x^{0}=t,x^{1},x^{2},x^{3}\}$
\begin{align}
\label{eq:metric}
\hat{g}=-N^{2}dt\otimes dt+g_{ij}(dx^{i}+Y^{i}dt)(dx^{j}+Y^{j}dt),
\end{align}
where $g:=g_{ij}dx^{i}\otimes dx^{j}$ is the induced Riemannian metric on $\Sigma_{t}$. We will denote the inverse metrics $\hat{g}^{-1}$ and $g^{-1}$ as  $\hat{g}^{-1}:=\hat{g}^{\mu\nu}\frac{\partial}{\partial x^{\mu}}\otimes \frac{\partial}{\partial x^{\nu}}$ and $g^{-1}:=g^{ij}\frac{\partial}{\partial x^{i}}\otimes \frac{\partial}{\partial x^{j}}$, respectively. In the analysis, the second fundamental form $k\in S^{0}_{2}(\Sigma_{t})$ will show up. It is defined in local coordinates as follows 
\begin{align}
k_{ij}=-\frac{1}{2}(L_{n}g)_{ij}=-\frac{1}{2N}(\partial_{t}g_{ij}-(L_{Y}g)_{ij}),
\end{align}
where $L$ denotes the Lie derivative operator. 

Our analysis holds in a geodesically convex domain $\mathcal{G}\subset M$. In a geodesically convex domain, frequently, we will use a null basis $\{l,\bar{l},\lambda_{1},\lambda_{2}\}$ of $T_{p}M$ at a point $p$. This is constructed by imposing the conditions 
\begin{align}
\label{eq:null1}
\hat{g}(l,l)=0,~\hat{g}(\bar{l},\bar{l})=0,~n=-l+\bar{l},
\end{align}
and following $\hat{g}(n,n)=-1$, we obtain
$\hat{g}(l,\bar{l})=\frac{1}{2}$. The two space-like vector fields $\lambda^{j}_{1}\partial_{j}$ and $\lambda^{j}_{2}\partial_{j}$ are such that 
\begin{align}
\label{eq:null2}
\hat{g}(\lambda_{i},\lambda_{j})=\delta_{ij},~~
\hat{g}(\lambda_{i},l)=\hat{g}(\lambda_{i},\bar{l})=0.
\end{align}
Simply counting the degrees of freedom and the number of equations involved, we see that this is a uniquely determined system and therefore there exists a unique null frame $\{l,\bar{l},\lambda_{1},\lambda_{2}\}$ in which the metric $\hat{g}$ is expressible as 
\begin{align}
\label{eq:null3}
\hat{g}=2(l\otimes \bar{l}+\bar{l}\otimes l)+\lambda_{1}\otimes \lambda_{1}+\lambda_{2}\otimes \lambda_{2}.
\end{align}
We will make use of this null basis frequently. Let us now mention the notations for the causal geometry. The mantle of the past (resp. future) light cone of a point $p$ is denoted simply by $C^{-}_{p}$ (resp. $C^{+}_{p}$) while the chronological past (resp. future), the solid interior, is denoted by $J^{-}_{p}$ (resp. $J^{+}_{p}$). The causal past (resp. future) of $p$ is denoted by $D^{-}_{p}$ (resp. $D^{+}_{p}$). When we write $C^{-}_{p}, D^{-}_{p}$ or $J^{-}_{p}$, we will mean that these sets are only defined up to the Cauchy hypersurface $\Sigma_{t_{0}}$ at $t=t_{0}$ on which the initial data is provided but do not contain the Cauchy hypersurface $\Sigma_{t_{0}}$. In addition, notice that $C^{-}_{p},D^{-}_{p},$ and $J^{-}_{p}$ do not contain the point $p$ as well. The 3 dimensional intersection of the past light cone of $p$ with a constant $t$ hypersurface $\Sigma_{t}$ is denoted by $S_{t}$, while its 2-dimensional intersection (a topological 2-sphere) is denoted by $\sigma^{t}_{p}$ (see Figure \ref{fig:fig1}) and $\sigma^{t_{0}}_{p}=\sigma_{p}$. Now let us consider a constant $t$ hypersurface $\Sigma_{t}$ such that $\Sigma_{t}\cap D^{-}_{p}\neq \emptyset$. The portion of $C^{-}_{p}$ ($D^{-}_{p}$ and $J^{-}_{p}$ resp.) lying above $\Sigma_{t}$ will be denoted by $C^{t}_{p}$ ($D^{t}_{p}$ and $J^{t}_{p}$ resp.). More generally, for a past null cone with vertex at $p$ and extending up to any constant time hypersurface $\Sigma_{t}$, the mantle of the cone, the causal past of $p$, and the chronological past of $p$ will be denoted by $C^{t}_{p}$, $D^{t}_{p}$, and $J^{t}_{p}$, respectively. 

Unless otherwise stated, we will work in geodesic normal coordinates about p. The details about the injectivity radius bound etc will be described in the fullness of time. Without loss of generality, we may simply set $p$ to 0, the coordinate origin. The portion of the light cone $C^{-}_{p}$ trapped between two constant $t$ hypersurfaces will be called the truncated light cone and is denoted by $C^{T}_{p}$.
The past solid cone $D^{-}_{p}$ may be parametrized by spherical coordinates as follows 
\begin{align}
x^{1}=r\sin\theta\cos\phi, x^{2}=r\sin\theta\sin\phi,x^{3}=r\cos\theta, x^{0}=t, \theta\in [0,\pi],\\\nonumber
\phi\in[0,2\pi],
\end{align}
where $r=t=0$ corresponds to the origin of the normal coordinate system i.e. $p=0$. Now we introduce the light-cone coordinates
\begin{align}
\label{eq:nullcoordinate}
u=t-r,~v=t+r,~\theta=\theta,~\phi=\phi.
\end{align}
The mantle of the past light cone $C^{-}_{p}$ of $p$ is essentially defined by $v=t+r=0$.
We will use these coordinates whenever necessary. We denote the geodesic squared distance between points $p$ and $q$ by $\Gamma(p,q)$ and accordingly the set $\{q|\Gamma(p,q)=0\}$ denotes the null cone through $p$. The lower branch of the hyperboloid $\{q|\Gamma(p,q)=-\delta\}$ is denoted by $\Gamma_{\delta}$. Using the light-cone coordinates $\{u,v,\theta,\phi\}$, the past null cone $C_{p}$ (of course here we mean the mantle of the cone) may be parametrized by $(u,\omega)\in (0,\tilde{u}]\times \mathbb{S}^{2}$. Since we are in the geodesically convex domain, the exponential map is a diffeomorphism from $C^{-}_{p}=C_{p}-\{p\}$ to $(0,\tilde{u}]\times \mathbb{S}^{2}$ i.e., $q\in C^{-}_{p}$ may be written as $q=\exp_{p}(-ul)$, where $l$ is the past directed null-vector passing through $p$. See the figure \ref{fig:fig1} for detail. when we say `on the cone $C^{-}_{p}$', we will always mean `on the the mantle of the cone $C^{-}_{p}$'.  

In terms of functions spaces we will make use of the Sobolev spaces $H^{1}$, $\dot{H}^{1}$ etc defined on the spacelike sub-manifold $\Sigma_{t}$. The homogeneous Sobolev space $\dot{H}^{1}$ is defined as follows: $f\in \dot{H}^{1}(\Sigma_{t})=>||f||_{\dot{H}^{1}(\Sigma_{t})}:=\sqrt{\int_{\Sigma_{t}}|\nabla f|^{2}_{g}\mu_{g}}<\infty$. For two positive functions $f(t)$ and $g(t)$, $f(t)\approx g(t)$ implies $C_{1}g(t)\leq f(t) \leq C_{2}g(t)$, $0<C_{1},C_{2}<\infty$, $f(t)\lesssim g(t)$, implies $f(t)\leq C g(t)$, for some $C$ s.t. $0<C<\infty$. The lower brach of the hyperboloid $\{q|\Gamma(p,q)=-\delta\}$ is denoted by $\Gamma_{\delta}$. We will consider that $\varphi\in\mathcal{S}$ i.e., in Schwartz class since $\mathcal{S}$ is dense in $\dot{H}^{1}$. The volume forms associated with $\hat{g}$ and $g$ are denoted by $\mu_{\hat{g}}$ and $\mu_{g}$, respectively. 

\section{Main Idea}
Here we briefly describe the idea of the proof. Firstly we derive the following integral equation for the scalar field $\varphi$ at an arbitrary point $p$ ($\equiv x$ in local coordinates) in terms of the integral of $\varphi$ (and its nonlinearity) over the mantle of the past light cone $C^{-}_{p}$ and the initial data 
\begin{align}
\label{eq:desired}
\varphi(x)=\int_{C^{-}_{p}}U(x,y)\frac{\varphi^{5}(y)}{u}\mu_{\hat{g}}(y)|_{C^{-}_{p}}+\frac{1}{2\pi}\int_{C^{-}_{p}}\hat\Box_{y}\nonumber U(x,y)\frac{\varphi(y)}{u}\mu_{\hat{g}}|_{C^{-}_{p}}\\\nonumber +~initial~data,
\end{align}
where $U(x,y)$ is a suitably defined symmetric bi-scalar satisfying $\lim_{x\to y}U(x,y)=1$, $\lim_{x\to y}\hat\Box_{x}U(x,y)=\frac{1}{6}R_{\hat{g}}(y)$, where $R_{\hat{g}}(y)$ is the scalar curvature of $\hat{g}$ evaluated at $y$. Assuming sufficient regularity of the background spacetimes, $\sup_{x\in D^{-}_{p}}\sup_{y\in D^{-}_{p}}|U(x,y)|\leq C$. Note here that for Minkowski space $U(x,y)=1,~ \hat\Box_{x}U(x,y)=0$ everywhere. Naturally, on a general curved spacetime, we can not get rid of the Huygens violating term containing $\hat\Box_{x}U(x,y)$ and therefore it requires control. However, when $x=y$ which is the most dangerous case, a simple calculation yields $|\hat\Box_{x}U(x,y)|$ equals to the scalar curvature of the spacetime. Therefore, if we assume that the spacetime is sufficiently regular then  $\sup_{x\in D^{-}_{p}}\sup_{y\in D^{-}_{p}}|\hat\Box_{x}U(x,y)|\leq C$. In other words this terms does not cause additional problems as long as the spacetime does not develop singular points. 

First, let us consider a constant time hypersurface $\Sigma_{t_{1}}$ such that $\Sigma_{t_{1}}\cap D^{-}_{p}\neq \emptyset$. As mentioned in the previous section, we denote the portion of $C^{-}_{p}$ ($D^{-}_{p}$ and $J^{-}_{p}$ resp.) lying above $\Sigma_{t_{1}}$ (i.e., for the part where $|t|<|t_{1}|$) by $C^{t_{1}}_{p}$ ($D^{t_{1}}_{p}$ and $J^{t_{1}}_{p}$ resp.). Now if we split the integral (\ref{eq:desired}) over $C^{-}_{p}$ into two parts: one over $C^{t_{1}}_{p}$ and one over $C^{-}_{p}-C^{t_{1}}_{p}$, then for a small but fixed $|t_{1}|$, the integral over $C^{-}_{p}-C^{t_{1}}_{p}$ is bounded by a constant $C(t_{1})$. The challenge is to control the integral over the top part (i.e., near the vertex). We will do so by choosing the height ($\approx |t_{1}|$) of the top part $C^{t_{1}}_{p}$ sufficiently small but a priori fixed. Notice that the integral over $C^{t_{1}}_{p}$ may be written as an integral over $C^{t_{1}}_{p}-C^{t_{2}}_{p}$ for $|t_{2}|<|t_{1}|$ and pass to the limit $t_{2}\to 0$. Our goal is to take one factor of $|\varphi|$ out of the nonlinear part of the integral over $C^{t_{1}}_{p}-C^{t_{2}}_{p}$ as $\sup_{x\in (C^{t_{1}}_{p}\cup \sigma_{t_{1}})-C^{t_{2}}_{p}}|\varphi(x)|$ and control the left over integral using available estimates. Now if we can show that $\sup_{x\in (C^{t_{1}}_{p}\cup\sigma_{t_{1}})-C^{t_{2}}_{p}}|\varphi(x)|~(\leq \sup_{x\in (D^{t_{1}}_{p}\cup S_{t_{1}})-D^{t_{2}}_{p}}|\varphi(x)|$) is finite then the remaining integral may be bounded in terms of energy after applying Cauchy-Scwartz and an important Hardy type inequality (for null hypersurface) that is to be derived  (observe that $(D^{t_{1}}_{p}\cup S_{t_{1}})-D^{t_{2}}_{p}$ is closed). Note that $p\notin D^{t_{1}}_{p}$ and therefore, we need to write the integral equation for $\varphi$ at a point lying in $(D^{t_{1}}_{p}\cup S_{t_{1}})-D^{t_{2}}_{p}$ ($|t_{2}|<|t_{1}|$) where the maximum is attained. Let us assume that the maximum is attained at $x^{'}$, $|\varphi(x^{'})|=A_{s}$ and $x^{'}\to p$ as $t_{2}\to 0$. Therefore, we may write the integral equation at $x^{'}$ and use the Cauchy-Scwartz inequality to yield
\begin{align}
A_{s} \leq C_{1} A_{s}\int_{C^{t_{1}}_{x^{'}}}\frac{\varphi^{4}}{u}\mu_{\hat{g}}(x)|_{C^{t_{1}}_{x^{'}}}+|u_{1}|\left(\int_{C^{t_{1}}_{x}}\varphi^{6}\mu_{\hat{g}}(x)|_{C^{t_{1}}_{x^{'}}}\right)^{1/3}+C_{2}\\\nonumber
\leq C_{1}A_{s}\left(\int_{C^{t_{1}}_{x^{'}}}\varphi^{6}(\mu_{\hat{g}}(x)|_{C^{t_{1}}_{x^{'}}})\right)^{1/2}\left(\int_{C^{t_{1}}_{x^{'}}}\frac{\varphi^{2}}{u^{2}}(\mu_{\hat{g}}(x)|_{C^{t_{1}}_{x^{'}}})\right)^{1/2}+C_{2}.
\end{align}
Here once again $C^{t_{1}}_{x^{'}}$ denotes the portion of the light cone $C^{-}_{x^{'}}$ that lies entirely within $C^{t_{1}}_{p}$.
We will show that either $\left(\int_{C^{t_{1}}_{x^{'}}}\varphi^{6}(\mu_{\hat{g}}(x)|_{C^{t_{1}}_{x^{'}}})\right)^{1/2}$ or $\left(\int_{C^{t_{1}}_{x^{'}}}\frac{\varphi^{2}}{u^{2}}(\mu_{\hat{g}}(x)|_{C^{t_{1}}_{x^{'}}})\right)^{1/2}$ is $\lesssim \delta$ for arbitrarily small $\delta>0$ if the height of the top part of the cone $C^{t_{1}}_{p}$ is chosen sufficiently small (i.e., making $|t_{1}|$ small but a priroi fixed, $|u_{1}|\lesssim |t_{1}|$). Now choosing $\delta$ sufficiently small, we obtain
\begin{align}
A_{s}\leq C\delta A_{s}+C_{2}\\\nonumber 
A_{s}\leq C.
\end{align}
Here $C$ depends on the initial energy in a harmless way. Now using boundedness of $\sup_{x\in    (D^{t_{1}}_{p}\cup S_{t_{1}})-D^{t_{2}}_{p}}|\varphi(x)|=|\varphi(x^{'})|=A_{s}$ and since the energy can not blow up in finite time, we observe from the integral equation (\ref{eq:desired}) that $|\varphi|$ can not blow up at $p$. Since $p$ is an arbitrary point in the globally hyperbolic spacetime $M$, we conclude that the point-wise norm of the wavefield $\varphi$ is bounded as long as the spacetime does not develop singularity/Cauchy horizon. Utilizing this $L^{\infty}$ bound, a continuity argument together with a contraction mapping argument based local existence theorem finishes the global existence proof. 
The main challenge is to show that the aforementioned two integrals are of the order $\delta$. Note that if one simply chooses sufficiently small initial energy, then just estimating the energy corresponding to the time-like normal $n$ is sufficient to control the $L^{\infty}$ norm of $\varphi$. However, since we are interested in arbitrarily large data, we need to use energy estimates associated with three additional vector fields which are no longer Killing/conformal Killing, but only so approximately. Use of the energies associated with these three aditional vector fields is crucial in bounding the aforementioned integrals.

Section 4 is devoted to deriving the desired integral equation (\ref{eq:desired}) for $\varphi$ which is one of the most important parts of the proof. In order to perform these aforementioned analyses, energy inequalities are indispensable. Therefore after deriving the integral equation, we derive the energy estimates associated with timelike approximate Killing and conformal Killing vector fields. Note that in order to control $\varphi$ at an interior point $x^{'}$ lying within $D^{t_{1}}_{p}$, we need to estimate the energy flux flowing (out) transversal to the mantle of the cone $C^{t_{1}}_{x^{'}}$. This however can not be obtained directly for the following reason. Firstly, we will consider the height of the exterior cone $C^{t_{1}}_{p}$ to be sufficiently small such that the energy flux flowing across $C^{t_{1}}_{p}$ is small. However, this does not imply that the energy flux flowing across $C^{t_{1}}_{x^{'}}$ is small. In fact, we will see that the full energy flux flowing across $C^{t_{1}}_{x^{'}}$ will never be small; only parts of it will be. Luckily the parts we require to control the nonlinearity in the integral equation will be small. In order to achieve this desired smallness, we will next use two additional estimates coming from two approximate conformal Killing vector fields. In section 5.1, we obtain an energy estimate associated with the unit timelike vector field $n$ which is orthogonal to the constant time hypersurfaces. This will establish the fact that if we choose the height of the exterior light cone $C^{t_{1}}_{p}$ sufficiently small, then the energy flux flowing across it can be made sufficiently small. Using this a priori estimate, in sections 5.3 and 5.4 we use the approximate inversion generator and the scaling vector field to derive extra estimates which will be used to control the energy flux flowing transversal to the interior cone $C^{t_{1}}_{x^{'}}$. In section 5.5, we derive an extremely important Hardy type inequality for a null hypersurface where curvature and its certain null derivatives appear as corrections to the flat space case. In section 5.6, using a re-scaled version of the scaling vector field, we obtain the desired smallness of the required parts of the energy flux flowing across the interior cone. Utilizing this final estimate and the inequality derived in section 5.5, we finish the proof of the boundedness of $\varphi(p)$. Lastly in section 6, we sketch a proof of global existence by making use of a priori point-wise bound on the wavefield $\varphi$ and a local existence theorem.         

A striking difference with the sub-critical case (to be presented in a forthcoming article by the current author and V. Moncrief) is that a simple application of Gr\"onwall's inequality, Holder's inequality, and the basic estimate of the energy $T(n,n)$ suffices to obtain an $L^{\infty}$ bound of $\varphi$ in terms of the initial energy in the case of the latter.    

\section{An integral equation for $\varphi$}
In this section we obtain the desired integral equation for the massless scalar field $\varphi$. Let us write the semi-linear wave equation in the natural covariant form after setting the coupling constant $\alpha$ to be 1 
\begin{align}
\label{eq:eom}
\nabla^{\mu}\nabla_{\mu}\varphi=\varphi^{5}.
\end{align}
Let us denote the co-variant spacetime Laplacian $\nabla^{\mu}\nabla_{\mu}=\hat{g}^{\mu\nu}\nabla_{\mu}\nabla_{\nu}$ by $P$ i.e., 
\begin{align}
P\varphi=\nabla^{\mu}\nabla_{\mu}\varphi.
\end{align}
One may obtain an integral equation for $\varphi$ at $p$ once the fundamental solution or the Green's function $G_{p,q}$ associated with the operator $P$ is available. The advanced Green's function $G^{+}_{p,q}$ is defined as follows 
\begin{align}
\nabla^{\mu}\nabla_{\mu}G^{+}_{p,q}=\delta(q)
\end{align}
and its support is contained in $D^{+}(q)$. Following Friedlander \cite{friedlander1976wave}, the advanced Green's function may be explicitly written in the following form 
\begin{align}
G^{+}_{q}(p)=\frac{1}{2\pi}U(p,q)\delta^{+}(\Gamma(p,q))+\frac{1}{2\pi}V^{+}(p,q),
\end{align}
where $\delta^{+}(\Gamma(p,q))$is the Dirac mass supported on the forward null cone of $q$ and defined as $\lim_{\delta\to 0}\delta^{+}(\Gamma(p,q)+\delta)$. $\Gamma(p,q)$ is the squared geodesic distance between $p$ and $q$. The symmetric bi-scalar $U(p,q)$ in local coordinates $(p,q)\equiv (x,y)$ may be expressed as follows 
\begin{align}
U(x,y):=\frac{|\det \partial^{2}\Gamma/\partial x^{\mu}\partial y^{\nu}|^{1/2}}{4|\det(\hat{g}_{\alpha\beta}(x))\det(\hat{g}_{\alpha\beta}(y))|^{1/4}}.
\end{align}
$V^{+}(p,q)$ is the solution of the following characteristic initial value problem 
\begin{align}
\label{eq:character}
PV^{+}(p,q)=0~\forall p\in D^{+}_{q}, V^{+}(p,q)=V_{0}~\forall p\in C^{+}(q),
\end{align}
where $V_{0}$ satisfies the transport type equation 
\begin{align}
\label{eq:transport}
2\hat{g}(\nabla\Gamma,\nabla V_{0})+(\hat\Box\Gamma-4)V_{0}=-PU,
\end{align}
where $\hat\Box$ is the ordinary spacetime Laplacian expressible in local coordinate as $\hat\Box:=\nabla^{\mu}\nabla_{\mu}=P$. Once the advanced Green's function is obtained, the integral equation for $\varphi\in C^{\infty}(\Omega)$ ($\Omega$ is a geodesically convex neighbourhood of a point $p$) in terms of data on an `initial' Cauchy hypersurface $\Sigma$ is given by the following theorem \cite{friedlander1976wave}. The detailed theory is developed in Friedlander's book \cite{friedlander1976wave} (which builds on the fundamental work of Hadamard, Riesz, Sobolev,
Choquet-Bruhat and others). Here we do not repeat the complete derivation of the intermediate integral equation. Starting from Friedlander's integral equation, we derive the final equation which will be of direct importance in obtaining the point-wise estimate. Interested readers are referred to chapter 5 of Friedlander's book \cite{friedlander1976wave}.\\
\textbf{Theorem \cite{friedlander1976wave}:} \textit{Let $\varphi\in C^{\infty}(\Omega)$ and assume that $p\in D^{+}(\Sigma)-\{\Sigma\}$. Then $\varphi$ at point $p$ ($\equiv x$ in local coordinates) is given by the following equation
\begin{align}
\varphi(x)=\frac{1}{2\pi}\int_{C^{-}_{p}}U(x,y)\varphi^{5}(y)\mu_{\Gamma}(y)+\frac{1}{2\pi}\int_{D^{-}_{p}\cap D^{+}(\Sigma)}V^{+}(x,y)\varphi^{5}(y)\mu_{\hat{g}}(y)\nonumber\\\nonumber
+\frac{1}{2\pi}\int_{S_{p}} {*}\left(V^{+}(x,y)\nabla_{y}\varphi(y)-\varphi(y)\nabla_{y} V^{+}(x,y)\right)\\\nonumber +\frac{1}{2\pi}\int_{\sigma_{p}}\left(2U(x,y)<\xi(y),\nabla_{y} \varphi(y)>+U(x,y)\Theta(y) \varphi(y)\right)d\sigma_{p}(y)\\\nonumber 
+\frac{1}{2\pi}\int_{\sigma_{p}}V^{+}(x,y)\varphi(y) d\sigma_{p}.
\end{align}
Here $\mu_{\Gamma}$ is a Leray form defined such that $d_{y}\Gamma(x,y)\wedge\mu_{\Gamma}(y)=\mu_{\hat{g}}(y)$ and $\Theta(y)$ is the dialation of $d\sigma_{p}$ along the bicharacteristics of the null hypersurface distinct from $C^{-}_{p}$ (let's denote this other null hypersurface by $T_{p}$) that contains $\sigma_{p}$ defined as 
\begin{align}
\Theta d\sigma_{p}:=\lim_{\delta\to 0}\frac{d\sigma_{p}-d\sigma^{'}_{p}}{\delta},
\end{align}
$*$ is the Hodge dual operator, $\sigma^{'}_{p}$ is the intersection of the pseudo-sphere $\{q;\Gamma(p,q)=-\delta\}$ and the null hypersurface $T_{p}$, and $\xi$ is tangent to the null generator of $T_{p}$ such that $\hat{g}(\xi,\nabla\Gamma)=1$.
}

Notice an important fact that the tail terms involving $V^{+}$ are the ones obstructing the Huygen's principle in a general curved spacetime. For the case of Minkowski space, these additional tail terms vanish (due to $V^{+}=0$) restoring the Huygen's principle for linear waves. One may in principle obtain a formal series solution of the characteristic initial value problem (\ref{eq:character}) assuming smoothness \cite{friedlander1976wave}. However, such solution is not very helpful towards obtaining the desired estimate. Motivated by Moncrief's treatment of the tensor wave equation for spacetime curvature \cite{moncrief2005integral}, we will use an integration by parts type argument to remove the tail terms instead. However, in doing so we will have to pay a price by taking two spacetime derivatives of the bi-scalar. Assuming sufficient regularity of the background spacetime metric, we will explicitly show that such a term does not create additional problems. The transformation of the tail term is obtained through the following series of calculations. 
\subsection{Cancellation of the integral over $S_{p}$ in the theorem}
Using the equation of motion $\nabla^{\mu}\nabla_{\mu}\varphi=\varphi^{5}$, write the following 
\begin{align}
\int_{D^{-}_{p}\cap D^{+}(\Sigma)}V^{+}(x,y)\varphi^{5}(y)\mu_{\hat{g}}=\int_{D^{-}_{p}\cap D^{+}(\Sigma)}V^{+}(x,y)\nabla^{\mu}\nabla_{\mu}\varphi(y)\mu_{\hat{g}}.
\end{align}
Now notice the following calculations 
\begin{align}
V^{+}(x,y)\nabla^{\mu}\nabla_{\mu}\varphi(y)=\nabla^{\mu}(V^{+}(x,y)\nabla_{\mu}\varphi(y))-\nabla^{\mu}V^{+}(x,y)\nabla_{\mu}\varphi(y)\\\nonumber 
=\nabla^{\mu}(V^{+}(x,y)\nabla_{\mu}\varphi(y))-\nabla^{\mu}(\varphi(y)\nabla_{\mu}V^{+}(x,y))\\\nonumber 
 +\varphi(y)\nabla^{\mu}\nabla_{\mu}V^{+}(x,y)\\\nonumber 
=\nabla^{\mu}\left(V^{+}(x,y)\nabla_{\mu}\varphi(y)-\varphi(y)\nabla_{\mu}V^{+}(x,y)\right)
\end{align}
where equation (\ref{eq:character}) i.e., $\nabla^{\mu}\nabla_{\mu}V^{+}=0$ throughout the causal domain of $y$, is used. Now since we have reduced it to a total covariant divergence term, we may use the Stokes' theorem to reduce the bulk-integral over $D^{-}_{p}\cap D^{+}(\Sigma)$ to an integral over the boundary $\partial (D^{-}_{p}\cap D^{+}(\Sigma))=C^{-}_{p}\cap S_{p}$. Therefore we have the following 
\begin{align}
\int_{D^{-}_{p}\cap D^{+}(\Sigma)}V^{+}(x,y)\varphi^{5}(y)\mu_{\hat{g}}=-\int_{S_{p}}~*\left(V^{+}(x,y)\nabla\varphi(y)\nonumber-\varphi(y)\nabla V^{+}(x,y)\right)\\\nonumber
 +\int_{C^{-}_{p}}~*\left(V^{+}(x,y)\nabla\varphi(y)\nonumber-\varphi(y)\nabla V^{+}(x,y)\right)
\end{align}
where the hypersurface $S_{p}$ is oriented such that the unit normal vector is future directed i.e., points toward $p$. Therefore, we note that the integral over $S_{p}$ in the theorem is cancelled point-wise by the term generated via integration by parts to yield 
\begin{align}
\varphi(x)=\frac{1}{2\pi}\int_{C^{-}_{p}}U(x,y)\varphi^{5}(y)\mu_{\Gamma}(y)\\\nonumber
+\frac{1}{2\pi}\int_{C^{-}_{p}} {*}\left(V^{+}(x,y)\nabla_{y}\varphi(y)-\varphi(y)\nabla_{y} V^{+}(x,y)\right)\\\nonumber +\frac{1}{2\pi}\int_{\sigma_{p}}\left(2U(x,y)<\xi(y),\nabla_{y} \varphi(y)>+U(x,y)\Theta(y) \varphi(y)\right)d\sigma_{p}(y)\\\nonumber 
+\frac{1}{2\pi}\int_{\sigma_{p}}V^{+}(x,y)\varphi(y) d\sigma_{p}.
\end{align}

Now notice that we have gotten rid of the bulk-integral involving the tail contribution $V^{+}$ and the remaining terms only involve the integrals over the mantle of the past light cone $C^{-}_{p}$ and its two dimensional intersection $\Sigma_{p}$ with $\Sigma$. However, this result is not satisfactory since one still needs to solve for the tail contribution $V^{+}$. As we mentioned earlier however one may construct a series solution for $V^{+}$ assuming analyticity and then approximate the solution in a suitable sense. We will nevertheless avoid such procedure all together. Instead we will make use of the transport equation (\ref{eq:transport}) and a reciprocity theorem to replace the integrals involving $V^{+}$ by terms that may be easily controlled. 

\subsection{Removal of `$\frac{1}{2\pi}\int_{C^{-}_{p}} {*}\left(V^{+}(x,y)\nabla_{y}\varphi(y)-\varphi(y)\nabla_{y} V^{+}(x,y)\right)+\frac{1}{2\pi}\int_{\sigma_{p}}V^{+}(x,y)\varphi(y) d\sigma_{p}$'}

In order to get rid of the tail terms in the integral equation, we will need the following reciprocity theorem. Notice that the operator $P=\nabla^{\mu}\nabla_{\mu}$ is self-adjoint i.e., $P=P^{\dag}$.\\
\textbf{Lemma}\cite{friedlander1976wave} \textit{Let $^{t}V^{+}$ (resp. $^{t}V^{-}$) and $V^{+}$ (resp. $V^{-}$) be the tails terms of the fundamental solutions of $P^{\dag}$ and $P$, respectively. Then the following holds 
\begin{align}
V^{+}(p,q)=~^{t}V^{-}(q,p), V^{-}(p,q)=~^{t}V^{+}(q,p). 
\end{align}
}
Following the previous lemma, the following holds
\begin{align}
V^{+}(p,q)=V^{-}(q,p), V^{-}(p,q)=V^{+}(q,p)
\end{align}
due to the fact that $P=P^{\dag}$ in the current context. Therefore, in the local coordinate expression, we will replace $V^{+}(x,y)$ by $V^{-}(y,x)$ in the following calculations. First note an important fact. Since $\nabla \Gamma\neq 0$ on $C^{-}_{p}-\{p\}$, we have $d\Gamma\wedge \mu_{\Gamma}=\mu_{\hat{g}}$, where $\mu_{\Gamma}$ is a Leray form on $C^{-}_{p}$. Using the definition of the Hodge dual, we may obtain the following for a $1-$form $v$ on $C^{-}_{p}-\{p\}$  
\begin{align}
*v(y)=<v(y),\nabla\Gamma(x,y)>\mu_{\Gamma}(y).
\end{align}
This holds precisely because $\nabla\Gamma$ is tangential to the null cone $C^{-}_{p}$ (see lemma 2.9.2 in Friedlander's book \cite{friedlander1976wave}). Therefore the term\\ $\frac{1}{2\pi}\int_{C^{-}_{p}} {*}\left(V^{+}(x,y)\nabla_{y}\varphi(y)-\varphi(y)\nabla_{y} V^{+}(x,y)\right)$ may be evaluated as follows 
\begin{align}
 \frac{1}{2\pi}\int_{C^{-}_{p}} {*}\left(V^{+}(x,y)\nabla_{y}\varphi(y)-\varphi(y)\nabla_{y} V^{+}(x,y)\right)\\\nonumber
=\frac{1}{2\pi}\int_{C^{-}_{p}} <\left(V^{+}(x,y)\nabla_{y}\varphi(y)-\varphi(y)\nabla_{y} V^{+}(x,y)\right),\nabla_{y}\Gamma(y)>\mu_{\Gamma}(y)\\\nonumber 
=\frac{1}{2\pi}\int_{C^{-}_{p}}<\nabla_{y}(V^{+}(x,y)\varphi(y))-2\varphi(y)\nabla_{y}V^{+}(x,y),\nabla_{y}\Gamma(y)>\mu_{\Gamma}(y)\\\nonumber 
=\frac{1}{2\pi}\int_{C^{-}_{p}}<\nabla_{y}(V^{-}(y,x)\varphi(y)),\nabla_{y}\Gamma(y)>\mu_{\Gamma}(y)\\\nonumber  -\frac{1}{2\pi}\int_{C^{-}_{p}}\varphi(y)<2\nabla_{y}V^{-}(y,x),\nabla_{y}\Gamma(y)>\mu_{\Gamma}(y).
\end{align}
Here we have used the reciprocity theorem (lemma). Now notice that in the second term of the above expression, only the tangential derivative of $V^{-}(y,x)$ appears and while restricted to $C^{-}_{p}$, we may use the transport equation (\ref{eq:transport}) to replace the second term by a lower order term in $V^{-}(y,x)|_{C^{-}_{p}}=V_{0}$. Doing so we obtain 
\begin{align}
\label{eq:split}
 \frac{1}{2\pi}\int_{C^{-}_{p}} {*}\left(V^{+}(x,y)\nabla_{y}\varphi(y)-\varphi(y)\nabla_{y} V^{+}(x,y)\right)\\\nonumber 
=\frac{1}{2\pi}\int_{C^{-}_{p}}<\nabla_{y}(V^{-}(y,x)\varphi(y)),\nabla_{y}\Gamma(y)>\mu_{\Gamma}(y)\\\nonumber  +\frac{1}{2\pi}\int_{C^{-}_{p}}\varphi(y)\left((\hat\Box_{y}\Gamma(x,y)-4)V_{0}(y,x)+PU(y,x)\right)\mu_{\Gamma}(y)\\\nonumber 
=I_{1}+I_{2}.
\end{align}
Here we make a series of coordinate transformations according to convenience. Note that $(x^{0},x^{1},x^{2},x^{3})$ denotes the geodesic normal coordinate system, while $(t,r,\theta,\varphi)$ denotes spherical  coordinates and $(u,v,\theta,\varphi)$ denotes spherical light-cone/null coordinates defined as follows
\begin{align}
u=t-r,v=t+r,\theta=\theta,\phi=\phi.
\end{align}
In a sense we perform $(x^{0},x^{1},x^{2},x^{3})\mapsto (t,r,\theta,\varphi)\mapsto (u,v,\theta,\varphi)$. Notice that $\hat{g}_{\mu\nu}|_{p}=\hat{g}_{\mu\nu}(0)=\eta_{\mu\nu}$ since the normal coordinate system is based at $p\equiv 0$. In addition the following two identities hold throughout the normal coordinate frame
\begin{align}
\hat{g}_{\mu\nu}(x)x^{\nu}=\hat{g}_{\mu\nu}(0)x^{\nu}=\eta_{\mu\nu}x^{\nu},\\
\Gamma[\hat{g}]^{\mu}_{\alpha\beta}x^{\alpha}x^{\beta}=0.
\end{align}
The second property essentially follows from the fact that the geodesics through $p$($\equiv 0$) are straight lines. For a complete proof, the reader is referred to the relevant sections of \cite{moncrief2005integral}.
Since in the integral equation we have $x=0$, we will only concern ourselves with $\Gamma(0,x):=\Gamma$.
Evaluating $\Gamma$ in the normal coordinates 
\begin{align}
\Gamma:=\hat{g}_{\mu\nu}(x)x^{\mu}x^{\nu}=\hat{g}_{\mu\nu}(p=0)x^{\mu}x^{\nu}=\eta_{\mu\nu}x^{\mu}x^{\nu}=-t^{2}+r^{2}=-uv.
\end{align}
We may immediately obtain 
\begin{align}
\nabla^{\mu}\Gamma\partial_{\mu}=2x^{\mu}\partial_{\mu}=2u\partial_{u}+2v\partial_{v},
\end{align}
and 
\begin{align}
d\Gamma:=\partial_{\mu}\Gamma dx^{\mu}=-vdu-udv.
\end{align}
The invariant volume form in $(u,v,\theta,\phi)$ coordinates is expressed as $\mu_{\hat{g}}:=\sqrt{-\det(\hat{g}_{\mu\nu}(u,v,\theta,\phi))}du\wedge dv\wedge d\theta\wedge d\phi$. On the other hand, the Leray form on $C^{-}_{p}$ is defined as $\mu_{\Gamma}$ on $C^{-}_{p}$ satisfying $d\Gamma\wedge \mu_{\Gamma}=\mu_{\hat{g}}$ and therefore noting $d\Gamma=-vdu-udv$, we may obtain a Leray form as the following 
\begin{align}
\mu_{\Gamma}=\frac{\sqrt{-\det(\hat{g}_{\mu\nu}(u,\theta,\phi))}}{u}du\wedge d\theta\wedge d\phi
\end{align}
since on $C^{-}_{p}$, one has $v=0$. Performing a series of calculations following the transformations $(x^{0},x^{1},x^{2},x^{3})\mapsto (t,r,\theta,\varphi)\mapsto (u,v,\theta,\varphi)$, we may explictly obtain the line element on $C^{-}_{p}$
\begin{align}
\label{eq:nullmetric}
ds^{2}|_{C^{-}_{p}}=-dudv+~^{2}V_{\theta}dvd\theta+~^{2}V_{\phi}dvd\phi+^{2}g_{AB}dx^{A}dx^{B}+\left(-\frac{1}{4}\hat{g}^{uu}\right.\\\nonumber 
\left.+\frac{1}{4}~^{2}g_{AB}V^{A}V^{B}\right)dv^{2},
\end{align}
where $^{2}g_{AB}$ is the metric induced on the 2-sphere at fixed $u$ and $v=0$ (i.e., on the cone $C^{-}_{p}$) and $^{2}V_{\theta}$ ($V^{A}$ resp.,$~A=1,2$) and $^{2}V_{\phi}$ are sections of $T^{*}\mathbb{S}^{2}$ ($T\mathbb{S}^{2}$ resp.) (here $\mathbb{S}^{2}$ is a topological sphere defined $u=$constant and $v=0$). Explicit calculations yield 
\begin{align}
\sqrt{-\det(\hat{g}_{\mu\nu}(u,\theta,\phi))}|_{C^{-}_{p}}=\frac{1}{2}\sqrt{\det(g_{AB}(u,\theta,\phi))}.
\end{align}
The invariant volume induced on the boundary $2-$sphere i.e., on $C^{-}_{p}\cap \Sigma=\sigma_{p}$ is as follows
\begin{align}
\label{eq:relation}
d\sigma_{p}=\sqrt{\det(g_{AB}(u,\theta,\phi))}|_{\partial C^{-}_{p}
=\sigma_{p}}d\theta\wedge d\phi\\\nonumber =2\sqrt{-\det(\hat{g}_{\mu\nu}(u,\theta,\phi))}|\sigma_{p}d\theta\wedge d\phi.
\end{align}
Now we go back to the integrals which were being evaluated. Let us consider $I_{1}$ first
\begin{align}
2\pi I_{1}=\int_{C^{-}_{p}}<\nabla_{y}(V^{-}(y,x)\varphi(y)),\nabla_{y}\Gamma(y)>\mu_{\Gamma}(y)\\\nonumber
=\int_{C^{-}_{p}}\nabla_{\mu}(\nabla^{\mu}\Gamma(x,y)V^{-}(y,x)\varphi(y))\mu_{\Gamma}-\int_{C^{-}_{p}}\nabla_{\mu}\nabla^{\mu}\Gamma(x,y)V^{-}(y,x)\varphi(y)\mu_{\Gamma}.
\end{align}
Now let us evaluate the first term explicitly 
\begin{align}
\nabla_{\mu}(\nabla^{\mu}\Gamma(x,y) V^{-}(y,x)\varphi)=\partial_{\mu}(\nabla^{\mu}\Gamma V^{-}(y,x)\varphi)+\Gamma^{\mu}_{\mu\alpha}\nabla^{\alpha}\Gamma V^{-}(y,x)\varphi
\end{align}
which utilizing $\nabla\Gamma:=\nabla^{\mu}\Gamma\partial_{\mu}=2u\partial_{u}+2v\partial_{v}$ becomes 
\begin{align}
\nabla_{\mu}(\nabla^{\mu}\Gamma(x,y) V^{-}(y,x)\varphi)=\partial_{u}(2uV^{-}(y,x)\varphi)+\partial_{v}(2vV^{-}(y,x)\varphi)\\\nonumber 
+\frac{1}{2}\hat{g}^{\mu\nu}\partial_{u}\hat{g}_{\mu\nu}(2u)V^{-}(y,x)\varphi+\frac{1}{2}\hat{g}^{\mu\nu}\partial_{v}\hat{g}_{\mu\nu}(2v)V^{-}(y,x)\varphi\\\nonumber 
=2V^{-}(y,x)\varphi+2u\partial_{u}(V^{-}(y,x)\varphi)+2V^{-}(y,x)\varphi+2v\partial_{v}(V^{-}(y,x)\varphi)\\\nonumber 
+\frac{1}{2}\hat{g}^{\mu\nu}\partial_{u}\hat{g}_{\mu\nu}(2u)V^{-}(y,x)\varphi+\frac{1}{2}\hat{g}^{\mu\nu}\partial_{v}\hat{g}_{\mu\nu}(2v)V^{-}(y,x)\varphi.
\end{align}
Now note that $v=0$ on $C^{-}_{p}$ and therefore the previous expression becomes
\begin{align}
\nabla_{\mu}(\nabla^{\mu}\Gamma(x,y) V^{-}(y,x)\varphi)=2V^{-}(y,x)\varphi+2u\partial_{u}(V^{-}(y,x)\varphi)\nonumber+2V^{-}(y,x)\varphi\\\nonumber 
+u\hat{g}^{\mu\nu}\partial_{u}\hat{g}_{\mu\nu}V^{-}(y,x)\varphi\\\nonumber 
=2u\partial_{u}(V^{-}(y,x)\varphi)+u\hat{g}^{\mu\nu}\partial_{u}\hat{g}_{\mu\nu}V^{-}(y,x)\varphi+4V^{-}(y,x)\varphi\\\nonumber 
=2u\frac{\partial_{u}(\sqrt{-\det(\hat{g}_{\mu\nu}(u,\theta,\phi))}V^{-}(y,x)\varphi)}{\sqrt{-\det (\hat{g}_{\mu\nu}(u,\theta,\phi))}}+4V^{-}(y,x)\varphi.
\end{align}
Therefore we obtain 
\begin{align}
2\pi I_{1}=\int_{C^{-}_{p}}\nabla_{\mu}(\nabla^{\mu}\Gamma(x,y)V^{-}(y,x)\varphi(y))\mu_{\Gamma}\nonumber-\int_{C^{-}_{p}}\nabla_{\mu}\nabla^{\mu}\Gamma(x,y)V^{-}(y,x)\varphi(y)\mu_{\Gamma}\\\nonumber
=\int_{C^{-}_{p}}u\partial_{u}(2\sqrt{-\det(\hat{g}_{\mu\nu}(u,\theta,\phi))}V^{-}(y,x)\varphi(y))\frac{du\wedge d\theta\wedge d\phi}{u}\\\nonumber 
+\int_{C^{-}_{p}}(4-\hat\Box \Gamma(x,y))V^{-}(y,x)\varphi(y) \mu_{\Gamma}(y)\\\nonumber 
=\int_{C^{-}_{p}}\partial_{u}(2\sqrt{-\det(\hat{g}_{\mu\nu}(u,\theta,\phi))}V^{-}(y,x)\varphi(y))du\wedge d\theta\wedge d\phi\\\nonumber 
+\int_{C^{-}_{p}}(4-\hat\Box \Gamma(x,y))V^{-}(y,x)\varphi(y) \mu_{\Gamma}(y)\\\nonumber 
=-\int_{C^{-}_{p}}2\sqrt{-\det(\hat{g}_{\mu\nu}(u,\theta,\phi))}V^{-}(y,x)\varphi d\theta\wedge d\phi\\\nonumber 
+\int_{\sigma_{p}}(4-\hat\Box_{y} \Gamma(x,y))V^{-}(y,x)\varphi(y) \mu_{\Gamma}(y)\\\nonumber 
=-\int_{\sigma_{p}}V^{-}(y,x)\varphi(y) d\sigma_{p}+\int_{\sigma_{p}}(4-\hat\Box_{y} \Gamma(x,y))V^{-}(y,x)\varphi(y) \mu_{\Gamma}(y).
\end{align}
Here we have used equation (\ref{eq:relation}) and the future direction is considered to be positive for the null normal vector of the boundary sphere $\sigma_{p}=\partial C^{-}_{p}$. 
Therefore, the integral $I_{1}+I_{2}$ now becomes (from equation (\ref{eq:split}))
\begin{align}
\frac{1}{2\pi}\int_{C^{-}_{p}} {*}\left(V^{+}(x,y)\nabla_{y}\varphi(y)-\varphi(y)\nabla_{y} V^{+}(x,y)\right)=I_{1}+I_{2}\\\nonumber 
=-\frac{1}{2\pi}\int_{\sigma_{p}}V^{-}(y,x)\varphi(y) d\sigma_{p}+\frac{1}{2\pi}\int_{\sigma_{p}}(4-\hat\Box_{y} \Gamma(x,y))V^{-}(y,x)\varphi(y) \mu_{\Gamma}\\\nonumber 
+\frac{1}{2\pi}\int_{C^{-}_{p}}\varphi(y)\left((\hat\Box_{y}\Gamma(y,x)-4)V_{0}(y,x)+PU(y,x)\right)\mu_{\Gamma}(y).
\end{align}
Now on $C^{-}_{p}$, we have $V^{-}(y,x)=V_{0}$ since data of $V^{-}(y,x)$ on $C^{-}_{p}$ is the initial data for the characteristic initial value problem (\ref{eq:character}). Using $V^{-}(y,x)|_{C^{-}_{p}}=V_{0}$, we see that the last two terms in the previous integral cancels point-wise yielding 
\begin{align}
\frac{1}{2\pi}\int_{C^{-}_{p}} {*}\left(V^{+}(x,y)\nabla_{y}\varphi(y)-\varphi(y)\nabla_{y} V^{+}(x,y)\right)\\\nonumber =-\frac{1}{2\pi}\nonumber\int_{\sigma_{p}}V^{-}(y,x)\varphi(y) d\sigma_{p}+\frac{1}{2\pi}\int_{C^{-}_{p}}PU(x,y)\varphi(y)\mu_{\Gamma}(y)\\\nonumber
=-\frac{1}{2\pi}\int_{\sigma_{p}}V^{+}(x,y)\varphi(y)d\sigma_{p}+\frac{1}{2\pi}\int_{C^{-}_{p}}PU(x,y)\varphi(y)\mu_{\Gamma}(y)
\end{align}
The tail contribution $\frac{1}{2\pi}\int_{C^{-}_{p}} {*}\left(V^{+}(x,y)\nabla_{y}\varphi(y)-\varphi(y)\nabla_{y} V^{+}(x,y)\right)\\+\frac{1}{2\pi}\int_{\sigma_{p}}V^{+}(x,y)\varphi(y) d\sigma_{p}$ therefore reduces to 
\begin{align}
\frac{1}{2\pi}\int_{C^{-}_{p}} {*}\left(V^{+}(x,y)\nabla_{y}\varphi(y)-\varphi(y)\nabla_{y} V^{+}(x,y)\right)\nonumber+\frac{1}{2\pi}\int_{\sigma_{p}}V^{+}(x,y)\varphi(y) d\sigma_{p}\\\nonumber 
=-\frac{1}{2\pi}\int_{\sigma_{p}}V^{+}(x,y)\varphi(y)d\sigma_{p}+\frac{1}{2\pi}\int_{C^{-}_{p}}PU(x,y)\varphi(y)\mu_{\Gamma}(y)+\frac{1}{2\pi}\int_{\sigma_{p}}V^{+}(x,y)\varphi(y)d\sigma_{p}\\\nonumber
=\frac{1}{2\pi}\int_{C^{-}_{p}}PU(x,y)\varphi(y)\mu_{\Gamma}(y)\\\nonumber 
=\frac{1}{2\pi}\int_{C^{-}_{p}}\hat\Box_{y} U(x,y)\varphi(y)\mu_{\Gamma}(y)
\end{align}
Putting everything together, we now obtain the desired integral equation for $\varphi$ which does not include the tails terms. The following theorem summarizes the result.\\
\textbf{Theorem 1:} \textit{Let $\varphi\in C^{\infty}(\Omega)$ and assume that $p\in D^{+}(\Sigma)-\{\Sigma\}$. Then $\varphi$ at point $p$ ($\equiv x$ in the local coordinate system) satisfies the following integral equation
\begin{align}
\label{eq:repn}
\varphi(x)=\frac{1}{2\pi}\int_{C^{-}_{p}}U(x,y)\varphi^{5}(y)\mu_{\Gamma}(y)+\frac{1}{2\pi}\int_{C^{-}_{p}}\hat\Box_{y} U(x,y)\varphi(y)\mu_{\Gamma}(y)\\\nonumber +\frac{1}{2\pi}\int_{\sigma_{p}}\left(2U(x,y)<\xi(y),\nabla_{y} \varphi(y)>+U(x,y)\Theta(y) \varphi(y)\right)d\sigma_{p}(y),
\end{align}
where $\mu_{\Gamma}$ and $\Theta$ are defined previously. 
} 

Notice an extremely important fact that even though the final expression we have obtained contains integrals over the mantle of the past light cone and its two dimensional intersection with the initial Cauchy hypersurface, it does not imply that Huygen's principle holds. One might just for the time being neglect the nonlinear term and consider the linear wave propagation, then 
the unknown $\varphi$ appears within the integral ($\frac{1}{2\pi}\int_{C^{-}_{p}}\hat\Box_{y} U(x,y)\varphi(y)\mu_{\Gamma}(y)$). If Huygen's principle were to hold, then at the linear level (with no nonlinear source term), the expression would only contain the Cauchy data. In Minkowski space, we have $U(x,y)\equiv 1=>\hat\Box_{y} U(x,y)\equiv 0$ and therefore this Huygen's violating term vanishes identically. For the present purpose of proving an $L^{\infty}$ control, this linear term does not cause any difficulty if we assume sufficient regularity of the background spacetime metric. Via explicit calculations, we will establish that under such circumstances, $\hat\Box_{y} U(x,y)$ is bounded.\\
Notice an important fact about integration on $C^{-}_{p}$. In the spherical null or lightcone coordinates $(u,v,\theta,\phi)$, the null cone $C^{-}_{p}$ is defined by $v=0$. The integral of a function $f$ on $C^{-}_{p}$ is written as 
\begin{align}
\int_{C^{-}_{p}}f \mu_{\hat{g}}|_{C^{-}_{p}}=\int_{\mathbb{S}^{2}}\int_{0}^{u_{1}}f\sqrt{-\det(\hat{g}_{\mu\nu}(u,\theta,\phi))}du\wedge d\theta\wedge d\phi
\end{align}
which upon using the fact that $\sqrt{-\det(\hat{g}_{\mu\nu}(u,\theta,\phi))}|_{C^{-}_{p}}=\frac{1}{2}\sqrt{\det(g_{AB}(u,\theta,\phi))}$ becomes 
\begin{align}
\int_{C^{-}_{p}}f \mu_{\hat{g}}|_{C^{-}_{p}}=\frac{1}{2}\int_{\mathbb{S}^{2}}\int_{0}^{u_{1}}f\sqrt{\det(g_{AB}(u,\theta,\phi))} du\wedge d\theta\wedge d\phi.
\end{align}
Now we will extract the conformal factor $u^{2}$ from $\sqrt{\det(g_{AB}(u,\theta,\phi))}$ to yield 
\begin{align}
\int_{C^{-}_{p}}f \mu_{\hat{g}}|_{C^{-}_{p}}=\frac{1}{2}\int_{\mathbb{S}^{2}}\int_{0}^{u_{1}}fu^{2}\sqrt{\det(\tilde{g}_{AB}(u,\theta,\phi))} du\wedge d\theta\wedge d\phi.
\end{align}
In the view of the fact that we are in a geodesically convex domain and assuming sufficient regularity of the spacetime metric, we will explicitly show in section 5.5 that $\sqrt{\det(\tilde{g}_{AB}(u,\theta,\phi))}$ is harmless (satisfies a point-wise $O(1)$ estimate at worst). For example, if $f(u)=\frac{1}{u^{\kappa}}$, then 
\begin{align}
\int_{C^{-}_{p}}f \mu_{\hat{g}}|_{C^{-}_{p}}=\frac{1}{2}\int_{\mathbb{S}^{2}}\int_{0}^{u_{1}}\nonumber\frac{u^{2}}{u^{\kappa}}\sqrt{\det(\tilde{g}_{AB}(u,\theta,\phi))}du\wedge d\theta\wedge d\phi\lesssim |u_{1}|^{3-\kappa},\kappa<3.
\end{align}
We will frequently use this type of estimate while performing integration on the cone $C^{-}_{p}$.

\begin{center}
\begin{figure}
\begin{center}
\includegraphics[width=13cm,height=17cm]{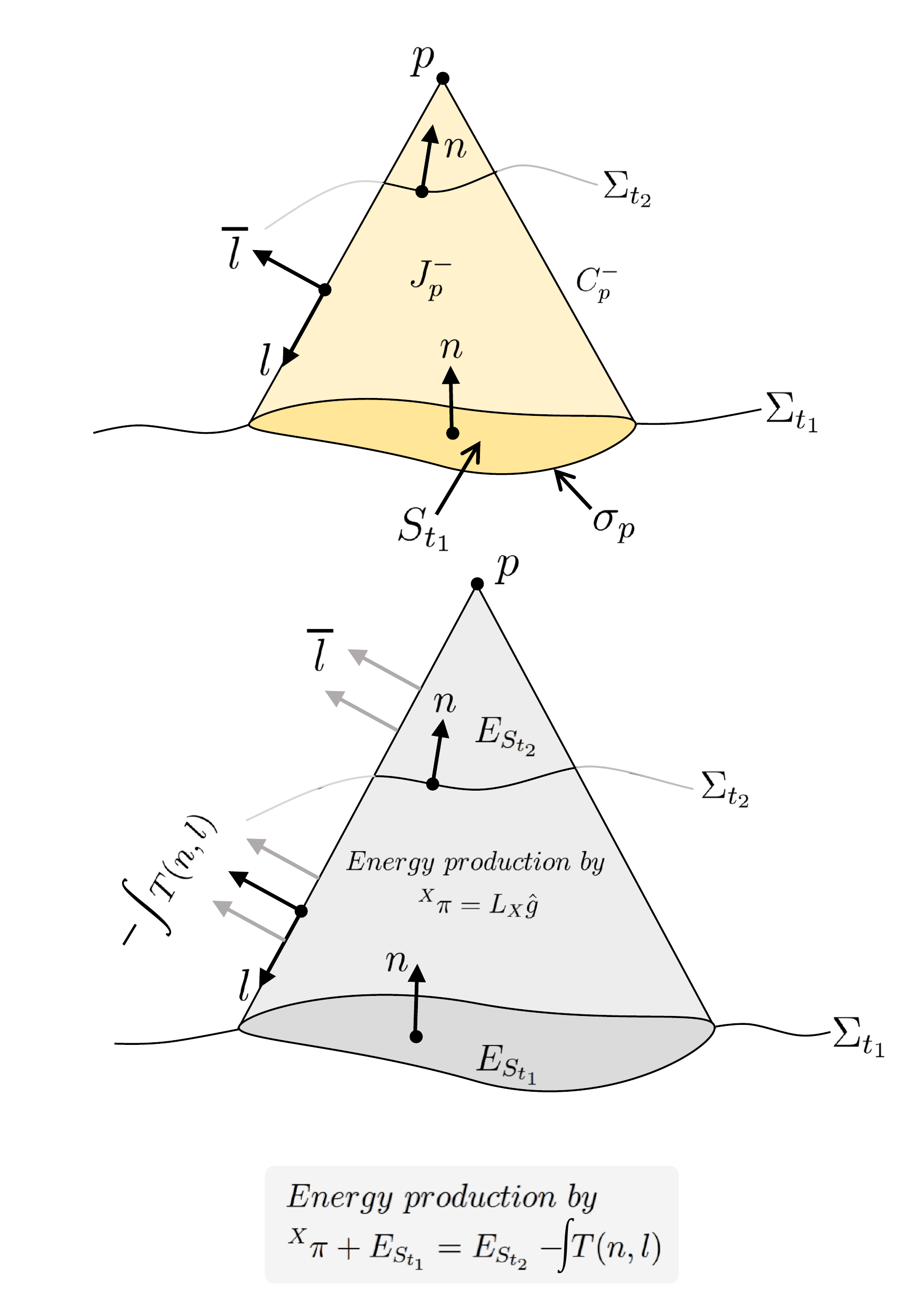}
\end{center}
\begin{center}
\caption{The null geometry associated to point $p\in M$ and energy conservation corresponding to the the vector field $X$. Clearly Energy at the initial hypersurface $S_{t_{1}} (E_{S_{t_{1}}})+$ energy production by the strain tensor $L_{X}\hat{g}=E_{S_{t_{2}}}+$ Energy flowing across the null cone. The coordinate chart is normal at $p(\equiv 0)$.}
\label{fig:fig1}
\end{center}
\end{figure}
\end{center}

\section{Energy Estimates}
In this section we derive an estimate of energy corresponding to the unit time like vector field $n$ orthogonal to the spacelike hypersurfaces $\Sigma$. In addition to this basic energy estimate, we also consider three additional quasi-local (to be defined later) time-like conformal Killing vector fields. As we mentioned previously, we need these three additional estimates only for this critical nonlinearity. For sub-critical nonlinearities (the nonlinear exponent may take value up to $5-\delta,~\delta>0$), the basic estimate of energy corresponding to $n=\frac{1}{N}(\partial_{t}-Y)$ and the integral equation derived in the previous section are sufficient to yield the desired spacetime $L^{\infty}$ estimate of $\varphi$. 
We start with the stress-energy tensor $T$ that is derivable from the action $S$ associated with the scalar field $\varphi$
\begin{align}
T_{\mu\nu}:=-\frac{2}{\sqrt{-\det(\hat{g})_{\alpha\beta}}}\frac{\delta S}{\delta \hat{g}^{\mu\nu}},
\end{align}
where $S$ is as follows 
\begin{align}
S:=\frac{1}{2}\int_{M}\left(-\hat{g}^{\mu\nu}\nabla_{\mu}\varphi\nabla_{\nu}\varphi-\frac{1}{3}\phi^{6}\right)\mu_{\hat{g}}.
\end{align}
An explicit calculation yields
\begin{align}
T_{\mu\nu}:=\nabla_{\mu}\varphi\nabla_{\nu}\varphi-\frac{1}{2}\nabla_{\alpha} \varphi\nabla^{\alpha}\varphi \hat{g}_{\mu\nu}-\frac{1}{6}\varphi^{6}\hat{g}^{\mu\nu},
\end{align}
Divergence of the stress-energy tensor vanishes by virtue of the equation of motion $\hat{g}^{\mu\nu}\nabla_{\mu}\nabla_{\nu}\varphi=\varphi^{5}$
\begin{align}
\nabla{\nu}T^{\mu\nu}=(\nabla_{\nu}\nabla^{\nu}\varphi)\nabla^{\mu}\varphi-\varphi^{5}\nabla^{\mu}\varphi=(\nabla_{\nu}\nabla^{\nu}\varphi-\varphi^{5})\nabla^{\mu}\varphi=0.
\end{align}
Now using this divergence free property of the stress-energy tensor (whenever the equation of motion is satisfied), we may derive several conservation laws. Let us consider a vector field $X$ and evaluate the following entity  
\begin{align}
\nabla_{\nu} (X_{\mu}T^{\mu\nu})=\nabla_{\nu}X_{\mu}T^{\mu\nu}+X_{\mu}\nabla_{\nu}T^{\mu\nu}=\nabla_{\mu}X_{\nu}T^{\mu\nu}
\end{align}
which upon integration over the truncated  past light cone of $p$ (see figure \ref{fig:fig1}) and an application of the Stokes theorem (assuming the domain to be Stokes regular) yields 
\begin{align}
\label{eq:conservation}
\int_{C^{-,t_{2},t_{1}}_{p}}T(X,l)\mu_{\hat{g}}|_{C^{-,t_{2},t_{1}}_{p}}+\int_{S_{t_{1}}}T(X,n)\mu_{\hat{g}}|_{S_{t_{1}}}-\int_{S_{t_{2}}}T(X,n)\mu_{\hat{g}}|_{S_{t_{2}}}\\\nonumber
=\int_{D^{-,t_{2},t_{1}}_{p}}\left(\nabla_{\mu}X_{\nu}T^{\mu\nu}\right)\mu_{\hat{g}}\\\nonumber 
=\frac{1}{2}\int_{C^{-,t_{2},t_{1}}_{p}}\left(\nabla_{\mu}X_{\nu}+\nabla_{\nu}X_{\mu}\right)T^{\mu\nu}\mu_{\hat{g}}
\end{align}
using symmetry of $T^{\mu\nu}$. Here $C^{-,t_{2},t_{1}}_{p}, S_{t_{1}}$, and $S_{t_{2}}$, and denote the mantle of the truncated causal past $D^{-,t_{2},t_{1}}_{p}$, the intersection of the causal past with the constant $t$ hypersurfaces $\Sigma_{t_{1}}$ and $\Sigma_{t_{2}}$, respectively. See the figure (\ref{fig:fig1}) for clarification. Notice an important fact that if $X$ is a Killing vector field, then the strain tensor $^{X}\pi_{\mu\nu}:=(L_{X}\hat{g})_{\mu\nu}=\nabla_{\mu}X_{\nu}+\nabla_{\nu}X_{\mu}$ vanishes and in those particular circumstances, we have a true conservation law.
\subsection{Elementary energy estimate using the vector field $`n=\frac{1}{N}(\partial_{t}-Y)$'}
The equation (\ref{eq:conservation}) holds for any general vector field $X$ (lying in a suitable function space). However, in order to construct positive (negative) definite energies, we will only focus on future (past) directed time-like vector fields. We first consider the unit future directed time-like vector field $n$ orthogonal to the constant $t$ hypersurfaces. The energy density associated with $n$ reads 
\begin{align}
\mathcal{E}:=T(n,n)
\end{align}
which may be explicitly evaluated to yield 
\begin{align}
\label{eq:energydensity}
\mathcal{E}=\frac{1}{2N^{2}}\left((\partial_{t}\varphi)^{2}-2Y^{i}\nabla_{i}\varphi\partial_{t}\varphi+Y^{i}Y^{j}\nabla_{i}\varphi\nabla_{j}\varphi\nonumber+N^{2}g^{ij}\nabla_{i}\varphi\nabla_{j}\varphi\right)\\\nonumber
+\frac{1}{6}\varphi^{6}\\\nonumber 
=\frac{1}{2}|L_{n}\varphi|^{2}+\frac{1}{2}|\nabla\varphi|^{2}_{g}+\frac{1}{6}\varphi^{6}\\
=\frac{1}{2}m^{2}+\frac{1}{2}|\nabla\varphi|^{2}_{g}+\frac{1}{6}\varphi^{6},
\end{align}
where we denote the velocity variable $L_{n}\varphi$ associated to $\varphi$ as $m$ (note that in case of gravity, this role is played by the second fundamental form in the Lagrangian language) and $g^{-1}:=g^{ij}\partial_{i}\otimes \partial_{j}$. Clearly $\hat{g}_{ij}=g_{ij}$ but $\hat{g}^{ij}=g^{ij}-\frac{Y^{i}Y^{j}}{N^{2}}$ (note the expression of metric (\ref{eq:metric}) in section 2). Note that one may decompose the equation of motion $\nabla^{\mu}\nabla_{\mu}\varphi=\varphi^{5}$ into two first order (in time) equations in terms of $\varphi$ and $m$ (the so called $3+1$ decomposition of the field equations). This may be obtained directly from the action 
\begin{align}
S=\frac{1}{2}\int_{M}\left(-\hat{g}^{\mu\nu}\nabla_{\mu}\varphi\nabla_{\nu}\varphi-\frac{1}{3}\phi^{6}\right)\mu_{\hat{g}}\\\nonumber
=\frac{1}{2}\int_{I\subset \mathbb{R}}\int_{\Sigma_{t}}\left(\frac{1}{N^{2}}(\partial_{t}\varphi-L_{Y}\varphi)^{2}-g^{ij}\nabla_{i}\varphi\nabla_{j}\varphi-\frac{1}{3}\varphi^{6}\right)N\mu_{g}d^{3}xdt\\\nonumber 
=\int_{I\subset \mathbb{R}}\int_{\Sigma_{t}}\left(m\partial_{t}\varphi-mL_{Y}\varphi-\frac{Nm^{2}}{2}-\frac{N}{2}g^{ij}\nabla_{i}\varphi\nabla_{j}\varphi-\frac{N}{6}\varphi^{6}\right)\mu_{g}d^{3}xdt.
\end{align}
Variation of $S$ with respect to $m$ and $\varphi$ yields the following coupled first order (in time) equations
\begin{align}
\label{eq:lagrange}
\partial_{t}\varphi=Nm+L_{Y}\varphi\\
\partial_{t}m=\nabla^{i}N\nabla_{i}\varphi+Ng^{ij}\nabla_{i}\nabla_{j}\varphi+L_{Y}m-N\varphi^{5}+m tr_{g}k,
\end{align}
where $k$ is the second fundamental form of the constant $t$ hypersurface $\Sigma_{t}$. We derived this set of equations because we want to study the gravitational dynamics coupled to critically nonlinear wave fields in the future. It is very straightforward to couple (in case of minimal coupling) this system with the Einstein equations. Noting $\mu_{\hat{g}}=N\mu_{g}$, the energy for this system in terms of $(m,\varphi)$ is defined as follows 
\begin{align}
E_{\Sigma_{t}}:=\int_{\Sigma_{t}}\left(\frac{1}{2}m^{2}+\frac{1}{2}|\nabla\varphi|^{2}_{g}+\frac{1}{6}\varphi^{6}\right)N\mu_{g}.
\end{align}
Clearly we observe that the $\dot{H}^{1}(\Sigma_{t})\times L^{2}(\Sigma_{t})$ norm of $(\varphi,m)$ is controlled by the energy. Obviously we have $\dot{H}^{1}(\Sigma_{t})\hookrightarrow L^{6}(\Sigma_{t})$ and therefore the $\varphi^{6}$ term is dominated by the second term in the energy expression. 

The equation (\ref{eq:conservation}) with $X=n$ yields 
\begin{align}
\int_{C^{-,t_{2},t_{1}}_{p}}T(n,l)\mu_{\hat{g}}|_{C^{-,t_{2},t_{1}}_{p}}+\int_{S_{t_{1}}}T(n,n)\mu_{\hat{g}}|_{S_{t_{1}}}-\int_{S_{t_{2}}}T(n,n)\mu_{\hat{g}}|_{S_{t_{2}}}\\\nonumber
=\int_{D^{-,t_{2},t_{1}}_{p}}\left(\nabla_{\mu}n_{\nu}T^{\mu\nu}\right)\mu_{\hat{g}}
\end{align}
i.e., 
\begin{align}
\label{eq:energy}
E_{S_{t_{2}}}-\int_{C^{-,t_{2},t_{1}}_{p}}T(n,l)\mu_{\hat{g}}|_{C^{-,t_{2},t_{1}}_{p}}
=E_{S_{t_{1}}}-\int_{t_{1}}^{t_{2}}\left(\int_{S_{t}}\left(\nabla_{\mu}n_{\nu}T^{\mu\nu}\right)N\mu_{g}\right) dt.
\end{align}
In the previous expression, we have the problematic term $\int_{t_{1}}^{t_{2}}\int_{S_{t}}\left(\nabla_{\mu}n_{\nu}T^{\mu\nu}\right)N\mu_{g}$ which may be written as the strain tensor of $n$ contracted with $T^{\mu\nu}$ due to symmetry of the later, that is,\\  $\int_{t_{1}}^{t_{2}}\int_{S_{t}}\left(\nabla_{\mu}n_{\nu}T^{\mu\nu}\right)N\mu_{g}=\frac{1}{2}\int_{t_{1}}^{t_{2}}\int_{S_{t}}\left((\nabla_{\mu}n_{\nu}+\nabla_{\nu}n_{\mu})T^{\mu\nu}\right)N\mu_{g}$. Since, the strain tensor of $n$ is essentially tied to the background spacetime, we need to somehow show that $T^{\mu\nu}$ is bounded component-wise and point-wise by the energy density. We use the following well known trick to verify that this is indeed the case
\begin{align}
|T^{\mu\nu}|=|\int_{0}^{1}\frac{\partial T^{\mu\nu}(s\varphi,s\partial \varphi)}{\partial s}ds|\leq\int_{0}^{1}\left(|\frac{\partial T^{\mu\nu}}{\partial\varphi}||\varphi|+|\frac{\partial T^{\mu\nu}}{\partial (\partial\varphi)}||\partial\varphi|\right)ds|.
\end{align}
Noting $|\frac{\partial T(s\varphi,s\partial\varphi)}{\partial\varphi}|\lesssim \varphi^{6}$ and $|\frac{\partial T(s\varphi,s\partial\varphi)}{\partial(\partial\varphi)}|\lesssim m^{2}+|\nabla\varphi|^{2}_{g}$, we may write 
\begin{align}
|T^{\mu\nu}|\lesssim \mathcal{E},
\end{align}
where the involved constants are harmless. Now if we further assume that the strain tensor $\nabla_{\mu}n_{\nu}+\nabla_{\nu}n_{\mu}$ associated to $n$ is bounded, then we may write the equation (\ref{eq:energy}) as an integral equation for $E$. Noting $\nabla_{\mu}n_{\nu}+\nabla_{\nu}n_{\mu}=L_{n}\hat{g}$, an explicit calculation for the strain tensor of $n$ yields 
\begin{align}
(L_{n}\hat{g})_{00}=2\nabla_{0}n_{0}=2Y^{k}\nabla_{k}N-2k_{ij}Y^{i}Y^{j}\\\nonumber 
(L_{n}\hat{g})_{0i}=(L_{n}\hat{g})_{i0}=\nabla_{0}n_{i}+\nabla_{i}n_{0}=\nabla_{i}N-2k_{ij}Y^{j}\\\nonumber 
(L_{n}\hat{g})_{ij}=\nabla_{i}n_{j}+\nabla_{j}n_{i}=\hat{g}(\nabla_{i}n,\partial_{j})+\hat{g}(\nabla_{j}n,\partial_{i})=-2k_{ij}.
\end{align}
In the globally hyperbolic background spacetime, we may further assume the following regularity estimate 
\begin{align}
\max(||Y||_{L^{\infty}(D^{-}_{p})}, ||\nabla N||_{L^{\infty}(D^{-}_{p})}, ||k||_{L^{\infty}(D^{-}_{p})})\leq C,
\end{align}
where $L^{\infty}(D^{-}_{p})$ denotes the spacetime point-wise norm. Therefore, (\ref{eq:energy}) may be written as follows 
\begin{align}
\label{eq:energy_inequality}
E_{S_{t_{2}}}-\int_{C^{-,t_{2},t_{1}}_{p}}T(n,l)\mu_{\hat{g}}|_{C^{-,t_{2},t_{1}}_{p}}\leq E_{S_{t_{1}}}+C\int_{t_{1}}^{t_{2}}E_{S_{t}}dt
\end{align}
where $0<C<\infty$ depends on spacetime $L^{\infty}$ norm of $N,N^{-1},g,g^{-1},k$, and $Y$. Note an extremely important fact that these assumed bounds on the spacetime entities certainly do not hold true in the gravitational problem (or when this scalar field is coupled to gravity). Instead one needs to control these terms simultaneously making the problem tremendously difficult. In a few special cases, one may control all the associated norms simultaneously by assuming a certain smallness condition on the data. Here we do not have to worry about such things. 

Now we need to evaluate the term $-\int_{C^{-,t_{2},t_{1}}_{p}}T(n,l)\mu_{\hat{g}}|_{C^{-,t_{2},t_{1}}_{p}}$ and show that this has a definite sign. Physically, it is not hard to see that this term is essentially the measure of energy flux flowing out through the null boundary. Now, for a physically reasonable matter source (i.e., one with positive definite energy), this flux term will always be positive since following causality the energy flux can not flow into the cone through the boundary (see the figure (\ref{fig:fig1}) for a physical description). However we will show explicitly that this term is indeed positive definite. Since the energy and energy flux integrals are diffeomorphism invariant, we may evaluate these in the null basis introduced in section (2) (\ref{eq:null1}-\ref{eq:null3}) 
\begin{align}
\hat{g}(l,l)=0,~\hat{g}(\bar{l},\bar{l})=0,~n=-l+\bar{l}, \hat{g}(\lambda_{i},\lambda_{j})=\delta_{ij},\\\nonumber
\hat{g}(\lambda_{i},l)=\hat{g}(\lambda_{i},\bar{l})=0.
\end{align}
Explicit calculations yield 
\begin{align}
T(n,l)=T(-l+\bar{l},l)=T(l,\bar{l})-T(l,l),\\\nonumber
T(l,\bar{l})=\hat{g}(l,\partial\varphi)\hat{g}(\bar{l},\partial\varphi)-\frac{1}{4}\hat{g}(\partial\varphi,\partial\varphi)-\frac{1}{12}\varphi^{6},\\\nonumber 
T(l,l)=\hat{g}(l,\partial\varphi)\hat{g}(l,\partial\varphi),\\
T(\bar{l},\bar{l})=\hat{g}(\bar{l},\partial\varphi)\hat{g}(\bar{l},\partial\varphi).
\end{align}
Now we split $\partial \varphi:=(\hat{g}^{\mu\nu}\partial_{\nu}\varphi)\partial_{\mu}\in$ sections$\{TM\}$ as follows 
\begin{align}
\partial\varphi=2\hat{g}(\partial\varphi,l)\bar{l}+2\hat{g}(\partial\varphi,\bar{l})l+\sum_{i=1}^{2}\hat{g}(\partial\varphi,\lambda_{i})\lambda_{i}
\end{align}
utilizing which we obtain 
\begin{align}
\hat{g}(\partial\varphi,\partial\varphi)=4\hat{g}(\partial \varphi,l)\hat{g}(\partial\varphi,\bar{l})+\sum_{i=1}^{2}|\hat{g}(\partial\varphi,\lambda_{i})|^{2}.
\end{align}
which yields the following expression for $T(l,\bar{l})$
\begin{align}
T(l,\bar{l})=-\frac{1}{4}|g(\lambda,\partial\varphi)|^{2}-\frac{1}{12}\varphi^{6}
\end{align}
Therefore, $T(n,l)$ has the following expression 
\begin{align}
-T(n,l)=|\hat{g}(l,\partial\varphi)|^{2}+\frac{1}{4}\sum_{i=1}^{2}|\hat{g}(\partial\varphi,\lambda_{i})|^{2}+\frac{1}{12}\varphi^{6}\geq 0.
\end{align}
This is a positive semi-definite entity and precisely expresses the flux going transversal to the lightcone, but, not along the light cone since, terms involving $\hat{g}(\partial\varphi,\bar{l})$ cancels out in the process. Using this inequality, we may conclude that the diffeomorphism invariant entity $\int_{C^{-,t_{2},t_{1}}_{p}}T(n,l)\mu_{\hat{g}}|_{C^{-,t_{2},t_{1}}_{p}}$ satisfies 
\begin{align}
-\int_{C^{-,t_{2},t_{1}}_{p}}T(n,l)\mu_{\hat{g}}|_{C^{-,t_{2},t_{1}}_{p}}\geq 0.
\end{align}
Therefore the energy inequality (\ref{eq:energy_inequality}) becomes 
\begin{align}
E_{S_{t_{2}}}\leq E_{S_{t_{1}}}+C\int_{t_{1}}^{t_{2}}E_{S_{t}}dt
\end{align}
which upon using Gr\"onwall's inequality yields \begin{align}
\label{eq:energy_bound}
E_{S_{t_{2}}}\leq E_{S_{t_{1}}}e^{C(t_{2}-t_{1})}.
\end{align}
Now if one goes back to the energy inequality (\ref{eq:energy_inequality}) and substitutes the energy bound (\ref{eq:energy_bound}), the following is obtained 
\begin{align}
E_{S_{t_{2}}}-\int_{C^{-,t_{2},t_{1}}_{p}}T(l,n)\mu_{\hat{g}}|_{C^{-,t_{2},t_{1}}_{p}} \leq E_{S_{t_{1}}}+CE_{S_{t_{1}}}\int_{t_{1}}^{t_{2}}e^{C(t-t_{1})}dt\\\nonumber
=E_{S_{t_{1}}}e^{C(t_{2}-t_{1})},
\end{align}
that is, 
\begin{align}
-\int_{C^{-,t_{2},t_{1}}_{p}}T(l,n)\mu_{\hat{g}}|_{C^{-,t_{2},t_{1}}_{p}}\leq E_{S_{t_{1}}}e^{C(t_{2}-t_{1})}-E_{S_{t_{2}}}.
\end{align}
We may set $t_{2}=0$ (and thus $t_{1}<0$) and write the previous inequality as 
\begin{align}
-\int_{C^{-,t_{2},t_{1}}_{p}}T(l,n)\mu_{\hat{g}}|_{C^{-,t_{2},t_{1}}_{p}}\leq E_{S_{t_{1}}}e^{-Ct_{1}}-E_{S_{0}}
\end{align}
Now we observe from (\ref{eq:energy_bound}) that energy $E$ can not blow up at the vertex i.e., at $t=0$. Therefore $0<E_{S_{0}}=E_{0}<\infty$. Taking $t_{1}\to 0$ (i.e., if we make the height of the light cone sufficiently small), we conclude 
\begin{align}
\lim_{t_{1}\to 0} -\int_{C^{t_{1}}_{p}}T(l,n)\mu_{\hat{g}}|_{C^{t_{1}}_{p}}=0
\end{align}
or more explicitly
\begin{align}
\label{eq:vanishing}
\lim_{t_{1}\to 0}\int_{C^{t_{1}}_{p}}\left(|\hat{g}(l,\partial\varphi)|^{2}+\frac{1}{4}\sum_{i=1}^{2}|\hat{g}(\partial\varphi,\lambda_{i})|^{2}+\frac{1}{12}\varphi^{6}\right)\mu_{\hat{g}}|_{C^{t_{1}}_{p}}=0.
\end{align}
Denoting positive continuous functions that vanish as one approaches $t_{1}=0$ by $zo(t_{1})$, we observe
\begin{align}
\label{eq:vanishing2}
\int_{C^{t_{1}}_{p}}|l(\varphi)|^{2}\mu_{\hat{g}}|_{C^{t_{1}}_{p}}=zo(t_{1}),\int_{C^{t_{1}}_{p}}|\hat{g}(\partial\varphi,\lambda_{i}|^{2}\mu_{\hat{g}}|_{C^{t_{1}}_{p}}=zo(t_{1})\\
\label{eq:vanishing3}
\int_{C^{t_{1}}_{p}}\varphi^{6}\mu_{\hat{g}}|_{C^{t_{1}}_{p}}=zo(t_{1}).
\end{align}
This essentially demonstrates the fact that the flux integral may be made arbitrarily small by reducing the height of the lightcone. 
We will make use of this important result in the next energy estimates. This however does not imply that the same estimate holds if we consider a light cone emanating from any point lying within causal past of $C^{t_{1}}_{p}$ and extending up to the hypersurface $S_{t_{1}}$. In order to obtain estimates on any interior cone (which we will require will be clarified in later sections), we will need to use the energies associated with the three additional approximate conformal Killing vector fields. Before moving to the next set of energy estimates, let us introduce some concepts which will prove to be useful. 
\subsection{Concept of geodesic normal charts and parallel propagated frames}
An important result of Riemannian geometry is that on any Riemannian or pseudo-Riemannian manifold, one can construct a geodesic normal coordinate chart on a neighbourhood of an arbitrary point $p$ (the exponential map from the tangent space $T_{p}M$ to the manifold $M$ is a diffeomorphism in the normal neighbourhood). The size of the geodesic normal neighbourhood (i.e., the injectivity radius) depends on the magnitude of the Riemann tensor in a suitable sense. There are concrete results about the relationship between the size of the Riemann curvature tensor (in a suitable function space of course) and the injectivity radius. The norm of the spacetime Riemann tensor may be defined in terms of the norm of the electric and magnetic components associated with the Weyl tensor (trace-free part of the Riemann tensor) and the norm of the Ricci tensor (trace part of the Riemann tensor). For example, one may simply construct a gauge invariant point-wise norm of the Riemann tensor simply by contracting with a spacetime Riemannian metric. Let $T$ be a future directed unit timelike vector field orthogonal to a family of spacelike hypersurfaces foliating the spacetime $M$. We may construct a Riemannian metric as follows 
\begin{align}
\zeta:=\hat{g}+2T\otimes T.
\end{align}
The Let $C^{-}_{p}\in M$ be a null cone with vertex at $p$. The point-wise norm of the Riemann curvature while restricted to $C^{-}_{p}$ may be defined as $||Riem||_{L^{\infty}(C^{-}_{p})}:=\sup_{x\in C^{-}_{p}}\sqrt{R_{\mu\nu\alpha\beta}R_{\delta\kappa\lambda\gamma}\zeta^{\mu\delta}\zeta^{\nu\kappa}\zeta^{\alpha\lambda}\zeta^{\beta\gamma}}$.
We will not present details about the relationship between the Riemann curvature and injectivity radius but instead refer to the theorem $1.1$ of \cite{chen2008injectivity}. \\
\textbf{Theorem}\cite{chen2008injectivity} \textit{Let $(M,\hat{g})$ be a time orientable Lorentzian `$3+1$' differentiable manifold. Consider an observer $(p,T)$ where $p\in M$ and $T$ is a future directed time-like unit vector belonging to $T_{p}M$. Assume that the exponential map $\exp_{p}$ is defined in a Riemannian ball $B_{T}(0,r)\subset T_{p}M$ and the Riemann curvature satisfies 
\begin{align}
\label{eq:riembound}
\sup_{\gamma}|Riem(\hat{g})|_{T_{\gamma}}\leq \frac{1}{r^{2}},
\end{align}
where supremum is taken over every $\hat{g}-$geodesic $\gamma$ initiating from a vector lying in $B_{T}(0,r)\subset T_{p}M$, then there exists a uniform constant $C$ such that the following is satisfied by the injectivity radius 
\begin{align}
\frac{inj_{\hat{g}}(M,p,T)}{r}\geq C \frac{Vol_{\hat{g}}(\mathcal{B}_{T}(p,Cr))}{r^{4}}.
\end{align}
Here $\mathcal{B}_{T}(p,r)=\exp_{p}(B_{T}(0,r))$.
}

This important theorem allows one to bound the injectivity radius in terms of the Riemann curvature and the volume of the Riemannian ball $B_{T}(0,r)\subset T_{p}M$. Let's set $r=1$. In a globally hyperbolic background spacetime, we assume that the Riemann curvature bound is $1$ i.e., $\sup_{\gamma}|Riem(\hat{g})|_{T_{\gamma}}\leq1$  and $Vol_{\hat{g}}(\mathcal{B}_{T}(p,C))$ is uniformly bounded from below.
Therefore the injectivity radius satisfies $inj_{\hat{g}}\gtrsim 1$, where the constant involved is uniform. Note that we may always make the magnitude of $Riem$ to be of the order $1$ by re-scaling as long as it has a definite lower (and upper) bound. In the local coordinates with $x(p)=0$, the following relations hold only at $p$ (recall that we are in a geodesic normal coordinate system which has $p$ as its origin)
\begin{align}
\hat{g}_{\mu\nu}|_{p}=\hat{g}_{\mu\nu}(0)=\eta_{\mu\nu}, \Gamma^{\mu}_{\alpha\beta}|_{p}=\Gamma^{\mu}_{\alpha\beta}(0)=0.
\end{align}
This is the so-called equivalence principle of general relativity i.e., at each point of the curved spacetime (so that gravity is present) one may construct an inertial frame (Minkowski metric). As used previously in section $3$, the following remarkable property holds throughout the normal neighbourhood of point $p$ i.e., not just at $p$
\begin{align}
\label{eq:normalprop}
\hat{g}_{\mu\nu}(x)x^{\nu}=\hat{g}_{\mu\nu}(0)x^{\nu}=\eta_{\mu\nu}x^{\nu}, \Gamma^{\mu}_{\alpha\beta}(x)x^{\alpha}x^{\beta}=0.
\end{align}
The second property is simply a consequence of the fact that the geodesics through $p$ ($x(p) \equiv 0$) are straight lines. We do not prove these properties here. For an elegant proof, reader is referred to the section $3$ of \cite{moncrief2005integral}. On this normal coordinate chart one may introduce the frame fields $e_{a}:=e^{\mu}_{a}\partial_{\mu}$ and the dual co-frame field $\theta^{b}:=\theta^{b}_{\mu}dx^{\mu}$. One way to construct such a frame (co-frame) field throughout the normal neighbourhood is to assign it at $p$ (e.g., $e_{a}|_{p}=\delta^{\mu}_{p}\partial_{\mu}$) and then parallel propagate by geodesics emanating from $p$. Since the parallel propagation preserves the duality $<e_{a},\theta^{b}>=\delta^{b}_{a}$, such a construction is possible. In the geodesic normal coordinate system, one may explicitly calculate the connection $1-$form $\omega^{a}_{\mu b}$, co-frame fields $\theta^{a}$ and the metric $\hat{g}_{\mu\nu}=\theta^{a}_{\mu}\theta^{b}_{\nu}\eta_{ab}$ in terms of the curvature 
\begin{align}
\label{eq:geodesic1}
\omega^{a}~_{b\mu}(x)=-\int_{0}^{1} \lambda x^{\nu}R^{a}~_{b\mu\nu}(\lambda x)d\lambda,\\
\label{eq:geodesic2}
\theta^{a}_{\mu}(x)=\theta^{a}_{\mu}(0)+\int_{0}^{1} \omega^{a}_{b\mu}(\lambda x)(\lambda x^{\nu}\theta^{b}_{\nu}(0))d\lambda,\\
\label{eq:geodesic3}
\hat{g}_{\mu\nu}(x)=\eta_{\mu\nu}-\left(2\eta_{ab}\int_{0}^{1}\int_{0}^{1}\lambda^{2}_{1}\lambda_{2}R^{b}~_{c\nu\alpha}(\lambda_{1}\lambda_{2}x)\theta^{c}_{\lambda}(0)\theta^{a}_{\mu}(0)\nonumber d\lambda_{1}d\lambda_{2}\right)x^{\alpha}x^{\lambda}\\ 
 +\left(\eta_{ab}\int_{0}^{1}\int_{0}^{1}\int_{0}^{1}\int_{0}^{1}\lambda^{2}_{1}\lambda_{2}\lambda^{2}_{3}\lambda_{4}R^{a}~_{p\mu\alpha}(\lambda_{1}\lambda_{2}x)\right.\\\nonumber 
 \left.R^{b}~_{q\nu\beta}(\lambda_{3}\lambda_{4}x)\theta^{p}_{\lambda}(0)\theta^{q}_{\delta}(0)d\lambda_{1}d\lambda_{2}d\lambda_{3}d\lambda_{4}\right)x^{\alpha}x^{\beta}x^{\lambda}x^{\delta}.
\end{align}
Once again, the interested readers are referred to \cite{moncrief2005integral} for the proof. Noting $|Riem(\hat{g})|\leq 1$ throughout the normal neighbourhood, one may observe that components of the spacetime metric $\hat{g}$ satisfy several estimates as one approaches $p$ ($x(p)\equiv 0$). In our application, the point $p$ will be the vertex of the cone $C^{-}_{p}$ under consideration and therefore we obtain the estimates for the components of the spacetime metric as we approach the vertex of the cone $C^{-}_{p}$. If we invoke the ADM form of the metric (\ref{eq:metric}) i.e.,
\begin{align}
\label{eq:metric_new}
\hat{g}=-N^{2}dt\otimes dt+g_{ij}(dx^{i}+Y^{i}dt)(dx^{j}+Y^{j}dt),
\end{align}
then as one approaches the vertex of the cone $C^{-}_{p}$ (i.e., $p$), the following point-wise estimates hold provided that the point-wise curvature is bounded ($|Riem|\leq 1$)
\begin{align}
\label{eq:estimates_normal}
-N^{2}+g(Y,Y)=-N^{2}+O(|x|^{4}), |g_{ij}Y^{j}|=O(|x|^{2}),g_{ij}=\eta_{ij}+O(|x|^{2}).
\end{align}
Here of course $\eta_{ij}=\delta_{ij}$. These elementary estimates will be extremely important to us as they will help us avoid a few brute-force calculations.\\
Now we define vector fields with respect to the point $p$ ($x(p)\equiv 0$) in its normal neighbourhood. In an arbitrary globally hyperbolic manifold, these vector fields will only make sense in the coordinate frame adapted to the neighbourhood of $p$ and therefore we will call these `\textit{quasi-local}' vector fields. Let us define three such vector fields 
\begin{align}
K:=\hat{g}(x,x)\partial_{t}-2x_{0}x^{\nu}\partial_{\nu},\\
S:=x^{\mu}\partial_{\mu},\\
R:=\frac{x^{\mu}}{x_{0}}\partial_{\mu}.
\end{align}
By the very definition these vector fields are only well defined in the normal neighbourhood of $p$ ($x(p)\equiv 0$). Throughout the normal neighbourhood of $p$, $K$ and $S$ denote the approximate inversion generator and approximate scaling vector field, respectively. The vector field $R$ is just a scaled version of $S$ by $1/x_{0}$. Let us first show that these vector fields are time-like within the null-cone $C^{-}_{p}$. An explicit calculation yields 
\begin{align}
\hat{g}(K,K)=\hat{g}(\hat{g}(x,x)\partial_{t}-2x_{0}x^{\nu}\partial_{\nu},\hat{g}(x,x)\partial_{t}-2x_{0}x^{\mu}\partial_{\mu})\\
=|\hat{g}(x,x)|^{2}g_{tt}-4\hat{g}(x,x)x_{0}x^{\nu}\hat{g}_{0\nu}+4x^{2}_{0}\hat{g}(x,x)\\
=|\hat{g}(x,x)|^{2}\hat{g}_{tt}\leq 0,\\\nonumber 
\hat{g}(S,S)=\hat{g}(x^{\mu}\partial_{\mu},x^{\nu}\partial_{\nu})\\\nonumber 
=\hat{g}(x,x)<0~on~D^{t_{1}}_{p}
\end{align}
since on a globally hyperbolic spacetime $\hat{g}_{tt}<0$ and $\hat{g}(x,x)<0$ inside of the light cone, $\hat{g}(x,x)=0$ on the light cone and $\hat{g}(x,x)>0$ outside the light cone.
The vector field $K$ is time-like everywhere except on the mantle of the light cone where it is null. The vector field $S$ is time-like within the lightcone and null on the mantle of the lightcone. Time-like characteristic of $R$ follows from that of $S$. We call the vector fields $K$ and $S$ approximate conformal Killing fields. The reason for such a terminology will soon become clear. Let us explicitly compute the strain tensors associated with $K$ and $S$. We denote the strain tensors of $K$ and $S$ by $^{K}\pi$ and $^{S}\pi$, respectively. Explicit calculations yield 
\begin{align}
\nabla_{\alpha}K^{\lambda}=\nabla_{\alpha}(\hat{g}_{\mu\nu}x^{\mu}x^{\nu}\delta^{\lambda}_{0}-2\hat{g}_{0\nu}x^{\nu}x^{\lambda})\\\nonumber
=\partial_{\alpha}(\hat{g}_{\mu\nu}x^{\mu}x^{\nu})\delta^{\lambda}_{0}+\hat{g}(x,x)\nabla_{\alpha}\delta^{\lambda}_{0}-2\hat{g}_{0\nu}x^{\lambda}\nabla_{\alpha}x^{\nu}-2\hat{g}_{0\nu}x^{\nu}\nabla_{\alpha}x^{\lambda}\\\nonumber
=2\hat{g}_{\alpha\nu}x^{\nu}\delta^{\lambda}_{0}-2\hat{g}_{0\nu}x^{\lambda}(\delta^{\nu}_{\alpha}+\Gamma^{\nu}~_{\alpha\mu}x^{\mu})-2\hat{g}_{0\nu}x^{\nu}(\delta^{\lambda}_{\alpha}+\Gamma^{\lambda}_{\alpha\mu}x^{\mu}),\\\nonumber
=2\hat{g}_{\alpha\nu}x^{\nu}\delta^{\lambda}_{0}-2\hat{g}_{0\alpha}x^{\lambda}-2\hat{g}_{0\nu}\delta^{\lambda}_{\alpha}x^{\nu}-2\hat{g}_{0\nu}\Gamma^{\nu}~_{\alpha\mu}x^{\mu}x^{\lambda}-2\hat{g}_{0\nu}\Gamma^{\lambda}~_{\alpha\mu}x^{\nu}x^{\mu},
\end{align}
where we have used the fact that $\hat{g}_{\mu\nu}x^{\mu}x^{\nu}=\eta_{\mu\nu}x^{\mu}x^{\nu}$ throughout the geodesic normal neighbourhood of $p$.
The trace $\nabla_{\alpha}K^{\alpha}$ is calculated as 
\begin{align}
\label{eq:trace}
\nabla_{\alpha}K^{\alpha}=-8\hat{g}_{0\nu}x^{\nu}-2\hat{g}_{0\nu}\Gamma^{\nu}~_{\alpha\mu}x^{\mu}x^{\alpha}-2\hat{g}_{0\nu}\Gamma^{\alpha}~_{\alpha\mu}x^{\mu}x^{\nu}\\\nonumber 
=-8x_{0}-2\hat{g}_{0\nu}\Gamma^{\alpha}~_{\alpha\mu}x^{\mu}x^{\nu}
\end{align}
since $\Gamma^{\nu}_{\alpha\mu}x^{\alpha}x^{\mu}=0$ throughout the normal coordinate system (\ref{eq:normalprop}). Therefore the strain tensor $^{K}\pi$ may be evaluated as 
\begin{align}
^{K}\pi^{\mu\nu}=\nabla^{\mu}K^{\nu}+\nabla^{\nu}K^{\mu}=-4x_{0}\hat{g}^{\mu\nu}-2\hat{g}_{0\alpha}\hat{g}^{\mu\beta}\Gamma^{\alpha}~_{\beta\lambda}x^{\lambda}x^{\nu}\\\nonumber  -2\hat{g}_{0\alpha}\hat{g}^{\mu\beta}\Gamma^{\nu}~_{\beta\lambda}x^{\alpha}x^{\lambda}-2\hat{g}_{0\alpha}\hat{g}^{\nu\beta}\Gamma^{\alpha}~_{\beta\lambda}x^{\lambda}x^{\mu}-2\hat{g}_{0\alpha}\hat{g}^{\nu\beta}\Gamma^{\mu}~_{\beta\lambda}x^{\alpha}x^{\lambda}\\\nonumber 
=\frac{1}{2}(\nabla_{\alpha}K^{\alpha})\hat{g}^{\mu\nu}+\hat{g}^{\mu\nu}\hat{g}_{0\beta}\Gamma^{\alpha}_{\alpha\lambda}x^{\lambda}x^{\beta}-2\hat{g}_{0\alpha}\hat{g}^{\mu\beta}\Gamma^{\alpha}~_{\beta\lambda}x^{\lambda}x^{\nu}\\\nonumber  -2\hat{g}_{0\alpha}\hat{g}^{\mu\beta}\Gamma^{\nu}~_{\beta\lambda}x^{\alpha}x^{\lambda}-2\hat{g}_{0\alpha}\hat{g}^{\nu\beta}\Gamma^{\alpha}~_{\beta\lambda}x^{\lambda}x^{\mu}-2\hat{g}_{0\alpha}\hat{g}^{\nu\beta}\Gamma^{\mu}~_{\beta\lambda}x^{\alpha}x^{\lambda}\\\nonumber 
=\frac{1}{2}(\nabla_{\alpha}K^{\alpha})\hat{g}^{\mu\nu}+\mathcal{ER}^{\mu\nu}_{K}.
\end{align}
Now notice an extremely important fact. If $K$ were to be a true conformal Killing vector field then $\mathcal{ER}^{\mu\nu}_{K}$ would vanish identically (which is the case in ordinary flat spacetime). However, using the equations (\ref{eq:geodesic1}-\ref{eq:geodesic3}), we will show that this error term vanishes at third order as one approaches the vertex of the cone $C^{-}_{p}$. Notice the following relation 
\begin{align}
\label{eq:geodesic4}
\Gamma^{\alpha}_{\mu\nu}x^{\nu}=\frac{1}{2}\hat{g}^{\beta\alpha}x^{\nu}\partial_{\nu}\hat{g}_{\mu\beta}
\end{align}
which may be further evaluated utilizing 
\begin{align}
\label{eq:geodesic5}
x^{\beta}\partial_{\beta}\hat{g}_{\mu\nu}=\eta_{ab}\left\{\theta^{b}_{\nu}(x)\left(\omega^{a}_{f\mu}(x)(x^{\gamma}\theta^{f}_{\gamma}(0)-\int_{0}^{1}[\omega^{a}_{f\mu}(\lambda x)(\lambda x^{\gamma}\theta^{f}_{\gamma}(0)]d\lambda\right)\right.\\\nonumber 
 \left.+ \theta^{a}_{\nu}(x)\left(\omega^{b}_{f\mu}(x)(x^{\gamma}\theta^{f}_{\gamma}(0)-\int_{0}^{1}[\omega^{b}_{f\mu}(\lambda x)(\lambda x^{\gamma}\theta^{f}_{\gamma}(0)]d\lambda\right)\right\}.
\end{align}
Now if we substitute the expression for the connection $1-$ form (\ref{eq:geodesic1}), then we obtain the following point-wise estimate for $|x^{\beta}\partial_{\beta}\hat{g}_{\mu\nu}|$
as one approaches the vertex of the cone $C^{-}_{p}$
\begin{align}
\label{eq:geodesic6}
|x^{\beta}\partial_{\beta}\hat{g}_{\mu\nu}|=O(|x|^{2})
\end{align}
and therefore the error term $\mathcal{ER}^{\mu\nu}_{K}$ satisfies 
\begin{align}
|\mathcal{ER}^{\mu\nu}_{K}|=|\hat{g}^{\mu\nu}g_{0\beta}\Gamma^{\alpha}_{\alpha\lambda}x^{\lambda}x^{\beta}-2g_{0\alpha}g^{\mu\beta}\Gamma^{\alpha}~_{\beta\lambda}x^{\lambda}x^{\nu}\\\nonumber  -2g_{0\alpha}g^{\mu\beta}\Gamma^{\nu}~_{\beta\lambda}x^{\alpha}x^{\lambda}-2g_{0\alpha}g^{\nu\beta}\Gamma^{\alpha}~_{\beta\lambda}x^{\lambda}x^{\mu}-2g_{0\alpha}g^{\nu\beta}\Gamma^{\mu}~_{\beta\lambda}x^{\alpha}x^{\lambda}|\\\nonumber 
=O(|x|^{3}).
\end{align}
Since the error term $\mathcal{ER}^{\mu\nu}_{K}$ which is obstructing the exact conformal Killing character of $K$ vanishes at third order as one approaches the vertex of $C^{-}_{p}$, we call $K$ an approximate quasi-local conformal Killing field. Notice an important fact that $K$ itself vanishes at second order as one approaches the vertex of $C^{-}_{p}$. Therefore it only makes sense to call it approximate conformal Killing since the error term vanishes at one order higher rate. This property will be extremely important while we derive the energy estimates. Therefore, the results derived in this section yield the following lemma.\\
\textbf{Lemma 1:} \textit{Let $p\in M$ such that in local coordinates $x(p)=0$ and $\mathcal{G}_{p}$ be its geodesic normal neighbourhood and $C^{-}_{p}$ be its past light cone. The quasi local vector field $K:=\hat{g}(x,x)\partial_{t}-2x_{0}x^{\nu}\partial_{\nu}$ adapted to $\mathcal{G}_{p}$ is an approximate conformal Killing vector field in a sense that its strain tensor $^{K}\pi$ satisfies 
\begin{align}
\label{eq:strainK}
^{K}\pi^{\mu\nu}=\frac{1}{2}(\nabla_{\mu}K^{\mu})\hat{g}^{\mu\nu}+\mathcal{ER}^{\mu\nu}_{K},
\end{align}
where $|\mathcal{ER}^{\mu\nu}_{K}|=O(|x|^{3})$ as one approaches the vertex $p$ of the light cone $C^{-}_{p}$.
}

Now let us consider the vector field $S$ and compute its strain tensor $^{S}\pi$. Explicit calculation yields 
\begin{align}
\nabla_{\mu}S^{\nu}=\nabla_{\mu}x^{\nu}=\delta^{\nu}_{\mu}+\Gamma^{\nu}~_{\mu\alpha}x^{\alpha}
\end{align}
the covariant divergence of which reads 
\begin{align}
\nabla_{\mu}S^{\mu}=4+\Gamma^{\mu}~_{\mu\alpha}x^{\alpha}.
\end{align}
The strain tensor reads 
\begin{align}
^{S}\pi^{\mu\nu}:(L_{S}\hat{g})^{\mu\nu}=\nabla^{\mu}S^{\nu}+\nabla^{\nu}S^{\mu}=2\hat{g}^{\mu\nu}+\hat{g}^{\mu\beta}\Gamma^{\nu}_{\beta\alpha}x^{\alpha}+\hat{g}^{\nu\beta}\Gamma^{\mu}_{\beta\alpha}x^{\alpha}\\\nonumber 
=\frac{1}{2}(\nabla_{\alpha}S^{\alpha})\hat{g}^{\mu\nu}-\frac{1}{2}(\Gamma^{\beta}~_{\beta\alpha}x^{\alpha})\hat{g}^{\mu\nu}+\hat{g}^{\mu\beta}\Gamma^{\nu}_{\beta\alpha}x^{\alpha}\\\nonumber 
 +\hat{g}^{\nu\beta}\Gamma^{\mu}_{\beta\alpha}x^{\alpha}\\\nonumber 
=\frac{1}{2}(\nabla_{\alpha}S^{\alpha})\hat{g}^{\mu\nu}+|\mathcal{ER}^{\mu\nu}_{S}|.
\end{align}
Now once again utilizing the equations in normal coordinates (\ref{eq:geodesic1}-\ref{eq:geodesic3}, \ref{eq:geodesic4}-\ref{eq:geodesic5}), we obtain 
\begin{align}
|\mathcal{ER}^{\mu\nu}_{S}|=O(|x|^{2})
\end{align}
as one approaches the vertex of the light cone $C^{-}_{p}$. Therefore we have proved the following lemma\\
\textbf{Lemma 2:} \textit{Let $p\in M$ such that in local coordinates $x(p)=0$ and let $\mathcal{G}_{p}$ be its geodesic normal neighbourhood and $C^{-}_{p}\subset \mathcal{G}_{p}$ be its past light cone. The quasi local vector field $S:=x^{\mu}\partial_{\mu}$ adapted to $\mathcal{G}_{p}$ is an approximate conformal Killing vector field in a sense that its strain tensor $^{S}\pi$ satisfies 
\begin{align}
^{S}\pi^{\mu\nu}=\frac{1}{2}(\nabla_{\alpha}S^{\alpha})\hat{g}^{\mu\nu}+\mathcal{ER}^{\mu\nu}_{S},
\end{align}
where $|\mathcal{ER}^{\mu\nu}_{S}|=O(|x|^{2})$ as one approaches the vertex $p$ of the light cone $C^{-}_{p}$.
}

Once again note an important fact that the error term obstructing the conformal Killing nature of the quasi-local vector field $S$ vanishes quadratically which is one order higher than the rate at which $S$ itself vanishes as one approaches $C^{t_{1}}_{p}$. Now we have the required machinery to move forward with deriving the estimates of the energies associated with $K$ and $S$.  
\subsection{Energy estimate using the quasi-local vector field $`K:=\hat{g}(x,x)\partial_{t}-2x_{0}x^{\nu}\partial_{\nu}$'}
We proceed in the standard way of obtaining the equation of conservation of energy associated with the vector field $K$. Instead of using the stress energy tensor, we go one step back and start with the equation of motion. The reason for doing so will become clear in the fullness of time. Noting $K(\varphi)=K^{\mu}\partial_{\mu}\varphi$, we multiply the equation of motion (\ref{eq:eom}) by $K(\varphi)$ and simplify the expression
\begin{align}
K(\varphi)\nabla^{\mu}\nabla_{\mu}\varphi=K(\varphi)\varphi^{5},\\\nonumber 
\nabla^{\mu}\left(K^{\nu}(\partial_{\mu}\varphi\partial_{\nu}\varphi-\frac{1}{2}\hat{g}_{\mu\nu}\hat{g}(\partial\varphi,\partial\varphi)-\hat{g}_{\mu\nu}\frac{\varphi^{6}}{6})\right)\\\nonumber
-\frac{1}{2}(\nabla^{\mu}K^{\nu}+\nabla^{\nu}K^{\mu})\left(\partial_{\mu}\varphi\partial_{\nu}\varphi-\frac{1}{2}\hat{g}_{\mu\nu}\hat{g}(\partial\varphi,\partial\varphi)-\frac{1}{6}\hat{g}_{\mu\nu}\varphi^{6} \right)=0.
\end{align}
Now we make use of the expression for the strain tensor $^{K}\pi^{\mu\nu}:=\nabla^{\mu}K^{\nu}+\nabla^{\nu}K^{\mu}$ (\ref{eq:strainK})
\begin{align}
\nabla^{\mu}\left(K^{\nu}(\partial_{\mu}\varphi\partial_{\nu}\varphi-\frac{1}{2}\hat{g}_{\mu\nu}\hat{g}(\partial\varphi,\partial\varphi)-\hat{g}_{\mu\nu}\frac{\varphi^{6}}{6})\right)\\\nonumber 
-\frac{1}{4}(\nabla_{\alpha}K^{\alpha}\hat{g}^{\mu\nu})\left(\partial_{\mu}\varphi\partial_{\nu}\varphi-\frac{1}{2}\hat{g}_{\mu\nu}\hat{g}(\partial\varphi,\partial\varphi)-\frac{1}{6}\hat{g}_{\mu\nu}\varphi^{6}\right)\\\nonumber 
-\frac{1}{2}\mathcal{ER}^{\mu\nu}_{K}\left(\partial_{\mu}\varphi\partial_{\nu}\varphi-\frac{1}{2}\hat{g}_{\mu\nu}\hat{g}(\partial\varphi,\partial\varphi)-\frac{1}{6}\hat{g}_{\mu\nu}\varphi^{6}\right)=0.
\end{align}
Now due to the estimate $|\mathcal{ER}^{\mu\nu}|=O(|x|^{3})$, we will observe that the term $\mathcal{ER}^{\mu\nu}_{K}\left(\partial_{\mu}\varphi\partial_{\nu}\varphi-\frac{1}{2}\hat{g}_{\mu\nu}\hat{g}(\partial\varphi,\partial\varphi)-\frac{1}{6}\hat{g}_{\mu\nu}\varphi^{6}\right)$ is harmless in the forthcoming analysis. However, we do have the problematic term\\ $\frac{1}{4}(\nabla_{\alpha}K^{\alpha}\hat{g}^{\mu\nu})\left(\partial_{\mu}\varphi\partial_{\nu}\varphi-\frac{1}{2}\hat{g}_{\mu\nu}\hat{g}(\partial\varphi,\partial\varphi)-\frac{1}{6}\hat{g}_{\mu\nu}\varphi^{6}\right)$ because $|\nabla_{\alpha}K^{\alpha}|=O(|x|)$ (\ref{eq:trace}). Therefore, we will add a counter term to the multiplier $K(\varphi)$ which cancels this problematic term point-wise. We multiply the equation of motion (\ref{eq:eom}) by $K(\varphi)-2x_{0}\varphi$ instead to yield 
\begin{align}
\label{eq:conserv}
\nabla^{\mu}\left(K^{\nu}(\partial_{\mu}\varphi\partial_{\nu}\varphi-\frac{1}{2}\hat{g}_{\mu\nu}\hat{g}(\partial\varphi,\partial\varphi)-\frac{\varphi^{6}}{6})-2x_{0}\varphi\nabla_{\mu}\varphi\right)\\\nonumber
-\nabla^{\mu}K^{\nu}(\partial_{\mu}\varphi\partial_{\nu}\varphi-\frac{1}{2}\hat{g}_{\mu\nu}\hat{g}(\partial\varphi,\partial\varphi)-\frac{\varphi^{6}}{6})+2x_{0}|\partial\varphi|^{2}\\\nonumber 
+2\nabla_{\mu}x_{0}\varphi\nabla^{\mu}\varphi
=-2x_{0}\varphi^{6}.
\end{align}
Notice that the equation (\ref{eq:conserv}) is nothing but the following 
\begin{align}
\nabla^{\mu}\left(T_{\mu\nu}K^{\nu}-2x_{0}\varphi\nabla_{\mu}\varphi\right)\nonumber-T_{\mu\nu}\nabla^{\mu}K^{\nu}+2x_{0}\hat{g}(\partial\varphi,\partial\varphi)+2(\nabla_{\mu}x_{0})\varphi\nabla^{\mu}\varphi\\\nonumber 
=-2x_{0}\varphi^{6}.
\end{align}
We may further evaluate the term $2\nabla_{\mu}x_{0}\varphi\nabla^{\mu}\varphi$ as follows 
\begin{align}
2\nabla_{\mu}x_{0}\varphi\nabla^{\mu}\varphi=2\nabla_{\mu}(g_{0\nu}x^{\nu})\varphi\nabla^{\mu}\varphi=2g_{0\nu}(\partial_{\mu}x^{\nu}+\Gamma^{\nu}_{\mu\alpha}x^{\alpha})\varphi\nabla^{\mu}\varphi\\\nonumber 
=2g_{0\mu}\varphi\nabla^{\mu}\varphi+2g_{0\nu}\Gamma^{\nu}_{\mu\alpha}x^{\alpha}\varphi\nabla^{\mu}\varphi=\nabla_{\mu}(\varphi^{2}\delta^{\mu}_{0})+2g_{0\nu}\Gamma^{\nu}_{\mu\alpha}x^{\alpha}\varphi\nabla^{\mu}\varphi
\end{align}
yielding
\begin{align}
\nabla^{\mu}\left(T_{\mu\nu}K^{\nu}-2x_{0}\varphi\nabla_{\mu}\varphi\right)\nonumber-T_{\mu\nu}\nabla^{\mu}K^{\nu}+2x_{0}\hat{g}(\partial\varphi,\partial\varphi)\\\nonumber 
+\nabla_{\mu}(\varphi^{2}\delta^{\mu}_{0})+2g_{0\nu}\Gamma^{\nu}_{\mu\alpha}x^{\alpha}\varphi\nabla^{\mu}\varphi=-2x_{0}\varphi^{6}.
\end{align}
Integrating the previous expression over the truncated light cone $C^{T}_{P}$ yields 
\begin{align}
\int_{S_{t_{1}}}[T(K,n)-2x_{0}\varphi m+\phi^{2}n_{0}]\mu_{\hat{g}}|_{S_{t_{1}}}-\nonumber\int_{S_{t_{2}}}[T(K,n)-2x_{0}\varphi m+\phi^{2}n_{0}]\mu_{\hat{g}}|_{S_{t_{2}}}\\\nonumber 
+\int_{C^{-,t_{2},t_{1}}_{p}}[T(K,l)-2x_{0}\varphi l(\varphi)+\phi^{2}l_{0}]\mu_{\hat{g}}|_{C^{-,t_{2},t_{1}}_{p}}=\int_{D^{-,t_{2},t_{1}}_{p}}[\nabla^{\mu}K^{\nu}T_{\mu\nu}-2x_{0}\hat{g}(\partial\varphi,\partial\varphi)]\mu_{\hat{g}}\\\nonumber -\int_{D^{-,t_{2},t_{1}}_{p}}2x_{0}\varphi^{6}\mu_{\hat{g}}-\int_{D^{-,t_{2},t_{1}}_{p}}2g_{0\nu}\Gamma^{\nu}_{\mu\alpha}x^{\alpha}\varphi\nabla^{\mu}\varphi.
\end{align}
Now noting that $\int_{S_{t}}(T(K,n)-2x_{0}\varphi m)\mu_{\hat{g}}|_{S_{t}}\lesssim t^{2}E_{S_{t}}$ throughout the domain of definition of $K$, we have the following 
\begin{align}
\lim_{t_{2}\to 0}\int_{S_{t_{2}}}[T(K,n)-2x_{0}\varphi m]\mu_{\hat{g}}|_{S_{t_{1}}}=0
\end{align}
and since $n_{0}=-N<0$ for a globally hyperbolic spacetime, $-\int_{S_{t_{2}}}\phi^{2}n_{0}\mu_{\hat{g}}|_{S_{t_{2}}}=\int_{S_{t_{2}}}\phi^{2}N\mu_{\hat{g}}|_{S_{t_{2}}}>0$ and therefore this term is harmless. Also notice an important fact that $x_{0}=-x^{0}=-t>0$ within the past light cone. Substituting $^{K}\pi:=\nabla^{\mu}K^{\nu}+\nabla^{\nu}K^{\mu}$ (\ref{eq:strainK}) in the previous expression with $t_{2}=0$ yields 
\begin{align}
\int_{S_{t_{1}}}[T(K,n)-2x_{0}\varphi m+\phi^{2}n_{0}]\mu_{\hat{g}}|_{S_{t_{1}}}\nonumber+\int_{C^{t_{1}}_{p}}[T(K,l)-2x_{0}\varphi l(\varphi)+\phi^{2}l_{0}]\mu_{\hat{g}}|_{C^{t_{1}}_{p}}\\\nonumber +\lim_{t_{2}\to 0}\int_{S_{t_{2}}}\varphi^{2}N\mu_{\hat{g}}|_{S_{t_{2}}}=\int_{D^{t_{1}}_{p}}[\frac{1}{2}(\nabla^{\mu}K^{\nu}+\nabla^{\nu}K^{\mu})T_{\mu\nu}-2x_{0}\hat{g}(\partial\varphi,\partial\varphi)]\mu_{\hat{g}}\\\nonumber
-\int_{D^{t_{1}}_{p}}2x_{0}\varphi^{6}\mu_{\hat{g}}
-\int_{D^{t_{1}}_{p}}2g_{0\nu}\Gamma^{\nu}_{\mu\alpha}x^{\alpha}\varphi\nabla^{\mu}\varphi\mu_{\hat{g}}\\\nonumber 
 =\int_{D^{t_{1}}_{p}}[(\frac{1}{4}(\nabla_{\mu}K^{\mu})\hat{g}^{\mu\nu}+\frac{1}{2}\mathcal{ER}^{\mu\nu}_{K})T_{\mu\nu}-2x_{0}\hat{g}(\partial\varphi,\partial\varphi)]\mu_{\hat{g}}-\int_{D^{t_{1}}_{p}}2x_{0}\varphi^{6}\mu_{\hat{g}}\\\nonumber 
-\int_{D^{t_{1}}_{p}}2g_{0\nu}\Gamma^{\nu}_{\mu\alpha}x^{\alpha}\varphi\nabla^{\mu}\varphi \mu_{\hat{g}}\\\nonumber 
=\int_{D^{t_{1}}_{p}}[(\frac{1}{4}(-8x_{0}+2\hat{g}_{0\nu}\Gamma^{\alpha}~_{\alpha\mu}x^{\mu}x^{\nu})\hat{g}^{\mu\nu}+\frac{1}{2}\mathcal{ER}^{\mu\nu}_{K})T_{\mu\nu}-2x_{0}\hat{g}(\partial\varphi,\partial\varphi)]\mu_{\hat{g}}\\\nonumber 
-\int_{D^{t_{1}}_{p}}2x_{0}\varphi^{6}\mu_{\hat{g}}
-\int_{D^{t_{1}}_{p}}2g_{0\nu}\Gamma^{\nu}_{\mu\alpha}x^{\alpha}\varphi\nabla^{\mu}\varphi \mu_{\hat{g}}\\\nonumber
=\int_{D^{t_{1}}_{p}}[2x_{0}\hat{g}(\partial\varphi,\partial\varphi)+\frac{4x_{0}}{3}\varphi^{6}+(2\hat{g}_{0\nu}\Gamma^{\alpha}~_{\alpha\mu}x^{\mu}x^{\nu}\hat{g}^{\mu\nu}+\frac{1}{2}\mathcal{ER}^{\mu\nu}_{K})T_{\mu\nu}\\\nonumber -2x_{0}\hat{g}(\partial\varphi,\partial\varphi)]\mu_{\hat{g}}
-\int_{D^{t_{1}}_{p}}2x_{0}\varphi^{6}\mu_{\hat{g}}
-\int_{D^{t_{1}}_{p}}2g_{0\nu}\Gamma^{\nu}_{\mu\alpha}x^{\alpha}\varphi\nabla^{\mu}\varphi \mu_{\hat{g}}\\\nonumber
=\int_{D^{t_{1}}_{p}}\left(-\frac{2x_{0}}{3}\varphi^{6}+(2\hat{g}_{0\nu}\Gamma^{\alpha}~_{\alpha\mu}x^{\mu}x^{\nu}\hat{g}^{\mu\nu}+\frac{1}{2}\mathcal{ER}^{\mu\nu}_{K})T_{\mu\nu}-2g_{0\nu}\Gamma^{\nu}_{\mu\alpha}x^{\alpha}\varphi\nabla^{\mu}\varphi\right)\mu_{\hat{g}},
\end{align}
where we have used the fact that $\hat{g}^{\mu\nu}T_{\mu\nu}=\hat{g}^{\mu\nu}(\partial_{\mu}\varphi\partial_{\nu}\varphi-\frac{1}{2}\hat{g}(\partial\varphi,\partial\varphi)\hat{g}_{\mu\nu}-\frac{1}{6}\varphi^{6}\hat{g}_{\mu\nu})=-\hat{g}(\partial\varphi,\partial\varphi)-\frac{4}{3}\varphi^{6}$. Now notice the following properties. For the moment, if we go to the spherical coordinate system $(t,r,\theta,\phi)$ with origin at $p$, then the quasi-local vector field $K$ may be written as follows \begin{align}
K=\hat{g}(x,x)\partial_{t}-2x_{0}x^{\nu}\partial_{\nu}=(t^{2}+r^{2})\partial_{t}+2tr\partial_{r}\\\nonumber 
=\frac{1}{2}\left((t+r)^{2}(\partial_{t}+\partial_{r})+(t-r)^{2}(\partial_{t}-\partial_{r})\right)
\end{align}
and noting that on $C^{t_{1}}_{p}$, $t+r=0$, we obtain
\begin{align}
K|_{C^{t_{1}}_{p}}=\frac{(t-r)^{2}}{2}(\partial_{t}-\partial_{r})=2t^{2}(\partial_{t}-\partial_{r}).
\end{align}
Now note that on $C^{t_{1}}_{p}$ $K$ is null due to the fact that $\hat{g}(K,K)=|\hat{g}(x,x)|^{2}g_{tt}$. Therefore on $C^{t_{1}}_{p}$, we have $K=-2t^{2}N l$, where the lapse $N>0$. This is due to the fact that $l$ is past directed in the definition (\ref{eq:null1}). Therefore the integral equation for the $K-$energy may be written as follows 
\begin{align}
\int_{S_{t_{1}}}[T(K,n)-2x_{0}\varphi m+\phi^{2}n_{0}]\mu_{\hat{g}}|_{S_{t_{1}}}\nonumber-\int_{C^{t_{1}}_{p}}2N t^{2}|\hat{g}(\partial\phi,l)|^{2}\mu_{\hat{g}}|_{C^{t_{1}}_{p}}\\\nonumber 
+\lim_{t_{2}\to 0}\int_{S_{t_{2}}}\varphi^{2}N\mu_{\hat{g}}|_{S_{t_{2}}}-\int_{D^{t_{1}}_{p}}\frac{2t}{3}\varphi^{6}\mu_{\hat{g}}=\int_{C^{t_{1}}_{p}}(2x_{0}\varphi l(\varphi)+\varphi^{2}l_{0})\mu_{\hat{g}}|_{C^{t_{1}}_{p}}\\\nonumber 
+\int_{D^{t_{1}}_{p}}\left((2\hat{g}_{0\nu}\Gamma^{\alpha}~_{\alpha\mu}x^{\mu}x^{\nu}\hat{g}^{\mu\nu}+\frac{1}{2}\mathcal{ER}^{\mu\nu}_{K})T_{\mu\nu}-2g_{0\nu}\Gamma^{\nu}_{\mu\alpha}x^{\alpha}\varphi\nabla^{\mu}\varphi\right)\mu_{\hat{g}},
\end{align}
where we have used that fact that $x_{0}=\hat{g}_{0\nu}x^{\nu}=\eta_{0\nu}x^{\nu}=-x^{0}=-t$ (\ref{eq:normalprop} ). Now Note that the terms
$\lim_{t_{2}\to 0}\int_{S_{t_{2}}}\varphi^{2}N\mu_{\hat{g}}|_{S_{t_{2}}}$ and $-\int_{D^{t_{1}}_{p}}\frac{2t}{3}\varphi^{6}\mu_{\hat{g}}$ are positive definite since $t<0$ in the past causal domain of $p$. Now we will show that the first integral i.e., $\int_{S_{t_{1}}}[T(K,n)-2x_{0}\varphi m+\phi^{2}n_{0}]\mu_{\hat{g}}|_{S_{t_{1}}}$ is positive definite modulo some error term which vanishes at a rate higher than quadratic as one approaches the vertex. Noting $n_{0}=-N$, let us evaluate the following expression explicitly 
\begin{align}
T(K,n)+2tm\varphi-N\varphi^{2}=\frac{1}{N}\left(\frac{\eta_{ij}x^{i}x^{j}+t^{2}}{2}(|\partial_{t}\varphi|^{2}+N^{2}\hat{g}^{ij}\partial_{i}\varphi\partial_{j}\varphi)\right.\\\left.+2t\partial_{t}\varphi x^{i}\partial_{i}\varphi+2t\varphi\partial_{t}\varphi-N^{2}\varphi^{2}-2t(Y^{i}\partial_{i}\varphi)(x^{j}\partial_{j}\varphi)-2t\varphi Y^{i}\partial_{i}\varphi\right)\\\nonumber 
+\frac{N}{6}(t^{2}+\eta_{ij}x^{i}x^{j})\varphi^{6}.
\end{align}
Now we may use the point-wise estimates (\ref{eq:estimates_normal}) to reduce this expression into the following form
\begin{align}
T(K,n)+2tm\varphi-N\varphi^{2}=\frac{1}{N}\left(\frac{1}{2}(|S(\varphi)|^{2}\nonumber+|\mathcal{B}(\varphi)|^{2})+2t\varphi\partial_{t}\varphi -\varphi^{2}\right)\\
 +\frac{N}{6}(t^{2}+\eta_{ij}x^{i}x^{j})\varphi^{6}+\mathcal{I},
\end{align}
where $|\int_{S_{t}}\mathcal{I} \mu_{\hat{g}}|_{S_{t}}|\lesssim t^{4}$. Here the constants involved depends on point-wise curvature (which is under control by assumption). $S$ and $\mathcal{B}$ denote the approximate scaling and the boost vector fields, respectively i.e., $\mathcal{B}^{\mu\nu}:=(\eta^{\lambda\mu}x^{\nu}-\eta^{\lambda\nu}x^{\mu})\partial_{\lambda}$. The term with ambiguous sign $2t\varphi\partial_{t}\varphi -\varphi^{2}$ may further be evaluted as follows 
\begin{align}
2t\varphi\partial_{t}\varphi-\varphi^{2}=2(S(\varphi)-x^{i}\partial_{i}\varphi)\varphi-\varphi^{2}\\\nonumber
=2\varphi S(\varphi)-x^{i}\partial_{i}\varphi^{2}-\varphi^{2}=2S(\varphi)\varphi-\partial_{i}(x^{i}\varphi^{2})+\varphi^{2}\partial_{i}x^{i}-\varphi^{2}\\\nonumber 
=2\varphi S(\varphi)+2\varphi^{2}-\partial_{i}(x^{i}\varphi^{2}).
\end{align}
One important thing to note here is that this term will complete a square producing a positive definite term in the expression of $T(K,n)+2tm\varphi-N\varphi^{2}$ modulo a total derivative term. Now the integral equation becomes 
\begin{align}
\label{eq:interm}
\int_{S_{t_{1}}}\frac{1}{N}\left(\frac{1}{2}(|S(\varphi)|^{2}\nonumber+|\mathcal{B}(\varphi)|^{2})+2\varphi S(\varphi)+2\varphi^{2}-\partial_{i}(x^{i}\varphi^{2})\right.\\ \left.+\frac{N^{2}}{6}(t_{1}^{2}+\eta_{ij}x^{i}x^{j})\varphi^{6}\right)N\mu_{g}+\int_{S_{t_{1}}}\mathcal{I}\mu_{\hat{g}}|_{S_{t_{1}}}-\int_{C^{t_{1}}_{p}}N t^{2}|\hat{g}(\partial\phi,l)|^{2}\mu_{\hat{g}}|_{C^{t_{1}}_{p}}\\\nonumber 
+\lim_{t_{2}\to 0}\int_{S_{t_{2}}}\varphi^{2}N\mu_{\hat{g}}|_{S_{t_{2}}}-\int_{D^{t_{1}}_{p}}\frac{2t}{3}\varphi^{6}\mu_{\hat{g}}=-\int_{C^{t_{1}}_{p}}(2x_{0}\varphi l(\varphi)+\varphi^{2}l_{0})\mu_{\hat{g}}|_{C^{t_{1}}_{p}}\\\nonumber 
+\int_{D^{t_{1}}_{p}}\left((2\hat{g}_{0\nu}\Gamma^{\alpha}~_{\alpha\mu}x^{\mu}x^{\nu}\hat{g}^{\mu\nu}+\frac{1}{2}\mathcal{ER}^{\mu\nu}_{K})T_{\mu\nu}-2g_{0\nu}\Gamma^{\nu}_{\mu\alpha}x^{\alpha}\varphi\nabla^{\mu}\varphi\right)\mu_{\hat{g}}.
\end{align}
Now noting $\partial_{i}(x^{i}\varphi^{2})=\nabla_{i}(x^{i}\varphi^{2})-\Gamma^{i}_{ij}x^{j}\varphi^{2}$, we may write the integral as follows $\int_{S_{t_{1}}}\partial_{i}(x^{i}\varphi^{2})\mu_{g}=\int_{S_{t_{1}}}\left(\nabla_{i}(x^{i}\varphi^{2})-\Gamma^{i}_{ij}x^{j}\varphi^{2}\right)\mu_{g}=\int_{\partial S_{t_{1}}}x^{i}\varphi^{2}\mathcal{N}_{i}\mu_{\partial{S}_{t_{1}}}-\int_{S_{t_{1}}}\Gamma^{i}_{ij}x^{j}\varphi^{2}\mu_{g}=I_{1}+I_{2}$. Now applying Ho\"lder to both these terms, we obtain 
\begin{align}
\label{eq:boundaryinequal}
|I_{1}|\lesssim |t_{1}|^{3}\int_{\partial S_{t_{1}}}\varphi^{2}\sqrt{\det \tilde{g}_{AB}(-t_{1},\theta,\phi)}d\theta\wedge d\phi\\\nonumber 
\lesssim t_{1}^{2}\left(t^{2}_{1}\int_{\partial S_{t_{1}}}\varphi^{4}\sqrt{\det \tilde{g}_{AB}(-t_{1},\theta,\phi)}d\theta\wedge d\phi\right)^{1/2}\\\nonumber
\lesssim t^{2}_{1}\left(\int_{C^{t_{1}}_{p}}\varphi^{6}\mu_{\hat{g}}|_{C^{t_{1}}_{p}}\right)^{1/4}\left((\int_{C^{t_{1}}_{p}}|l(\varphi)|^{2}\mu_{\hat{g}}|_{C^{t_{1}}_{p}})^{1/2}+(\int_{C^{t_{1}}_{p}}\varphi^{6}\mu_{\hat{g}}|_{C^{t_{1}}_{p}})^{1/6}\right)^{1/2}\\\nonumber 
\lesssim t^{2}_{1}zo(t_{1})
\end{align}
since $\int_{C^{t_{1}}_{p}}\varphi^{6}\mu_{\hat{g}}|_{C^{t_{1}}_{p}}=zo(t_{1})$. Here we have used the inequality (\ref{eq:boundaryinequalprevious}) (to be proven in section $5.5$) and $\tilde{g}_{AB}$ is defined to be the induced Riemannian metric on the topological sphere $\partial S_{t_{1}}$ after extracting the conformal factor $\approx t^{2}_{1}$). Similarly 
\begin{align}
|I_{2}|\leq \left(\int_{S_{t_{1}}}\varphi^{6}\mu_{g}\right)^{1/3}\left(\int_{S_{t_{1}}}|\Gamma[g]^{i}_{ij}x^{j}|^{3/2}\mu_{g}\right)^{2/3}\lesssim |t_{1}|^{3}(E_{S_{t_{1}}})^{1/3},
\end{align}
where we have used the estimates (\ref{eq:geodesic4}-\ref{eq:geodesic6}) and the identity $\Gamma[g]^{i}_{ij}=\Gamma[\hat{g}]^{i}_{ij}-\frac{1}{N}k_{ij}Y^{i}$. Another remaining term which does not satisfy a straightforward estimate is the term $-\int_{C^{t_{1}}_{p}}(2x_{0}\varphi l(\varphi)+\varphi^{2}l_{0})\mu_{\hat{g}}|_{C^{t_{1}}_{p}}$. This term also does not have a definite sign. We may however use H\"older's inequality (noting $x_{0}=-t$) to obtain the following 
\begin{align}
\int_{C^{t_{1}}_{p}}t\varphi l(\varphi)\mu_{\hat{g}}|_{C^{t_{1}}_{p}} \leq t^{2}_{1}\int_{C^{t_{1}}_{p}}(\frac{\varphi^{2}}{t^{2}_{1}}+|l(\varphi)|^{2})\mu_{\hat{g}}|_{C^{t_{1}}_{p}}\\\nonumber 
\leq t_{1}^{2}\left((\int_{C^{t_{1}}_{p}}\varphi^{6}\mu_{\hat{g}}|_{C^{t_{1}}_{p}})^{1/3}(\frac{1}{t^{3}_{1}}\int_{C^{t_{1}}_{p}}\mu_{\hat{g}}|_{C^{t_{1}}_{p}})^{2/3}+|l(\varphi)|^{2})\mu_{\hat{g}}|_{C^{t_{1}}_{p}}\right)\\\nonumber
\lesssim t_{1}^{2}zo(t_{1}),
\end{align}
\begin{align}
\int_{C^{t_{1}}_{p}}N t^{2}|\hat{g}(\partial\phi,l)|^{2}\mu_{\hat{g}}|_{C^{t_{1}}_{p}}\lesssim t^{2}_{1}zo(t_{1}),
\end{align}
and 
\begin{align}
\int_{C^{t_{1}}_{p}}\varphi^{2}l_{0}\mu_{\hat{g}}|_{C^{t_{1}}_{p}}\lesssim \int_{C^{t_{1}}_{p}}t^{2}_{1}\frac{\varphi^{2}}{t^{2}_{1}}\mu_{\hat{g}}|_{C^{t_{1}}_{p}}\leq t_{1}^{2}\int_{C^{t_{1}}_{p}}\frac{\varphi^{2}}{t^{2}_{1}}\mu_{\hat{g}}|_{C^{t_{1}}_{p}}\lesssim t^{2}_{1}zo(t_{1}).
\end{align}
Here we have used (\ref{eq:vanishing2}-\ref{eq:vanishing3}).
Now notice some additional estimates (harmless)
\begin{align}
\int_{D^{t_{1}}_{p}}(2\hat{g}_{0\nu}\Gamma^{\alpha}~_{\alpha\mu}x^{\mu}x^{\nu}\hat{g}^{\mu\nu}T_{\mu\nu})\mu_{\hat{g}}|_{C^{t_{1}}_{p}}\lesssim t_{1}^{4}E_{S_{t_{1}}},\\\nonumber 
\int_{D^{t_{1}}_{p}}\mathcal{ER}^{\mu\nu}_{K}T_{\mu\nu}\mu_{\hat{g}}|_{C^{t_{1}}_{p}}\lesssim t_{1}^{4}E_{S_{t_{1}}},\\\nonumber 
\int_{D^{t_{1}}_{p}}2g_{0\nu}\Gamma^{\nu}_{\mu\alpha}x^{\alpha}\varphi\nabla^{\mu}\varphi\mu_{\hat{g}}|_{C^{t_{1}}_{p}}\lesssim t_{1}^{4}E_{S_{t_{1}}}.
\end{align}
Utilizing these estimates, we may reduce equation (\ref{eq:interm}) into the following form
\begin{align}
\int_{S_{t_{1}}}\left(\frac{1}{2}((S(\varphi)+2\varphi)^{2}\nonumber+|\mathcal{B}(\varphi)|^{2})+\frac{N^{2}}{6}(t_{1}^{2}+\eta_{ij}x^{i}x^{j})\varphi^{6}\right)N\mu_{g}\\\nonumber +\lim_{t_{2}\to 0}\int_{S_{t_{2}}}\varphi^{2}\mu_{\hat{g}}|_{S_{t_{2}}}-\int_{D^{t_{1}}_{p}}\frac{2t}{3}\varphi^{6}\mu_{\hat{g}}
\lesssim t_{1}^{2}zo(t_{1}).
\end{align}
Now noting $N^{2}=1+O(|t_{1}|^{2})$ and $t_{1}^{2}+\eta_{ij}x^{i}x^{j}\geq t_{1}^{2}$, and all the terms are positive, we have the following estimate 
\begin{align}
\int_{S_{t_{1}}}\varphi^{6}\mu_{g}\lesssim zo(t_{1}).
\end{align}
In order for the units to be consistent, all the involved constants are assumed to have suitable units.
The results obtained so far yield the following lemma\\
\textbf{Lemma 3:} \textit{Let $p\in M$ be such that in local coordinates $x(p)=0$ and $\mathcal{G}$ be its geodesic normal neighbourhood and $C^{t_{1}}_{p}\subset \mathcal{G}$ be its past light cone extending up to the constant time hypersurface $t_{1}$. Utilizing the quasi-local approximate conformal Killing vector field $K$, we obtain the following estimate 
\begin{align}
\int_{S_{t_{1}}}\varphi^{6}\mu_{g}\lesssim zo(t_{1}).
\end{align}
}
The physical significance of this estimate is that the non-linearity cannot focus energy. This is a rough indication of why the global existence should hold. This estimate will be crucial in the later stages of the argument. 

\subsection{Energy estimate using the quasi-local vector field `$S=x^{\mu}\partial_{\mu}$'}
Using the estimate obtained from the energy associated with the approximate inversion generator $K$, we move on to derive an energy estimate associated with the scaling vector field $S:=x^{\mu}\partial_{\mu}$. We will once again utilize the fact that this is only an approximate conformal Killing field. Notice an important fact that the integral equation for $K$ in the previous section contained a term $^{K}\pi^{\mu\nu}T_{\mu\nu}$ which was of $O(|x|)$ and therefore needed cancellation. In order to do so we introduced an additional multiplier. Here we will do the identical operation to take care of the term $^{S}\pi^{\mu\nu}T_{\mu\nu}$. Multiplying the equation of motion (\ref{eq:eom}) by $S(\varphi)+\varphi$  yields 
\begin{align}
\nabla^{\mu}\left(T_{\mu\nu}S^{\nu}+\varphi\nabla_{\mu}\varphi\right)-T_{\mu\nu}\nabla^{\mu}S^{\nu}-\hat{g}(\nabla\varphi,\nabla\varphi)\\\nonumber 
=\varphi^{6}
\end{align}
Which upon integrating over the truncated light cone $C^{T}_{p}$ yields 
\begin{align}
\int_{S_{t_{1}}}[T(S,n)+\varphi m]\mu_{\hat{g}}|_{S_{t_{1}}}-\nonumber\int_{S_{t_{2}}}[T(S,n)+\varphi m]\mu_{\hat{g}}|_{S_{t_{2}}}\\\nonumber +\int_{C^{T}_{p}}[T(S,l)+\varphi l(\varphi)]\mu_{\hat{g}}|_{C^{T}_{p}}=\int_{D^{T}_{p}}[\nabla^{\mu}S^{\nu}T_{\mu\nu}+\hat{g}(\partial\varphi,\partial\varphi)]\mu_{\hat{g}} +\int_{D^{T}_{p}}\varphi^{6}\mu_{\hat{g}}.
\end{align}
Here notice an important fact that since $t<0$ in the past light cone, $S=x^{\mu}\partial_{\mu}$ is actually past directed and therefore $T(S,n)<0$. However, this difference does not matter since the equation of motion (\ref{eq:eom}) is clearly invariant if one performs $t\mapsto -t$. Since $\int_{S_{t}}(T(S,n)+\varphi m)\mu_{\hat{g}}|_{S_{t}}\lesssim |t|\mathcal{E}$, we have using the boundedness of energy 
\begin{align}
\lim_{t_{2}\to 0}\int_{S_{t_{2}}}[T(S,n)+\varphi m]\mu_{\hat{g}}|_{S_{t_{2}}}=0.
\end{align}
Therefore the integral equation over the full past null cone and interior becomes 
\begin{align}
&&\int_{S_{t_{1}}}[T(S,n)+\varphi m]\mu_{\hat{g}}|_{S_{t_{1}}}+\int_{C^{t_{1}}_{p}}[T(S,l)+\varphi l(\varphi)]\mu_{\hat{g}}|_{C^{t_{1}}_{p}}\\\nonumber 
&=&\int_{D^{t_{1}}_{p}}[\nabla^{\mu}S^{\nu}T_{\mu\nu}+\hat{g}(\partial\varphi,\partial\varphi)]\mu_{\hat{g}} +\int_{D^{t_{1}}_{p}}\varphi^{6}\mu_{\hat{g}}\\\nonumber
&=&\int_{D^{t_{1}}_{p}}[\frac{1}{2}(\nabla^{\mu}S^{\nu}+\nabla^{\nu}S^{\mu})T_{\mu\nu}+\hat{g}(\partial\varphi,\partial\varphi)]\mu_{\hat{g}} +\int_{D^{t_{1}}_{p}}\varphi^{6}\mu_{\hat{g}}\\\nonumber
&=&\int_{D^{t_{1}}_{p}}[\frac{1}{2}(\frac{1}{2}(\nabla_{\alpha}S^{\alpha})\hat{g}^{\mu\nu}+\mathcal{ER}^{\mu\nu}_{S})T_{\mu\nu}+\hat{g}(\partial\varphi,\partial\varphi)]\mu_{\hat{g}} +\int_{D^{t_{1}}_{p}}\varphi^{6}\mu_{\hat{g}}\\\nonumber
&=&\int_{D^{t_{1}}_{p}}\left(\frac{1}{4}\nabla_{\alpha}S^{\alpha}\hat{g}^{\mu\nu}T_{\mu\nu}+\hat{g}(\partial\varphi,\partial\varphi)\right)\mu_{\hat{g}}+\int_{D^{t_{1}}_{p}}\frac{1}{2}\mathcal{ER}^{\mu\nu}_{S}T_{\mu\nu}\mu_{\hat{g}}+\int_{D^{t_{1}}_{p}}\varphi^{6}\mu_{\hat{g}}\\\nonumber 
&=&\int_{D^{t_{1}}_{p}}\left(\frac{1}{4}(4+\Gamma^{\beta}~_{\beta\alpha}x^{\alpha})\hat{g}^{\mu\nu}T_{\mu\nu}+\hat{g}(\partial\varphi,\partial\varphi)\right)\mu_{\hat{g}}+\int_{D^{t_{1}}_{p}}\frac{1}{2}\mathcal{ER}^{\mu\nu}_{S}T_{\mu\nu}\mu_{\hat{g}}+\int_{D^{t_{1}}_{p}}\varphi^{6}\mu_{\hat{g}}\\\nonumber 
&=&-\int_{D^{t_{1}}_{p}}\frac{2}{3}\varphi^{6}\mu_{\hat{g}}+\int_{D^{t_{1}}_{p}}(\frac{1}{4}\Gamma^{\beta}~_{\beta\alpha}x^{\alpha}\hat{g}^{\mu\nu}+\frac{1}{2}\mathcal{ER}^{\mu\nu}_{S})T_{\mu\nu}\mu_{\hat{g}}+\int_{D^{t_{1}}_{p}}\varphi^{6}\mu_{\hat{g}}\\\nonumber 
&=&\frac{1}{3}\int_{D^{t_{1}}_{p}}\varphi^{6}\mu_{\hat{g}}+\int_{D^{t_{1}}_{p}}(\frac{1}{4}\Gamma^{\beta}~_{\beta\alpha}x^{\alpha}\hat{g}^{\mu\nu}+\frac{1}{2}\mathcal{ER}^{\mu\nu}_{S})T_{\mu\nu}\mu_{\hat{g}},
\end{align}
that is, the final equation becomes 
\begin{align}
\int_{S_{t_{1}}}[T(S,n)+\varphi m]\mu_{\hat{g}}|_{S_{t_{1}}}+\int_{C^{t_{1}}_{p}}[T(S,l)+\varphi l(\varphi)]\mu_{\hat{g}}|_{C^{t_{1}}_{p}}\\\nonumber
=\frac{1}{3}\int_{D^{t_{1}}_{p}}\varphi^{6}\mu_{\hat{g}}+\int_{D^{t_{1}}_{p}}(\frac{1}{4}\Gamma^{\beta}~_{\beta\alpha}x^{\alpha}\hat{g}^{\mu\nu}+\frac{1}{2}\mathcal{ER}^{\mu\nu}_{S})T_{\mu\nu}\mu_{\hat{g}}.
\end{align}
Noting that $S$ is past time-like within the null cone $C^{t_{1}}_{p}$ and $n$ is future time-like, $T(S,n)<0$ in the interior of $C^{t_{1}}_{p}$. Therefore rearranging the terms we obtain
\begin{align}
-\int_{S_{t_{1}}}T(S,n)\mu_{\hat{g}}|_{S_{t_{1}}}+\frac{1}{3}\int_{D^{t_{1}}_{p}}\varphi^{6}\mu_{\hat{g}}=\int_{C^{t_{1}}_{p}}[T(S,l)+\varphi l(\varphi)]\mu_{\hat{g}}|_{C^{t_{1}}_{p}}\\\nonumber 
+\int_{S_{t_{1}}}\varphi m\mu_{\hat{g}}|_{S_{t_{1}}}-\int_{D^{t_{1}}_{p}}(\frac{1}{4}\Gamma^{\beta}~_{\beta\alpha}x^{\alpha}\hat{g}^{\mu\nu}+\frac{1}{2}\mathcal{ER}^{\mu\nu}_{S})T_{\mu\nu}\mu_{\hat{g}}.
\end{align}
Now noting that $l$ is a past directed null vector and $\partial_{t}-\partial_{r}$ is a future directed null vector on $C^{t_{1}}_{p}$ (and only on $C^{t_{1}}_{p}$), we may write $\partial_{t}-\partial_{r}=-N l$ (of course the lapse $N>0$). Then $S=x^{\mu}\partial_{\mu}=t\partial_{t}+r\partial_{r}=\frac{1}{2}((t+r)(\partial_{t}+\partial_{r})+(t-r)\\(\partial_{t}-\partial_{r}))=\frac{(t-r)}{2}(\partial_{t}-\partial_{r})=t(\partial_{t}-\partial_{r})=-tN l$ on $C^{t_{1}}_{p}$ since $t+r=0$ on $C^{t_{1}}_{p}$. Noting that $t<0$ in the causal past of $p$, $T(S,l)|_{C^{t_{1}}_{p}}=-tN |l(\varphi)|^{2}>0$ and therefore we have \begin{align}
-\int_{S_{t_{1}}}T(S,n)\mu_{\hat{g}}|_{S_{t_{1}}}+\frac{1}{3}\int_{D^{t_{1}}_{p}}\varphi^{6}\mu_{\hat{g}}+\int_{C^{t_{1}}_{p}}tN |l(\varphi)|^{2}\mu_{\hat{g}}|_{C^{t_{1}}_{p}}\\\nonumber 
=-\int_{C^{t_{1}}_{p}}\varphi l(\varphi)\mu_{\hat{g}}|_{C^{t_{1}}_{p}}+\int_{S_{t_{1}}}\varphi m\mu_{\hat{g}}|_{S_{t_{1}}}-\int_{D^{t_{1}}_{p}}(\frac{1}{4}\Gamma^{\beta}~_{\beta\alpha}x^{\alpha}\hat{g}^{\mu\nu}+\frac{1}{2}\mathcal{ER}^{\mu\nu}_{S})T_{\mu\nu}\mu_{\hat{g}}
\end{align}
Now we will show that the right hand side of the previous equation behaves like $|t_{1}|zo(t_{1})$ by simply using H\"older's inequality and the result (\ref{eq:vanishing})
\begin{align}
\int_{S_{t_{1}}}\varphi m\mu_{\hat{g}}|_{S_{t_{1}}} \leq \left(\int_{S_{t_{1}}}\varphi^{2}\mu_{\hat{g}}|_{S_{t_{1}}}\right)^{1/2}\left(\int_{S_{t_{1}}}m^{2}\mu_{\hat{g}}|_{S_{t_{1}}}\right)^{1/2}\\\nonumber
 \leq\left(\int_{S_{t_{1}}}\varphi^{6}\mu_{\hat{g}}|_{S_{t_{1}}}\right)^{1/6}\left(\int_{S_{t_{1}}}\mu_{\hat{g}}|_{S_{t_{1}}}\right)^{1/3}\left(\int_{S_{t_{1}}}m^{2}\mu_{\hat{g}}|_{S_{t_{1}}}\right)^{1/2}\\\nonumber 
\lesssim|t_{1}|\left(\int_{S_{t_{1}}}\varphi^{6}\mu_{\hat{g}}|_{S_{t_{1}}}\right)^{1/6}\left(\int_{S_{t_{1}}}m^{2}\mu_{\hat{g}}|_{S_{t_{1}}}\right)^{1/2}\\\nonumber
\lesssim |t_{1}|zo(t_{1}),
\end{align}
where we have used lemma 3 in the last step. The next term satisfies
\begin{align}
\int_{C^{t_{1}}_{p}}\varphi l(\varphi)\mu_{\hat{g}}|_{C^{t_{1}}_{p}} \leq \left(\int_{C^{t_{1}}_{p}}\varphi^{2}\mu_{\hat{g}}|_{C^{t_{1}}_{p}}\right)^{1/2}\left(\int_{C^{t_{1}}_{p}}|l(\varphi)|^{2}\mu_{\hat{g}}|_{C^{t_{1}}_{p}}\right)^{1/2}\\\nonumber
 \leq\left(\int_{C^{t_{1}}_{p}}\varphi^{6}\mu_{\hat{g}}|_{C^{t_{1}}_{p}}\right)^{1/6}\left(\int_{C^{t_{1}}_{p}}\mu_{\hat{g}}|_{C^{t_{1}}_{p}}\right)^{1/3}\left(\int_{C^{t_{1}}_{p}}|l(\varphi)|^{2}\mu_{\hat{g}}|_{C^{t_{1}}_{p}}\right)^{1/2}\\\nonumber 
\lesssim|t_{1}|\left(\int_{C^{t_{1}}_{p}}\varphi^{6}\mu_{\hat{g}}|_{C^{t_{1}}_{p}}\right)^{1/6}\left(\int_{C^{t_{1}}_{p}}|l(\varphi)|^{2}\mu_{\hat{g}}|_{C^{t_{1}}_{p}}\right)^{1/2}\\\nonumber
\lesssim |t_{1}|zo(t_{1}),
\end{align}
where the last line follows from (\ref{eq:vanishing}). Using (\ref{eq:vanishing})
\begin{align}
\int_{C^{t_{1}}_{p}}N t|\hat{g}(\partial\phi,l)|^{2}\mu_{\hat{g}}|_{C^{t_{1}}_{p}}\lesssim |t_{1}|zo(t_{1}).
\end{align}
Lastly we have the following straightforward estimate 
\begin{align}
\int_{D^{t_{1}}_{p}}(\frac{1}{4}\Gamma^{\beta}~_{\beta\alpha}x^{\alpha}\hat{g}^{\mu\nu}+\frac{1}{2}\mathcal{ER}^{\mu\nu}_{S})T_{\mu\nu}\mu_{\hat{g}}\lesssim |t_{1}|^{3}E_{S_{t_{1}}}.
\end{align}
Therefore noting the positivity of the term $\frac{1}{3}\int_{D^{t_{1}}_{p}}\varphi^{6}\mu_{\hat{g}}$, we have the following estimate  
\begin{align}
-\int_{S_{t_{1}}}T(S,n)\mu_{\hat{g}}|_{S_{t_{1}}}\lesssim |t_{1}|zo(t_{1}).
\end{align}
The results of this section yield the following lemma\\
\textbf{Lemma 4:} \textit{Let $p\in M$ be such that in local coordinates $x(p)=0$ and let $\mathcal{G}$ be its geodesic normal neighbourhood and $C^{t_{1}}_{p}\subset \mathcal{G}$ be its past light cone extending up to the constant time hypersurface $t_{1}$. Utilizing the quasi-local approximate homothetic Killing vector field $S:x^{\mu}\partial_{\mu}$, we obtain the following estimate for the positive entity $-\int_{S_{t_{1}}}T(S,n)\mu_{g}$
\begin{align}
-\int_{S_{t_{1}}}T(S,n)\mu_{\hat{g}}|_{S_{t_{1}}}\lesssim |t_{1}|zo(t_{1}).
\end{align}
}
This lemma will be crucial in obtaining the last estimate which will finish the proof of an $L^{\infty}$ bound of $\varphi$.  

\subsection{Elementary calculations for vector fields and an important inequality}
Before proceeding with the energy estimate associated with the quasi-local vector field $R:=\frac{x^{\mu}}{x_{0}}\partial_{\mu}$, 
we need to perform a series of elementary calculations to represent $\partial_{t}$ and $x^{i}\partial_{i}=\partial_{r}$ in terms of $l$ and $\bar{l}$ throughout the causal past of $p$ in its normal neighbourhood. In a general curved spacetime, we obviously know that $\partial_{t}$ is not necessarily orthogonal to the constant $t$ hypersurfaces. We therefore expand $\partial_{t}$ and $\partial_{r}$ in the null basis as follows 
\begin{align}
\label{eq:expansion1}
\partial_{t}=al+b\bar{l}+c\lambda_{1}+d\lambda_{2},\\
\label{eq:expansion2}
\partial_{r}=a^{'}l+b^{'}\bar{l}+c^{'}\lambda_{1}+d^{'}\lambda_{2}
\end{align}
where $l,\bar{l}$ are the null fields which together with $\{\lambda_{i}\}_{i=1}^{2}$ constitute the null-frame for the tangent space at a point $p\in M$. $(a,b,c,d)$ and $(a^{'},b^{'},c^{'},d^{'})$ may be calculated using the known relations involving the available vector fields assuming $||Riem||_{L^{\infty}}\leq 1$
\begin{align}
\label{eq:expanse1}
a=-N+O(|x|), b=N+O(|x|), c=O(|x|), d=O(|x|)\\
\label{eq:expanse2}
a^{'}=\frac{\sqrt{g_{ij}x^{i}x^{j}}}{r}+O(|x|),b^{'}=\frac{\sqrt{g_{ij}x^{i}x^{j}}}{r}+O(|x|), c^{'}=O(|x|),\\\nonumber d^{'}=O(|x|).
\end{align}
We do not present the lengthy formulas for all the terms but only their leading order behaviours. The leading order behaviours of $(a,b,c,d)$ may be seen more directly by using the expression for $\partial_{t}$ (\ref{eq:shift})
\begin{align}
\partial_{t}=Nn+Y=N(-l+\bar{l})+Y
\end{align}
and noting that $|Y|=O(|x|^{2})$ for $||Riem||_{L^{\infty}}\leq 1$ (\ref{eq:estimates_normal}).

In this section we will establish a few important inequalities. Let us parametrize the past light cone of a point $I\in M$ which is defined by $v=0$ (note \ref{eq:nullcoordinate}) by the spherical null coordinates (based at I) $(u,\theta,\phi)$. Let $\varphi$ be the scalar field. We first see that the following inequality holds
\begin{align}
\int_{C^{-}_{I}}\frac{\varphi^{2}}{u^{2}}\mu_{\hat{g}}|_{C^{-}_{I}}\lesssim \int_{C^{-}_{I}}|\partial_{u}\varphi|^{2}\mu_{\hat{g}}|_{C^{-}_{I}}\nonumber+|u_{1}|\int_{\mathbb{S}^{2}}\varphi^{2}(u_{1},\theta,\varphi)\sqrt{\det(\tilde{g}_{AB}(u_{1},\theta,\phi)}\\d\theta\wedge d\phi
+|u_{1}|\left(\int_{C^{-}_{I}}\varphi^{6}\mu_{\hat{g}}|_{C^{-}_{I}}\right)^{1/3},
\end{align}
where $\sqrt{\det{\tilde{g}}_{AB}(u_{1},\theta,\phi)}d\theta\wedge d\phi$ is the re-scaled (after extracting the conformal factor $u^{2}$) volume form on the boundary sphere $\mathbb{S}^{2}=\partial C^{-}_{I}$ defined by $u=u_{1}$ (actually $u_{1}$ is equal to $2t$ on $C^{-}_{I}$). This inequality follows from elementary calculus. We note that the following holds   
\begin{align}
\partial_{u}(\sqrt{u}\varphi)=\sqrt{u}\partial_{u}\varphi+\frac{\varphi}{2\sqrt{u}}
\end{align}
which after squaring becomes 
\begin{align}
|\partial_{u}\varphi|^{2}=|\frac{1}{\sqrt{u}}\partial_{u}(\sqrt{u}\varphi)-\frac{\varphi}{2u}|^{2}\\\nonumber 
\geq \frac{\varphi^{2}}{4u^{2}}-\frac{\partial_{u}(u\varphi^{2})}{2u^{2}}.
\end{align}
Now multiplying both sides with $\sqrt{-\det{\hat{g}}_{\mu\nu}(u,v,\theta,\phi)}|_{C^{-}_{I}}du\wedge d\theta\wedge d\phi=\sqrt{-\det{\hat{g}}_{\mu\nu}(u,\theta,\phi)}du\wedge d\theta\wedge d\phi=\mu_{\hat{g}}|_{C^{-}_{I}}$ and integrating, we get
\begin{align}
\int_{C^{-}_{I}}\frac{\varphi^{2}}{4u^{2}}\mu_{\hat{g}}|_{C^{-}_{I}}\leq \int_{C^{-}_{I}}|\partial_{u}\varphi|^{2}\nonumber \mu_{\hat{g}}|_{C^{-}_{I}}+\int_{C^{-}_{I}}\frac{\partial_{u}(u\varphi^{2})}{2u^{2}}\sqrt{-\det{\hat{g}}_{\mu\nu}(u,\theta,\phi)}du\wedge d\theta\wedge d\phi.
\end{align}
Now, we have explicitly shown in section (4.2) that at least on $C^{-}_{I}$, the following relation holds 
\begin{align}
\sqrt{-\det(\hat{g}_{\mu\nu}(u,\theta,\phi))}=\frac{1}{2}\sqrt{\det(g_{AB}(u,\theta,\phi))},
\end{align}
where $g_{AB}(u,\theta,\phi)$ is the Riemannian $2-$metric on the topological sphere defined by $u=$constant, $v=0$ ($A,B=\theta,\phi$).
Therefore, the previous inequality is equivalent to the following 
\begin{align}
\label{eq:inequality_medium}
\int_{C^{-}_{I}}\frac{\varphi^{2}}{4u^{2}}\mu_{\hat{g}}|_{C^{-}_{I}}\leq \int_{C^{-}_{I}}|\partial_{u}\varphi|^{2}\nonumber \mu_{\hat{g}}|_{C^{-}_{I}}+\int_{C^{-}_{I}}\frac{\partial_{u}(u\varphi^{2})}{4u^{2}}\sqrt{\det(g_{AB}(u,\theta,\phi))}du\wedge d\theta\wedge d\phi\\\nonumber 
=\int_{C^{-}_{I}}|\partial_{u}\varphi|^{2}\nonumber \mu_{\hat{g}}|_{C^{-}_{I}}+\int_{C^{-}_{I}}\partial_{u}(\frac{\varphi^{2}}{4u}\sqrt{\det(g_{AB}(u,\theta,\phi))})du\wedge d\theta\wedge d\phi\\\nonumber 
+\int_{C^{-}_{I}}\frac{\varphi^{2}}{2u^{2}}\sqrt{\det(g_{AB}(u,\theta,\phi))})du\wedge d\theta\wedge d\phi-\int_{C^{-}_{I}}\frac{\varphi^{2}}{8u}g^{AB}\partial_{u}g_{AB}\sqrt{\det(g_{AB}(u,\theta,\phi))})\\\nonumber 
du\wedge d\theta\wedge d\phi.
\end{align}
Here we note that the last two terms of the previous line are potentially dangerous and need attention. Notice that in the case of Minkowski space $g^{AB}\partial_{u}g_{AB}=\frac{4}{u}$ and therefore, the last two terms cancel each other. In the present context however, there will be extra terms generated due to non-vanishing background curvature. Since we are assuming a point-wise bound on the curvature, the error term is harmless. Let us explicitly show that the term $g^{AB}\partial_{u}g_{AB}$ is equal to the trace of a certain null second fundamental form of a $u=$constant, $v=0$ topological sphere $\mathbb{S}^{2}$. From the expression of the metric on $C^{-}_{I}$ (\ref{eq:nullmetric}), we observe that $\hat{g}_{u\theta}=\hat{g}(\partial_{u},\partial_{\theta})=0=\hat{g}_{u\phi}=\hat{g}(\partial_{u},\partial_{\phi})$ and therefore $\partial_{u}\perp \partial_{\theta}$ and $\partial_{u}\perp \partial_{\phi}$. Notice the following calculations 
\begin{align}
\partial_{u}g_{AB}=\partial_{u}g(\partial_{A},\partial_{B})=\hat{g}(\nabla_{\partial_{u}}\partial_{A},\partial_{B})+\hat{g}(\partial_{A},\nabla_{\partial_{u}}\partial_{B})\\\nonumber 
=\hat{g}(\nabla_{\partial_{A}}\partial_{u},\partial_{B})+\hat{g}(\partial_{A},\nabla_{\partial_{B}}\partial_{u})=2\kappa_{AB},
\end{align}
where we have used the fact that the connection $\nabla$ is torsion free, $[\partial_{u},\partial_{A}]=[\partial_{u},\partial_{B}]=0$, and $\partial_{u}$ is null on $C^{-}_{I}$. Therefore we obtain
\begin{align}
g^{AB}\partial_{u}g_{AB}=2 tr\kappa.
\end{align}
Via explicit calculations in the geodesic normal coordinate system, we will estimate $2tr\kappa$. First note that we are in a geodesic normal coordinate system $(x^{0}=t,x^{1},x^{2},x^{3})$ where the spacetime metric is expressed in terms of the lapse function ($N$), the shift vector field ($Y$), and the Riemannian metric ($g$) induced on constant $t$ space-like hypersurface (\ref{eq:metric},\ref{eq:metric_new}) and each of these entities differs from their respective Minkowski space values by additional curvature terms. Notice that the spherical null coordinate system $(u,v,\theta,\phi)$ is defined as follows (where the associated coordinates take values from their respective domains of definition) 
\begin{align}
t=\frac{u+v}{2}, x^{1}=\frac{v-u}{2}\sin\theta\cos\phi, x^{2}\nonumber=\frac{v-u}{2}\sin\theta\sin\phi, x^{3}=\frac{v-u}{2}\cos\theta.
\end{align}
We compute $g_{AB}$ ($A,B=\theta,\phi$) in terms of $g_{ij}=\hat{g}(\frac{\partial}{\partial x^{i}},\frac{\partial}{\partial x^{j}})$ ($i,j=1,2,3$) explicitly as follows 
\begin{align}
g_{\theta\theta}=\frac{\partial x^{i}}{\partial\theta}\frac{\partial x^{j}}{\partial\theta}g_{ij}, g_{\phi\phi}=\frac{\partial x^{i}}{\partial\phi}\frac{\partial x^{j}}{\partial\phi}g_{ij}, g_{\theta\phi}=\frac{\partial x^{i}}{\partial\theta}\frac{\partial x^{j}}{\partial\phi}g_{ij}. 
\end{align}
Now in the view of (\ref{eq:geodesic3}) and (\ref{eq:metric_new}), 
\begin{align}
\label{eq:metric_new2}
g_{ij}=\delta_{ij}-\left(2\eta_{ab}\int_{0}^{1}\int_{0}^{1}\lambda^{2}_{1}\lambda_{2}R^{b}~_{cj\alpha}(\lambda_{1}\lambda_{2}x)\theta^{c}_{\lambda}(0)\theta^{a}_{i}(0)\nonumber d\lambda_{1}d\lambda_{2}\right)x^{\alpha}x^{\lambda}\\ 
 +\left(\eta_{ab}\int_{0}^{1}\int_{0}^{1}\int_{0}^{1}\int_{0}^{1}\lambda^{2}_{1}\lambda_{2}\lambda^{2}_{3}\lambda_{4}R^{a}~_{pi\alpha}(\lambda_{1}\lambda_{2}x)\right.\\\nonumber 
 \left.R^{b}~_{qj\beta}(\lambda_{3}\lambda_{4}x)\theta^{p}_{\lambda}(0)\theta^{q}_{\delta}(0)d\lambda_{1}d\lambda_{2}d\lambda_{3}d\lambda_{4}\right)x^{\alpha}x^{\beta}x^{\lambda}x^{\delta}
\end{align}
and therefore on $C^{-}_{I}$ 
\begin{align}
g_{\theta\theta}=\frac{u^{2}}{4}+\mathcal{A}, g_{\phi\phi}=\frac{u^{2}}{4}\sin^{2}\theta+\mathcal{B}, g_{\theta\phi}=\mathcal{C},
\end{align}
where $\mathcal{A}$, $\mathcal{B}$, and $\mathcal{C}$ satisfy the following point-wise estimate 
\begin{align}
|A|\lesssim ||Riem||_{L^{\infty}}u^{4}(t,x)+||Riem||^{2}_{L^{\infty}}u^{6}(t,x),\\\nonumber
|B|\lesssim ||Riem||_{L^{\infty}}u^{4}(t,x)\nonumber+||Riem||^{2}_{L^{\infty}}u^{6}(t,x),\\\nonumber
|C|\lesssim ||Riem||_{L^{\infty}}u^{4}(t,x)+||Riem||^{2}_{L^{\infty}}u^{6}(t,x).
\end{align}
Note here that we only need a point-wise bound of certain null components of the curvature. However, since we are working on a curved background spacetime and assuming $||R||_{L^{\infty}}\lesssim 1$, replacing certain null components by full curvature is harmless. Now we can extract the conformal factor $u^{2}$ factor from the metric $g_{AB}$ and write $g_{AB}=u^{2}\tilde{g}_{AB}$. Obviously, $\tilde{g}_{AB}$ is also a metric and satisfies $\tilde{g}^{AB}\tilde{g}_{AB}=2$. Now we can explicitly compute $2tr\kappa=g^{AB}\partial_{u}g_{AB}$ as follows 
\begin{align}
2tr\kappa=g^{AB}\partial_{u}g_{AB}=u^{-2}\tilde{g}^{AB}\partial_{u}(u^{2}\tilde{g}_{AB})=\frac{4}{u}+\tilde{g}^{AB}\partial_{u}\tilde{g}_{AB},
\end{align}
where $|\tilde{g}^{AB}\partial_{u}\tilde{g}_{AB}|$ is to be estimated. Now notice an extremely important fact. The geodesics through the origin $I$ are straight lines. In addition, on $C^{-}_{I}$, $\partial_{u}$ is null and therefore the integral curves of $\partial_{u}$ are parallel to the null geodesic generators of $C^{-}_{I}$ and therefore are straight lines passing through the vertex $I$. On the other hand, in the coordinates $\{x^{\mu}\}$, $x^{\mu}(\lambda)=x^{\mu}\cdot \lambda$ ($\lambda\in[0,1]$ is the affine parameter) is a null geodesic on $C^{-}_{I}$. From the expression (\ref{eq:metric_new2}), we clearly observe that certain null-component of the curvature (not all of its components) and its square are integrated along the null generators. Let us denote these integrals by $\mathcal{I}^{1}_{null}$ and $\mathcal{I}^{2}_{null}$, respectively i.e., 
\begin{align}
\eta_{ab}\int_{0}^{1}\int_{0}^{1}\lambda^{2}_{1}\lambda_{2}R^{b}~_{cj\alpha}(\lambda_{1}\lambda_{2}x)\tilde{l}^{\alpha}\tilde{l}^{\lambda}\theta^{c}_{\lambda}(0)\theta^{a}_{i}(0)\nonumber d\lambda_{1}d\lambda_{2}u^{2}(t,x)=\mathcal{I}^{1}_{null} u^{2}(t,x),\\\nonumber 
\eta_{ab}\int_{0}^{1}\int_{0}^{1}\int_{0}^{1}\int_{0}^{1}\lambda^{2}_{1}\lambda_{2}\lambda^{2}_{3}\lambda_{4}R^{a}~_{pi\alpha}(\lambda_{1}\lambda_{2}x)\tilde{l}^{\alpha}\tilde{l}^{\beta}\tilde{l}^{\lambda}\tilde{l}^{\delta}R^{b}~_{qj\beta}(\lambda_{3}\lambda_{4}x)\theta^{p}_{\lambda}(0)\theta^{q}_{\delta}(0)\prod_{k=1}^{4}d\lambda_{k})\\\nonumber 
u^{4}(t,x)=\mathcal{I}^{2}_{null}u^{4}(t,x),
\end{align}
where $\tilde{l}=-\partial_{t}+\frac{x^{i}}{t}\partial_{i}$ is a null vector field on $C^{-}_{I}$.
Therefore, $\tilde{g}^{AB}\partial_{u}\tilde{g}_{AB}$ is estimated as 
\begin{align}
|\tilde{g}^{AB}\partial_{u}\tilde{g}_{AB}|\lesssim ||\mathcal{I}^{1}_{null}||_{L^{\infty}}|u(t,x)|+||\mathcal{I}^{2}_{null}||_{L^{\infty}}|u^{3}(t,x)|\\\nonumber+||(\partial_{t}-\frac{x^{i}}{t}\partial_{i}) \mathcal{I}^{1}_{null}||_{L^{\infty}}u^{2}(t,x)
+||(\partial_{t}-\frac{x^{i}}{t}\partial_{i}) \mathcal{I}^{2}_{null}||_{L^{\infty}}u^{4}(t,x).
\end{align}
Here $\partial_{u}$ while expressed in coordinates $(t,x^{1},x^{2},x^{3})$ reads $\partial_{t}-\frac{x^{i}}{t}\partial_{i}$. The constants involved in the associated estimates have suitable dimensions to make everything dimensionally consistent.
Notice that we can assume a uniform upper bound on $||\mathcal{I}^{1}_{null}||_{L^{\infty}}$, $||\partial \mathcal{I}^{1}_{null}||_{L^{\infty}}$, $||\mathcal{I}^{2}_{null}||_{L^{\infty}}$, and $||\partial\mathcal{I}^{2}_{null}||_{L^{\infty}}$ only in the case when we are working on a curved \textit{background}. Such freedom will be lost while studying the gravity problem (see \cite{klainerman2005causal} for the difficulty associated with controlling the point-wise behaviour of $tr \kappa$ in vacuum Einsteinian spacetimes when the curvature has limited regularity). Denoting $||\mathcal{I}^{1}_{null}||_{L^{\infty}}|u(t,x)|+||\mathcal{I}^{2}_{null}||_{L^{\infty}}|u^{3}(t,x)|+||(\partial_{t}-\frac{x^{i}}{t}\partial_{i}) \mathcal{I}^{1}_{null}||_{L^{\infty}}u^{2}(t,x)
+||(\partial_{t}-\frac{x^{i}}{t}\partial_{i}) \mathcal{I}^{2}_{null}||_{L^{\infty}}u^{4}(t,x)$ by $\mathcal{K}$ (and therefore $\mathcal{K}$ satisfies $|\mathcal{K}|\lesssim |u(t,x)|$ in view of the global hyperbolicity), we notice that
\begin{align}
|2 tr\kappa-\frac{4}{u}|\lesssim \mathcal{K}.
\end{align}
Therefore, the inequality of interest (\ref{eq:inequality_medium})
\begin{align}
\label{eq:nextinequalityimpt}
\int_{C^{-}_{I}}\frac{\varphi^{2}}{4u^{2}}\mu_{\hat{g}}|_{C^{-}_{I}}\leq \int_{C^{-}_{I}}|\partial_{u}\varphi|^{2}\nonumber \mu_{\hat{g}}|_{C^{-}_{I}}+\int_{C^{-}_{I}}\partial_{u}(\frac{\varphi^{2}}{4u}\sqrt{\det(g_{AB}(u,\theta,\phi))})du\wedge d\theta\wedge d\phi\\\nonumber
+\int_{C^{-}_{I}}\frac{\varphi^{2}}{u}\mathcal{K}\sqrt{\det(g_{AB}(u,\theta,\phi))}du\wedge d\theta\wedge d\phi\\
=\int_{C^{-}_{I}}|\partial_{u}\varphi|^{2}\nonumber \mu_{\hat{g}}|_{C^{-}_{m}}+\int_{C^{-}_{I}}\partial_{u}(\frac{\varphi^{2}}{4u}\sqrt{\det(g_{AB}(u,\theta,\phi))})du\wedge d\theta\wedge d\phi\\ 
+2\int_{C^{-}_{I}}\frac{\varphi^{2}}{u}\mathcal{K}\sqrt{-\det(\hat{g}_{\mu\nu}(u,\theta,\phi))}du\wedge d\theta\wedge d\phi,
\end{align}
where we have once again used $\sqrt{-\det(\hat{g}_{\mu\nu}(u,\theta,\phi))}=\frac{1}{2}\sqrt{\det(g_{AB}(u,\theta,\phi))}$ only on $C^{-}_{I}$.
Now applying Holder on the last term of (\ref{eq:nextinequalityimpt}) we have 
\begin{align}
\int_{C^{-}_{I}}\frac{\varphi^{2}}{u}\mathcal{K}\sqrt{-\det(\hat{g}_{\mu\nu}(u,\theta,\phi))}du\wedge d\theta\wedge d\phi\\\nonumber 
\lesssim u_{1}^{2}\left(\int_{C^{-}_{I}}\varphi^{6}\mu_{\hat{g}}|_{C^{-}_{I}}\right)^{1/3}.
\end{align}
Therefore we have the desired inequality after integrating the total derivative term in (\ref{eq:nextinequalityimpt}) and using the fact that $\sqrt{\det g_{AB}}=u^{2}\sqrt{\det \tilde{g}_{AB}}$
\begin{align}
\label{eq:inter}
\int_{C^{-}_{I}}\frac{\varphi^{2}}{u^{2}}\mu_{\hat{g}}|_{C^{-}_{I}}\lesssim \int_{C^{-}_{I}}|\partial_{u}\varphi|^{2}\mu_{\hat{g}}|_{C^{-}_{I}}\nonumber+|u_{1}|\int_{\mathbb{S}^{2}}\varphi^{2}(u_{1},\theta,\varphi)\sqrt{\det(\tilde{g}_{AB})}d\theta\wedge d\phi\\
+u^{2}_{1}\left(\int_{C^{-}_{I}}\varphi^{6}\mu_{\hat{g}}|_{C^{-}_{I}}\right)^{1/3},
\end{align}
where the constant involved only depends on the background geometry and of the order $1$ by assumption of global hyperbolicity. Let us now obtain a second inequality which will be of importance. Now consider that the intersection of causal past $D^{-}_{m}$ of $m$ with the $t=t_{1}$ hypersurface be $S_{t_{1}}$. We need to finally  
estimate the following term  
\begin{align}
|u_{1}|\int_{\mathbb{S}^{2}}\varphi^{2}(u_{1},\theta,\varphi)\sqrt{\det(\tilde{g}_{AB})(u_{1},\theta,\phi)}d\theta\wedge d\phi.
\end{align}
If one for now goes back to the spherical coordinates i.e., $(u,v,\theta,\phi)\mapsto (t,r,\theta,\phi)$, then it is obvious that $u_{1}=(t-r)|_{\partial C^{-}_{m}}=2t|_{\partial C^{-}_{m}}=-2r|_{\partial C^{-}_{m}}$ (since on $C^{-}_{m}$, $v=t+r=0$). Therefore
\begin{align}
|u_{1}|\int_{\mathbb{S}^{2}}\varphi^{2}(u_{1},\theta,\varphi)\sqrt{\det(\tilde{g}_{AB})(u_{1},\theta,\phi)}d\theta\wedge d\phi\\\nonumber 
\approx r_{1}\int_{\mathbb{S}^{2}}\varphi^{2}(r_{1},\theta,\varphi)\sqrt{\det(\tilde{g}_{AB})(r_{1},\theta,\phi)}d\theta\wedge d\phi\\\nonumber 
=|t_{1}|\int_{\mathbb{S}^{2}}\varphi^{2}(-t_{1},\theta,\varphi)\sqrt{\det(\tilde{g}_{AB})(-t_{1},\theta,\phi)}d\theta\wedge d\phi
\end{align}
Now consider the following calculations over the topological sphere $\mathbb{S}^{2}$ (we denote the volume form on this $\mathbb{S}^{2}$ by $\sqrt{\det(\tilde{g}_{AB})(r,\theta,\phi)}$, where $\tilde{g}_{AB}$ is the 2-metric after extracting the conformal factor $r^{2}$ via explicit calculations similar to the one presented previously; notice that these spheres foliate the space-like topological ball $S_{t_{1}}$) 
\begin{align}
\frac{\partial}{\partial r}\left(r^{2}\int_{\mathbb{S}^{2}}\varphi^{4}(r,\theta,\varphi)\sqrt{\det(\tilde{g}_{AB})(r,\theta,\phi)}d\theta\wedge d\phi\right)\\\nonumber 
=2r\int_{\mathbb{S}^{2}}\varphi^{4}(r,\theta,\varphi)\sqrt{\det(\tilde{g}_{AB})(r,\theta,\phi)}d\theta\wedge d\phi\\\nonumber 
+4r^{2}\int_{\mathbb{S}^{2}}\varphi^{3}\partial_{r}\varphi \sqrt{\det(\tilde{g}_{AB})(r,\theta,\phi)}d\theta\wedge d\phi\\\nonumber 
+\frac{r^{2}}{2}\int_{\mathbb{S}^{2}}\varphi^{4}\tilde{g}^{AB}\partial_{r}\tilde{g}_{AB}\sqrt{\det(\tilde{g}_{AB})(r,\theta,\phi)}d\theta\wedge d\phi.
\end{align}
 Now if we integrate this entity over $r$ from $0$ to $r_{1}$ the three dimensional integral becomes an integral over $S_{t_{1}}$ and application of Cauchy-Scwartz yields
 \begin{align}
 \label{eq:phi4}
 r_{1}^{2}\int_{\mathbb{S}^{2}}\varphi^{4}(r_{1},\theta,\varphi)\sqrt{\det(\tilde{g}_{AB})(r_{1},\theta,\phi)}d\theta\wedge d\phi\\\nonumber 
 =2\int_{S_{t_{1}}}r\varphi^{4}\sqrt{\det(\tilde{g}_{AB})(r,\theta,\phi)}dr\wedge d\theta\wedge d\phi\\\nonumber 
 +4\int_{S_{t_{1}}}\varphi^{3}\partial_{r}\varphi r^{2}\sqrt{\det(\tilde{g}_{AB})(r,\theta,\phi)}d\theta\wedge d\phi\\\nonumber 
 +\frac{1}{2}\int_{\mathbb{S}^{2}}\varphi^{4}\tilde{g}^{AB}\partial_{r}\tilde{g}_{AB}r^{2}\sqrt{\det(\tilde{g}_{AB})(r,\theta,\phi)}dr\wedge d\theta\wedge d\phi\\\nonumber 
 \lesssim \left(\int_{S_{t_{1}}}\varphi^{6}\mu_{\hat{g}}|_{S_{t_{1}}}\right)^{1/2}\left(\int_{S_{t_{1}}}\frac{\varphi^{2}}{r^{2}}\mu_{\hat{g}}|_{S_{t_{1}}}\right)^{1/2}+\left(\int_{S_{t_{1}}}\varphi^{6}\mu_{\hat{g}}|_{S_{t_{1}}}\right)^{1/2}\\\nonumber \left(\int_{S_{t_{1}}}(\partial_{r}\varphi)^{2}\mu_{\hat{g}}|_{S_{t_{1}}}\right)^{1/2}+r^{2}_{1}\left(\int_{S_{t_{1}}}\varphi^{6}\mu_{\hat{g}}|_{S_{t_{1}}}\right)^{2/3}\\\nonumber 
 =\left(\int_{S_{t_{1}}}\varphi^{6}\mu_{\hat{g}}|_{S_{t_{1}}}\right)^{1/2}\left(\left(\int_{S_{t_{1}}}\frac{\varphi^{2}}{r^{2}}\mu_{\hat{g}}|_{S_{t_{1}}}\right)^{1/2}+\left(\int_{S_{t_{1}}}(\partial_{r}\varphi)^{2}\mu_{\hat{g}}|_{S_{t_{1}}}\right)^{1/2}\right.\\\nonumber 
 \left.+r^{2}_{1}\left(\int_{S_{t_{1}}}\varphi^{6}\mu_{\hat{g}}|_{S_{t_{1}}}\right)^{1/6}\right).
 \end{align}
Here once again $|\tilde{g}^{AB}\partial_{r}\tilde{g}_{AB}|$ can be bounded by curvature components and the radial derivative of the integral of curvature and its square over $\mathbb{S}^{2}$. The calculations are similar to the one we performed in the null case. Therefore, we do not repeat the same here. Now analogous calculations as in the previous case may be performed on the topological ball $S_{t_{1}}$ instead of the cone $C^{-}_{I}$ to yield 
\begin{align}
\int_{S_{t_{1}}}\frac{\varphi^{2}}{r^{2}}\mu_{\hat{g}}|_{S_{t_{1}}}\lesssim \int_{S_{t_{1}}}|\partial_{r}\varphi|^{2}\mu_{\hat{g}}|_{S_{t_{1}}}\nonumber+|r_{1}|\int_{\mathbb{S}^{2}}\varphi^{2}(r_{1},\theta,\varphi)\sqrt{\det(g_{ab})}d\theta\wedge d\phi\\
+r^{2}_{1}\left(\int_{S_{t_{1}}}\varphi^{6}\mu_{\hat{g}}|_{S_{t_{1}}}\right)^{1/3}.
\end{align}
Now since $\partial C^{-}_{I}=\partial S_{t_{1}}=\mathbb{S}^{2}$, the topological sphere,  we use the inequality (\ref{eq:phi4}) and the previous inequality becomes
\begin{align}
\int_{S_{t_{1}}}\frac{\varphi^{2}}{r^{2}}\mu_{\hat{g}}|_{S_{t_{1}}}\lesssim \int_{S_{t_{1}}}|\partial_{r}\varphi|^{2}\mu_{\hat{g}}|_{S_{t_{1}}}\nonumber+\left(\int_{S_{t_{1}}}\varphi^{6}\mu_{\hat{g}}|_{S_{t_{1}}}\right)^{1/4}\left(\left(\int_{S_{t_{1}}}\frac{\varphi^{2}}{r^{2}}\mu_{\hat{g}}|_{S_{t_{1}}}\right)^{1/2}\right.\\\nonumber
 \left.+\left(\int_{S_{t_{1}}}(\partial_{r}\varphi)^{2}\mu_{\hat{g}}|_{S_{t_{1}}}\right)^{1/2}+r^{2}_{1}\left(\int_{S_{t_{1}}}\varphi^{6}\mu_{\hat{g}}|_{S_{t_{1}}}\right)^{1/6}\right)^{1/2}\\
 +r^{2}_{1}\left(\int_{S_{t_{1}}}\varphi^{6}\mu_{\hat{g}}|_{S_{t_{1}}}\right)^{1/3},
\end{align}
which yields by iteration and $r_{1}=-t_{1}$
\begin{align}
\int_{S_{t_{1}}}\frac{\varphi^{2}}{r^{2}}\mu_{\hat{g}}|_{S_{t_{1}}}\lesssim  \int_{S_{t_{1}}}|\partial_{r}\varphi|^{2}\mu_{\hat{g}}|_{S_{t_{1}}}+\left(\int_{S_{t_{1}}}\varphi^{6}\mu_{\hat{g}}|_{S_{t_{1}}}\right)^{1/3}\nonumber+t^{2}_{1}\left(\int_{S_{t_{1}}}\varphi^{6}\mu_{\hat{g}}|_{S_{t_{1}}}\right)^{1/3},
\end{align}
where the involved constant may involve a positive power of $|t_{1}|$ and is therefore harmless. In addition notice that these implicit constants have dimensions such that the each inequality here is dimensionally \textit{consistent}. 
Substituting this result back into the inequality (\ref{eq:phi4}) yields 
\begin{align}
\label{eq:boundary2}
r_{1}^{2}\int_{\mathbb{S}^{2}}\varphi^{4}(r_{1},\theta,\varphi)\sqrt{\det(\tilde{g}_{AB})(r_{1},\theta,\phi)}d\theta\wedge d\phi\\\nonumber 
\lesssim \left(\int_{S_{t_{1}}}\varphi^{6}\mu_{\hat{g}}|_{S_{t_{1}}}\right)^{1/2}\left((\int_{S_{t_{1}}}|\partial_{r}\varphi|^{2}\mu_{\hat{g}}|_{S_{t_{1}}})^{1/2}+(\int_{S_{t_{1}}}\varphi^{6}\mu_{\hat{g}}|_{S_{t_{1}}})^{1/6}\right).
\end{align}
Therefore we finally obtain the following inequality after substituting (\ref{eq:boundary2}) into (\ref{eq:inter}), which will be of tremendously important in the final analysis
\begin{align}
\label{eq:maininequality}
\int_{C^{-}_{I}}\frac{\varphi^{2}}{u^{2}}\mu_{\hat{g}}|_{C^{-}_{I}}\lesssim \int_{C^{-}_{I}}|\partial_{u}\varphi|^{2}\mu_{\hat{g}}|_{C^{-}_{I}}\nonumber+u^{2}_{1}\left(\int_{C^{-}_{I}}\varphi^{6}\mu_{\hat{g}}|_{C^{-}_{I}}\right)^{1/3}\\
+\left(\int_{S_{t_{1}}}\varphi^{6}\mu_{\hat{g}}|_{S_{t_{1}}}\right)^{1/4}\left((\int_{S_{t_{1}}}g(\partial\varphi,\partial\varphi)\mu_{\hat{g}}|_{S_{t_{1}}})^{1/4}+(\int_{S_{t_{1}}}\varphi^{6}\mu_{\hat{g}}|_{S_{t_{1}}})^{1/12}\right)\\\nonumber 
\lesssim \int_{C^{-}_{I}}|l(\varphi)|^{2}\mu_{\hat{g}}|_{C^{-}_{I}}\nonumber+u^{2}_{1}\left(\int_{C^{-}_{I}}\varphi^{6}\mu_{\hat{g}}|_{C^{-}_{I}}\right)^{1/3}\\
+\left(\int_{S_{t_{1}}}\varphi^{6}\mu_{\hat{g}}|_{S_{t_{1}}}\right)^{1/4}\left((\int_{S_{t_{1}}}g(\partial\varphi,\partial\varphi)\mu_{\hat{g}}|_{S_{t_{1}}})^{1/4}+(\int_{S_{t_{1}}}\varphi^{6}\mu_{\hat{g}}|_{S_{t_{1}}})^{1/12}\right)
\end{align}
since $|\partial_{r}\varphi|^{2}\lesssim g^{ij}\partial_{i}\varphi\partial_{j}\varphi$ and $\partial_{u}=-Nl$ on $C^{-}_{I}$, $N=O(1)$ modulo point-wise curvature which is bounded by $1$. This may be verified by a simple calculation. Note that we are in the normal coordinate based at the vertex of the cone $C^{-}_{I}$ and therefore the optical function $\Gamma=\hat{g}_{\mu\nu}x^{\mu}x^{\nu}=-t^{2}+r^{2}$ vanishes i.e., $t^{2}-r^{2}=0$. A calculation yields $N=\sqrt{g(x,x)}/r$ holds only on the mantle of the cone. Therefore, on the mantle $C^{-}_{I}$, $l=-\frac{1}{2N}(\partial_{t}-\partial_{r})$ which is quite obvious since $\partial_{t}-\partial_{r}$ is null on $C^{-}_{I}$. On the other hand, in coordinate $(u,v,\theta,\phi)$ we have $\partial_{u}=\frac{1}{2}(\partial_{t}-\partial_{r})$ or $\partial_{u}=-Nl$
only on $C^{-}_{I}$. In addition to the previous inequality, one may repeat the exact same calculations for $r_{1}^{2}\int_{\mathbb{S}^{2}}\varphi^{4}(r_{1},\theta,\varphi)\sqrt{\det(\tilde{g}_{AB})(r_{1},\theta,\phi)}d\theta\wedge d\phi$ over $\partial S_{t_{1}}$ as the boundary of $C^{-}_{I}$ since $\partial C^{-}_{I}=\partial S_{t_{1}}=\mathbb{S}^{2}$, the topological sphere, to yield the following inequality
\begin{align}
\label{eq:boundaryinequalprevious}
r_{1}^{2}\int_{\mathbb{S}^{2}}\varphi^{4}(r_{1},\theta,\varphi)\sqrt{\det(\tilde{g}_{AB})(r_{1},\theta,\phi)}d\theta\wedge d\phi\\\nonumber 
\lesssim \left(\int_{C^{-}_{I}}\varphi^{6}\mu_{\hat{g}}|_{C^{-}_{I}}\right)^{1/2}\left((\int_{C^{-}_{I}}|l(\varphi)|^{2}\mu_{\hat{g}}|_{C^{-}_{I}})^{1/2}+(\int_{C^{-}_{I}}\varphi^{6}\mu_{\hat{g}}|_{C^{-}_{I}})^{1/6}\right).
\end{align}
This inequality was actually used in (\ref{eq:boundaryinequal}).

\subsection{Energy estimate using the quasi-local vector field `$R=\frac{x^{\mu}}{x_{0}}\partial_{\mu}$'}
Proceeding the same way as before that is, multiplying the equation of motion by $R(\varphi)+\frac{\varphi}{x_{0}}$ followed by elementary manipulation, we obtain 
\begin{align}
\label{eq:estimateR}
\nabla^{\mu}(R^{\nu}T_{\mu\nu}+\frac{\varphi}{x_{0}}\nabla_{\mu}\varphi)-\nabla^{\mu}R^{\nu}T_{\mu\nu}-\frac{1}{x_{0}}\nabla_{\mu}\varphi\nabla^{\mu}\varphi
+\frac{1}{x^{2}_{0}}\varphi\nabla^{\mu}\varphi g_{0\mu}\\\nonumber =\frac{1}{x_{0}}\varphi^{6}.
\end{align}
Now we evaluate some of the terms explicitly. The strain tensor $^{R}\pi_{\mu\nu}$ is explicitly computed to be 
\begin{align}
\nabla_{\mu}R_{\nu}+\nabla_{\nu}R_{\mu}=\frac{2\hat{g}_{\mu\nu}}{x_{0}}-\frac{\hat{g}_{0\mu}x_{\nu}+\hat{g}_{0\nu}x_{\mu}}{x^{2}_{0}}+\frac{\hat{g}_{\nu\beta}\Gamma^{\beta}~_{\mu\alpha}x^{\alpha}+\hat{g}_{\mu\beta}\Gamma^{\beta}_{\nu\alpha}x^{\alpha}}{x_{0}}.
\end{align}
Writing $\nabla^{\mu}R^{\nu}T_{\mu\nu}=\frac{1}{2}(\nabla^{\mu}R^{\nu}+\nabla^{\nu}R^{\mu})T_{\mu\nu}$ and substituting $T_{\mu\nu}$ and $^{R}\pi_{\mu\nu}$ yields
\begin{align}
\nabla^{\mu}R^{\nu}T_{\mu\nu}=\frac{1}{2}(\nabla_{\mu}R_{\nu}+\nabla_{\nu}R_{\mu})(\nabla^{\mu}\varphi\nabla^{\nu}\varphi-\frac{1}{2}g(\partial\varphi,\partial\varphi)\hat{g}^{\mu\nu}\nonumber-\frac{1}{6}\varphi^{6}\hat{g}^{\mu\nu})\\\nonumber 
=\frac{1}{2}\left(\frac{2\hat{g}_{\mu\nu}}{x_{0}}-\frac{\hat{g}_{0\mu}x_{\nu}+\hat{g}_{0\nu}x_{\mu}}{x^{2}_{0}}+\frac{\hat{g}_{\nu\beta}\Gamma^{\beta}~_{\mu\alpha}x^{\alpha}+\hat{g}_{\mu\beta}\Gamma^{\beta}_{\nu\alpha}x^{\alpha}}{x_{0}}\right)\\\nonumber \left(\nabla^{\mu}\varphi\nabla^{\nu}\varphi-\frac{1}{2}g(\partial\varphi,\partial\varphi)\hat{g}^{\mu\nu}\nonumber-\frac{1}{6}\varphi^{6}\hat{g}^{\mu\nu}\right)\\\nonumber 
=-\frac{1}{x_{0}}\left(\nabla_{\mu}\varphi\nabla^{\mu}\varphi+\frac{2}{3}\varphi^{6}\right)-\frac{1}{x^{2}_{0}}\nabla_{0}\varphi x^{\nu}\nabla_{\nu}\varphi+\frac{1}{2x_{0}}\hat{g}(\partial\varphi,\partial\varphi)+\frac{1}{6x_{0}}\varphi^{6}\\\nonumber
+\frac{1}{2}\left(\frac{\hat{g}_{\nu\beta}\Gamma^{\beta}~_{\mu\alpha}x^{\alpha}+\hat{g}_{\mu\beta}\Gamma^{\beta}_{\nu\alpha}x^{\alpha}}{x_{0}}\right)T^{\mu\nu}
\\\nonumber 
=-\frac{1}{2x_{0}}\nabla_{\mu}\varphi\nabla^{\mu}\varphi-\frac{1}{2x_{0}}\varphi^{6}-\frac{1}{x^{2}_{0}}\nabla_{0}\varphi x^{\nu}\nabla_{\nu}\varphi+\mathcal{L}_{\mu\nu}T^{\mu\nu},
\end{align}
where we have denoted the term $\left(\frac{\hat{g}_{\nu\beta}\Gamma^{\beta}~_{\mu\alpha}x^{\alpha}+\hat{g}_{\mu\beta}\Gamma^{\beta}_{\nu\alpha}x^{\alpha}}{x_{0}}\right)$ by $\mathcal{L}_{\mu\nu}$. Notice that 
\begin{align}
|\mathcal{L}_{\mu\nu}|=O(|x|).
\end{align}
The term $-\nabla^{\mu}R^{\nu}T_{\mu\nu}-\frac{1}{x_{0}}\nabla_{\mu}\varphi\nabla^{\mu}\varphi+\frac{1}{x^{2}_{0}}\varphi\nabla^{\mu}\varphi g_{0\mu}$ in the equation (\ref{eq:estimateR}) becomes
\begin{align}
-\nabla^{\mu}R^{\nu}T_{\mu\nu}-\frac{1}{x_{0}}\nabla_{\mu}\varphi\nabla^{\mu}\varphi+\frac{1}{x^{2}_{0}}\varphi\nabla^{\mu}\varphi g_{0\mu}\\\nonumber 
=\frac{1}{2x_{0}}\nabla_{\mu}\varphi\nabla^{\mu}\varphi+\frac{1}{2x_{0}}\varphi^{6}+\frac{1}{x^{2}_{0}}\nabla_{0}\varphi x^{\nu}\nabla_{\nu}\varphi-\mathcal{L}_{\mu\nu}T^{\mu\nu}
-\frac{1}{x_{0}}\nabla_{\mu}\varphi\nabla^{\mu}\varphi+\frac{1}{x^{2}_{0}}\varphi\nabla_{0}\varphi\\\nonumber 
=-\frac{1}{2x_{0}}\nabla_{\mu}\varphi\nabla^{\mu}\varphi+\frac{1}{x^{2}_{0}}\nabla_{0}\varphi x^{\nu}\nabla_{\nu}\varphi+\frac{1}{x^{2}_{0}}\varphi\nabla_{0}\varphi+\frac{1}{2x_{0}}\varphi^{6}-\mathcal{L}_{\mu\nu}T^{\mu\nu}\\\nonumber 
=-\frac{1}{2x_{0}}\nabla_{0}\varphi\nabla^{0}\varphi-\frac{1}{2x_{0}}\nabla_{i}\varphi\nabla^{i}\varphi+\frac{1}{x_{0}}\nabla_{0}\varphi\nabla^{0}\varphi+\frac{x^{i}}{x^{2}_{0}}\nabla_{0}\varphi\nabla_{i}\varphi+\frac{1}{x^{2}_{0}}\varphi\nabla_{0}\varphi\\\nonumber 
+\frac{1}{2x_{0}}\varphi^{6}-\mathcal{L}_{\mu\nu}T^{\mu\nu}\\\nonumber 
=\frac{1}{2x_{0}}\nabla_{0}\varphi\nabla^{0}\varphi-\frac{1}{2x_{0}}\nabla_{i}\varphi\nabla^{i}\varphi+\frac{x^{i}}{x^{2}_{0}}\nabla_{0}\varphi\nabla_{i}\varphi+\frac{1}{x^{2}_{0}}\varphi\nabla_{0}\varphi+\frac{1}{2x_{0}}\varphi^{6}\\\nonumber 
-\mathcal{L}_{\mu\nu}T^{\mu\nu}.
\end{align}
Now the term $\frac{1}{x^{2}_{0}}\varphi\nabla_{0}\varphi$ may be further reduced through the following calculation
\begin{align}
\frac{1}{x^{2}_{0}}\varphi\nabla_{0}\varphi=\nabla_{0}(\frac{\varphi^{2}}{2x^{2}_{0}})-\varphi^{2}\nabla_{0}(\frac{1}{2x^{2}_{0}})=\nabla_{0}(\frac{\varphi^{2}}{2x^{2}_{0}})+\frac{\varphi^{2}g_{00}}{x^{3}_{0}}\\\nonumber 
=\delta^{\mu}_{0}\nabla_{\mu}(\frac{\varphi^{2}}{x^{2}_{0}})+\frac{\varphi^{2}g_{00}}{x^{3}_{0}}.
\end{align}
Therefore the term $-\nabla^{\mu}R^{\nu}T_{\mu\nu}-\frac{1}{x_{0}}\nabla_{\mu}\varphi\nabla^{\mu}\varphi+\frac{1}{x^{2}_{0}}\varphi\nabla^{\mu}\varphi g_{0\mu}$ has the following final form 
\begin{align}
-\nabla^{\mu}R^{\nu}T_{\mu\nu}-\frac{1}{x_{0}}\nabla_{\mu}\varphi\nabla^{\mu}\varphi+\frac{1}{x^{2}_{0}}\varphi\nabla^{\mu}\varphi g_{0\mu}\\\nonumber 
=\frac{1}{2x_{0}}\nabla_{0}\varphi\nabla^{0}\varphi-\frac{1}{2x_{0}}\nabla_{i}\varphi\nabla^{i}\varphi+\frac{x^{i}}{x^{2}_{0}}\nabla_{0}\varphi\nabla_{i}\varphi+\frac{1}{2x_{0}}\varphi^{6}
+\delta^{\mu}_{0}\nabla_{\mu}(\frac{\varphi^{2}}{2x^{2}_{0}})\\\nonumber +\frac{\varphi^{2}g_{00}}{x^{3}_{0}}-\mathcal{L}_{\mu\nu}T^{\mu\nu}.
\end{align}
Substituting this expression into the equation (\ref{eq:estimateR}) yields 
\begin{align}
\nabla^{\mu}(R^{\nu}T_{\mu\nu}+\frac{\varphi}{x_{0}}\nabla_{\mu}\varphi)-\nabla^{\mu}R^{\nu}T_{\mu\nu}-\frac{1}{x_{0}}\nabla_{\mu}\varphi\nabla^{\mu}\varphi\\\nonumber 
+\frac{1}{x^{2}_{0}}\varphi\nabla^{\mu}\varphi g_{0\mu}
=\frac{1}{x_{0}}\varphi^{6}\\\nonumber 
\nabla^{\mu}(R^{\nu}T_{\mu\nu}+\frac{\varphi}{x_{0}}\nabla_{\mu}\varphi+\hat{g}_{0\mu}\frac{\varphi^{2}}{2x_{0}^{2}})=\frac{1}{2x_{0}}\varphi^{6}-\frac{1}{2x_{0}}\nabla_{0}\varphi\nabla^{0}\varphi+\frac{1}{2x_{0}}\nabla_{i}\varphi\nabla^{i}\varphi\\\nonumber 
-\frac{x^{i}}{x^{2}_{0}}\nabla_{0}\varphi\nabla_{i}\varphi-\frac{\varphi^{2}g_{00}}{x^{3}_{0}}+\mathcal{L}_{\mu\nu}T^{\mu\nu}.
\end{align}

%\begin{center}
%\begin{figure}
%\begin{center}
%\includegraphics[width=13cm,height=60cm,keepaspectratio,keepaspectratio]{lightcone3.eps}
%\end{center}
%\begin{center}
%\caption{In order to estimate integrals over the mantle of the interior null cone $C^{t_{1}}_{q}$, we integrate over the domain $D^{t_{1}}_{p}-J^{t_{1}}_{q}$. Clearly the null geodesics passing through $q$ are not straight lines in general.}
%\label{fig:fig2}
%\end{center}
%\end{figure}
%\end{center}

\begin{center}
\begin{figure}
\begin{center}
\includegraphics[width=13cm,height=20cm,keepaspectratio,keepaspectratio]{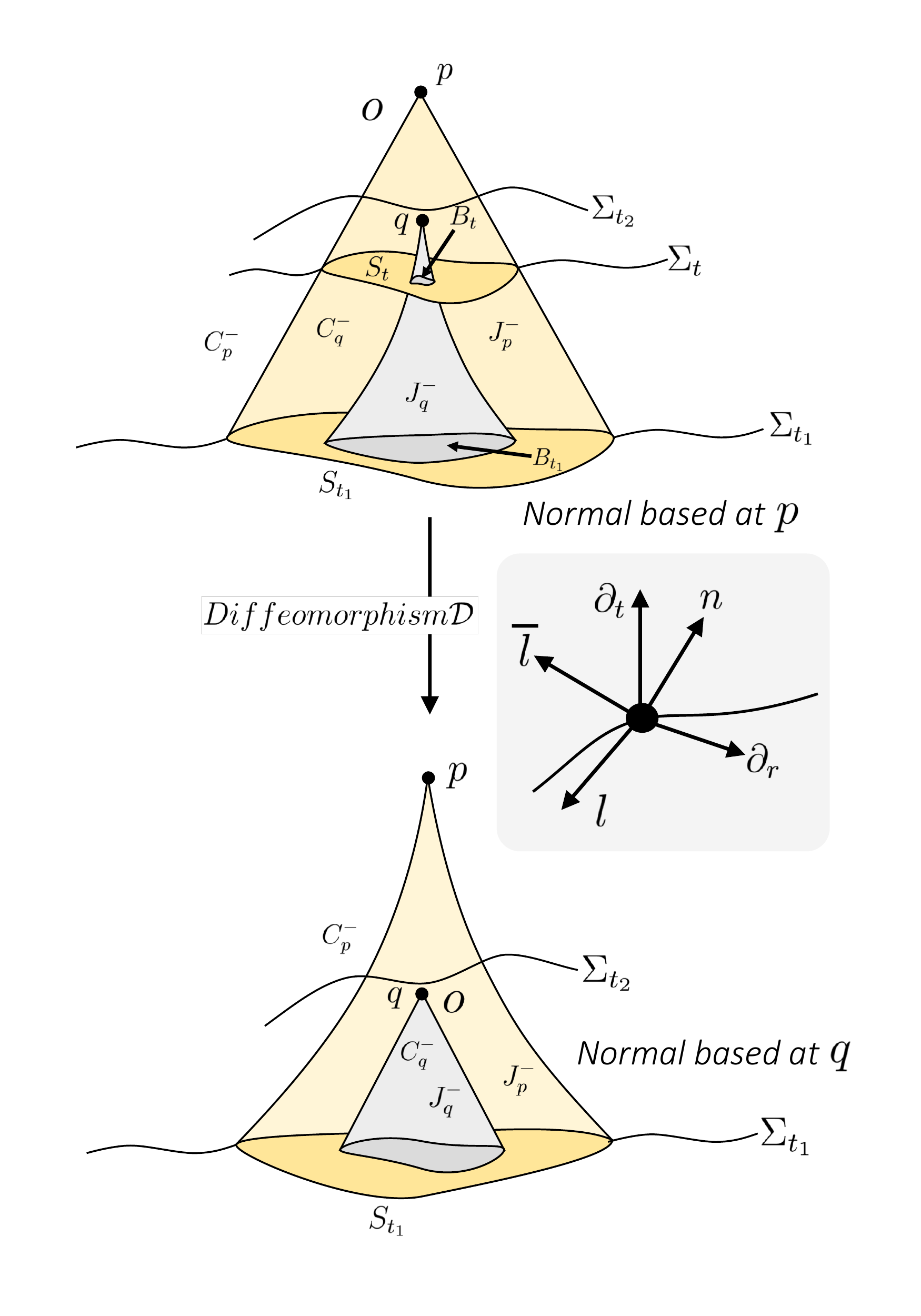}
\end{center}
\begin{center}
\caption{In the top figure, we integrate over the domain $D^{t_{1}}_{p}-J^{t_{1}}_{q}$. The grey shaded portion of $S_{t_{1}}$ is denoted by $B_{t_{1}}$. 
If $|\varphi|$ attains its supremum within the causal past$D^{t_{1}}_{p}$ at $q$, then we move to a coordinate system which is normal based at $q$ and utilize the estimates on the diffeomorphism invariant integrals. For convenience, we still denote the intersection of the solid null cone of $p$ and the initial hypersurface by $S_{t_{1}}$.}
\label{fig:fig3}
\end{center}
\end{figure}
\end{center}

Now we integrate over the domain $D^{t_{1}}_{p}-J^{t}_{q}-D^{t_{2}}_{p}$ (we set $t_{2}\to 0$) as shown in figure (\ref{fig:fig3}), to obtain 
\begin{align}
\int_{S_{t_{1}}-B_{t_{1}}}\left(T(R,n)+\frac{\varphi}{x_{0}}m+n_{0}\frac{\varphi^{2}}{2x^{2}_{0}}\right)\mu_{\hat{g}}|_{S_{t_{1}}}-\int_{S_{t_{2}}}\left(T(R,n)+\nonumber\frac{\varphi}{x_{0}}m+n_{0}\frac{\varphi^{2}}{2x^{2}_{0}}\right)\mu_{\hat{g}}|_{S_{t_{2}}}\\\nonumber
-\int_{C^{t_{1}}_{q}}\left(T(R,l)+\frac{\varphi}{x_{0}}l(\varphi)+l_{0}\frac{\varphi^{2}}{2x^{2}_{0}}\right)\mu_{\hat{g}}|_{C^{t_{1}}_{q}} 
+\int_{C^{T}_{p}}\left(T(R,l)+\frac{\varphi}{x_{0}}l(\varphi)+l_{0}\frac{\varphi^{2}}{2x^{2}_{0}}\right)\mu_{\hat{g}}|_{C^{T}_{p}}\\\nonumber
=\int_{D^{T}_{p}-J^{t_{1}}_{q}}\left(-\frac{1}{2x_{0}}\nabla_{0}\varphi\nabla^{0}\varphi+\frac{1}{2x_{0}}\nabla_{i}\varphi\nabla^{i}\varphi\nonumber-\frac{x^{i}}{x^{2}_{0}}\nabla_{0}\varphi\nabla_{i}\varphi+\frac{1}{2x_{0}}\varphi^{6}\right.\\\nonumber
\left.-\frac{\varphi^{2}g_{00}}{x^{3}_{0}}+ \mathcal{L}_{\mu\nu}T^{\mu\nu}\right)\mu_{\hat{g}}.
\end{align}
Now noting $R=S/x_{0}$ and utilizing the results of lemmas $3$ and $4$, we have the following 
\begin{align}
\lim_{t_{2}\to0}\int_{S_{t_{2}}}\left(T(R,n)+\nonumber\frac{\varphi}{x_{0}}m+n_{0}\frac{\varphi^{2}}{2x^{2}_{0}}\right)\mu_{\hat{g}}|_{S_{t_{2}}}=0.
\end{align}
The previous energy equation reduces to 
\begin{align}
\int_{S_{t_{1}}-B_{t_{1}}}\left(T(R,n)+\frac{\varphi}{x_{0}}m+n_{0}\frac{\varphi^{2}}{2x^{2}_{0}}\right)\mu_{\hat{g}}|_{S_{t_{1}}}
-\int_{C^{t_{1}}_{q}}\left(T(R,l)\nonumber+\frac{\varphi}{x_{0}}l(\varphi)+l_{0}\frac{\varphi^{2}}{2x^{2}_{0}}\right)\mu_{\hat{g}}|_{C^{t_{1}}_{q}} 
\\\nonumber
+\int_{C^{t_{1}}_{p}}\left(T(R,l)+\frac{\varphi}{x_{0}}l(\varphi)+l_{0}\frac{\varphi^{2}}{2x^{2}_{0}}\right)\mu_{\hat{g}}|_{C^{t_{1}}_{p}}\\\nonumber
=\int_{D^{t_{1}}_{p}-J^{t_{1}}_{q}}\left(-\frac{1}{2x_{0}}\nabla_{0}\varphi\nabla^{0}\varphi+\frac{1}{2x_{0}}\nabla_{i}\varphi\nabla^{i}\varphi\nonumber-\frac{x^{i}}{x^{2}_{0}}\nabla_{0}\varphi\nabla_{i}\varphi+\frac{1}{2x_{0}}\varphi^{6}\right.\\\nonumber
\left.-\frac{\varphi^{2}g_{00}}{x^{3}_{0}}+ \mathcal{L}_{\mu\nu}T^{\mu\nu}\right)\mu_{\hat{g}}.
\end{align}
Now we will use lemmas $3$ and $4$ together with some elementary inequalities to control certain terms of the integral over $C^{t_{1}}_{q}$. First we concentrate on the integrals over $S_{t_{1}}-B_{t_{1}}$ and $C^{t_{1}}_{p}$. Notice that $T(R,n)=\frac{1}{x_{0}}T(S,n)$ according to the definition. Therefore using lemmas $3$ and $4$ we have 
\begin{align}
-\int_{S_{t_{1}}-B_{t_{1}}}T(R,n)\mu_{\hat{g}}|_{S_{t_{1}}}=-\frac{1}{|t_{1}|}\int_{S_{t_{1}}-B_{t_{1}}}T(S,n)\mu_{\hat{g}}|_{S_{t_{1}}}=zo(t_{1}),\\\nonumber
\frac{1}{|t_{1}|}\int_{S_{t_{1}}-B_{t_{1}}}\varphi m\mu_{\hat{g}}|_{S_{t_{1}}}\lesssim \frac{1}{|t_{1}|}\left(\int_{S_{t_{1}}-B_{t_{1}}}\varphi^{6}\mu_{\hat{g}}|_{S_{t_{1}}}\right)^{1/6}\left(\int_{S_{t_{1}}-B_{t_{1}}}\mu_{\hat{g}}|_{S_{t_{1}}}\right)^{1/3}\\
 \left(\int_{S_{t_{1}}-B_{t_{1}}}m^{2}\mu_{\hat{g}}|_{S_{t_{1}}}\right)^{1/2}= zo(t_{1}),\\\nonumber
\frac{1}{|t_{1}|^{2}}\int_{S_{t_{1}}-B_{t_{1}}}N\varphi^{2}\mu_{\hat{g}}|_{S_{t_{1}}}\lesssim\frac{1}{|t_{1}|^{2}}\left(\int_{S_{t_{1}}-B_{t_{1}}}\varphi^{6}\mu_{\hat{g}}|_{S_{t_{1}}}\right)^{1/3}\left(\int_{S_{t_{1}}-B_{t_{1}}}\mu_{\hat{g}}|_{S_{t_{1}}}\right)^{2/3}\\
=zo(t_{1}).
\end{align}
Now we focus on the integral over $C^{t_{1}}_{p}$. Once again noting $S|_{C^{t_{1}}_{p}}=t(\partial_{t}-\partial_{r})=-tN l$, with lapse $N>0$ the first term becomes (using $x_{0}=-t$ in the normal neighbourhood)
\begin{align}
\int_{C^{t_{1}}_{p}}\frac{1}{x_{0}}T(S,l)\mu_{\hat{g}}|_{C^{t_{1}}_{p}}=\int_{C^{t_{1}}_{p}}N |l(\varphi)|^{2}\mu_{\hat{g}}|_{C^{t_{1}}_{p}}=zo(t_{1}).
\end{align}
Now for the second and third terms in the $C^{t_{1}}_{p}$ integral, we invoke the following inequality (\ref{eq:maininequality}).  
\begin{align}
\label{eq:midequation}
\int_{C^{t_{1}}_{p}}\frac{\varphi^{2}}{t^{2}}\mu_{\hat{g}}|_{C^{t_{1}}_{p}}\lesssim \int_{C^{t_{1}}_{p}}|l(\varphi)|^{2}\mu_{\hat{g}}|_{C^{t_{1}}_{p}}\nonumber+|t_{1}|\left(\int_{C^{t_{1}}_{p}}\varphi^{6}\mu_{\hat{g}}|_{C^{t_{1}}_{p}}\right)^{1/3}\\
+\left(\int_{S_{t_{1}}}\varphi^{6}\mu_{\hat{g}}|_{S_{t_{1}}}\right)^{1/4}\left((\int_{S_{t_{1}}}g(\partial\varphi,\partial\varphi)\mu_{\hat{g}}|_{S_{t_{1}}})^{1/4}+(\int_{S_{t_{1}}}\varphi^{6}\mu_{\hat{g}}|_{S_{t_{1}}})^{1/12}\right)\nonumber 
\end{align}
Utilizing this inequality we have (recalling $x_{0}=-t$) 
\begin{align}
\int_{C^{t_{1}}_{p}}\frac{\varphi}{x_{0}}l(\varphi)\mu_{\hat{g}}|_{C^{t_{1}}_{p}}\leq \left( \int_{C^{t_{1}}_{p}}|l(\varphi)|^{2}\mu_{\hat{g}}|_{C^{t_{1}}_{p}}\right)^{1/2}\nonumber\left(\int_{C^{t_{1}}_{p}}\frac{\varphi^{2}}{t^{2}}\mu_{\hat{g}}|_{C^{t_{1}}_{p}}\right)^{1/2}= zo(t_{1}).
\end{align}
Since $l_{0}=\hat{g}(l,\partial_{t})=b/2=O(1)$, therefore
\begin{align}
\int_{C^{t_{1}}_{p}}l_{0}\frac{\varphi^{2}}{2x_{0}^{2}}\mu_{\hat{g}}|_{C^{t_{1}}_{p}}=zo(t_{1})
\end{align}
by using inequality (\ref{eq:maininequality}).
On $C^{t_{1}}_{q}$ however, $R$ is not null (it is in fact timelike). We will express $R$ explicitly in terms of $(l,\bar{l},\lambda_{1},\lambda_{2})$. Note that $l$ is past directed and therefore $l_{0}>0$. Now we will show that  $\mathcal{I}_{a}=\int_{C^{t_{1}}_{q}}(\frac{\varphi}{x_{0}}l(\varphi)+l_{0}\frac{\varphi^{2}}{2x^{2}_{0}})\mu_{\hat{g}}|_{C^{t_{1}}_{q}}=\mathcal{I}_{b}+zo(t_{1})$, where $\mathcal{I}_{b}\geq0$. From (\ref{eq:expansion1}-\ref{eq:expansion2}), we obtain 
\begin{align}
\label{eq:ldecomp}
l=\frac{1}{(ab^{'}-a^{'}b)}(b^{'}\partial_{t}-b\partial_{r}+(c^{'}b-cb^{'})\lambda_{1}+(d^{'}b-db^{'})\lambda_{2}).
\end{align}
Now focus on the integral $\int_{C^{t_{1}}_{q}}\frac{\varphi}{x_{0}}l(\varphi)\mu_{\hat{g}}|_{C^{t_{1}}_{q}}$ which may be written as follows 
\begin{align}
\int_{C^{t_{1}}_{q}}\frac{\varphi}{x_{0}}l(\varphi)\mu_{\hat{g}}|_{C^{t_{1}}_{q}}=\frac{1}{2}\int_{C^{t_{1}}_{q}}\frac{1}{x_{0}}l(\varphi^{2})\mu_{\hat{g}}|_{C^{t_{1}}_{q}}\\\nonumber
=\frac{1}{2}\int_{C^{t_{1}}_{q}}l(\frac{\varphi^{2}}{x_{0}})\mu_{\hat{g}}|_{C^{t_{1}}_{q}} -\frac{1}{2}\int_{C^{t_{1}}_{q}}\varphi^{2}l(\frac{1}{x_{0}})\mu_{\hat{g}}|_{C^{t_{1}}_{q}}
\end{align}
Now $l$ is parallel to the null cone $C^{t_{1}}_{q}$ and therefore one can integrate the first term by parts to obtain a boundary term and an additional term (trace of certain null second fundamental form of the topological spheres foliating $C^{t_{1}}_{q}$). Notice that the volume form $\mu_{\hat{g}}|_{C^{t_{1}}_{q}}$ is equiavlent to $(t-t_{q})^{2}\mu_{\mathbb{S}^{2}}$, where $\mu_{\mathbb{S}^{2}}$ is the volume form of the standard unit sphere and $t_{q}$ is the time coordinate of $q$. For the second term, we use the decomposition (\ref{eq:ldecomp}) and $x_{0}=-t$ to yield
\begin{align}
\int_{C^{t_{1}}_{q}}\varphi^{2}l(\frac{1}{x_{0}})\mu_{\hat{g}}|_{C^{t_{1}}_{q}}=\int_{C^{t_{1}}_{q}}\frac{b^{'}\varphi^{2}}{t^{2}(ab^{'}-a^{'}b)}\mu_{\hat{g}}|_{C^{t_{1}}_{q}}
\end{align}
since $\partial_{r}t=\frac{x^{i}}{r}\partial_{i}t=0=\lambda_{1}(t)=\lambda_{2}(t)$. Therefore, $\mathcal{I}_{a}$ satisfies ($l_{0}=\hat{g}(l,\partial_{t})=\frac{b}{2}=\frac{N}{2}+O(t)$) \begin{align}
\mathcal{I}_{a}\approx \int_{\mathbb{S}^{2}_{t_{1}}}\frac{\varphi^{2}}{x_{0}}\mu_{\hat{g}}|_{\mathbb{S}^{2}_{t_{1}}}+\int_{C^{t_{1}}_{q}}\left(\frac{b}{2}-\frac{b^{'}}{(ab^{'}-a^{'}b)}\right)\frac{\varphi^{2}}{2t^{2}}\mu_{\hat{g}}|_{C^{t_{1}}_{q}}\\\nonumber 
+\int_{C^{t_{1}}_{q}}\frac{b^{'}}{(ab^{'}-a^{'}b)}\frac{\varphi^{2}}{t(t-t_{q})}\mu_{\hat{g}}|_{C^{t_{1}}_{q}}.
\end{align}
%Now we consider two cases: $Kt_{q}\leq t-t_{q}\leq 0$ and $t_{1}\leq t-t_{q}< Kt_{q}$ for some large $K>0$. 
Now in the view of (\ref{eq:expanse1}-\ref{eq:expanse2}), $\frac{b^{'}}{(ab^{'}-a^{'}b)}<0$ (this is also obvious from the fact that $l$ is past directed) which yields 
\begin{align}
l_{0}-\frac{b^{'}}{(ab^{'}-a^{'}b)}=\hat{g}(l,\partial_{t})-\frac{b^{'}}{(ab^{'}-a^{'}b)}=\frac{b}{2}-\frac{b^{'}}{(ab^{'}-a^{'}b)}>0.
\end{align}
%and therefore, for the first case $Kt_{q}\leq t-t_{q}\leq 0$, we only need to estimate the negative part. The negative part $\int_{C^{t_{1}}_{q}}\frac{b^{'}}{(ab^{'}-a^{'}b)}\frac{\varphi^{2}}{t(t-t_{q})}\mu_{\hat{g}}|_{C^{t_{1}}_{q}}$ is estimated as follows \begin{align}
%\int_{C^{t_{1}}_{q}}\frac{b^{'}}{(ab^{'}-a^{'}b)}\frac{\varphi^{2}}{t(t-t_{q})}\mu_{\hat{g}}|_{C^{t_{1}}_{q}}\\\nonumber 
%\lesssim \sup_{t\in[t_{1},t_{q}]}(||t|-|t_{q}||\int_{\mathbb{S}^{2}_{||t|-|t_{q}||}}\varphi^{2}\mu_{\hat{g}}|_{\mathbb{S}^{2}_{||t|-|t_{q}|}})\ln(K+1)
%\end{align}
%which in the view of the inequality (\ref{eq:boundary2}) and lemma 3 is $\lesssim \delta$ for sufficiently small $|t_{1}|$.
Now when $t_{q}\to 0$ then the cone $C^{t_{1}}_{q}$ approaches $C^{t_{1}}_{p}$ and therefore one would expect that $\mathcal{I}_{a}$ should satisfy an estimate of type $zo(t_{1})$. This is indeed the case. Notice in view of the estimates (\ref{eq:expanse1}-\ref{eq:expanse2}), $\frac{b}{2}-\frac{b^{'}}{(ab^{'}-a^{'}b)}=\frac{1}{N}+O(t)$ and $\frac{b^{'}}{(ab^{'}-a^{'}b)}=-\frac{1}{2N}+O(t)$ and therefore in the limit $t_{q}\to 0$ the dangerous leading order terms cancel each other. More explicitly the last two terms combine to yield (using Holder for the $O(t)$ terms) 
\begin{align}
\int_{C^{t_{1}}_{q}}\left(\frac{b}{2}-\frac{b^{'}}{(ab^{'}-a^{'}b)}\right)\frac{\varphi^{2}}{2t^{2}}\mu_{\hat{g}}|_{C^{t_{1}}_{q}}
+\int_{C^{t_{1}}_{q}}\frac{b^{'}}{(ab^{'}-a^{'}b)}\frac{\varphi^{2}}{t(t-t_{q})}\mu_{\hat{g}}|_{C^{t_{1}}_{q}}\\\nonumber
=-\int_{C^{t_{1}}_{q}}\frac{t_{q}\varphi^{2}}{Nt^{2}(t-t_{q})}\mu_{\hat{g}}|_{C^{t_{1}}_{q}}+\mathcal{R},
\end{align}
where $\mathcal{R}\lesssim |t_{1}|\left(\int_{C^{t_{1}}_{q}}\varphi^{6}\mu_{\hat{g}}|_{C^{t_{1}}_{q}}\right)^{1/3}$. From this expression, it is obvious that $t_{q}= 0$ cancels the first potentially dangerous term leaving only the harmless term $\mathcal{R}$. Now consider the second case when $t_{q}>0$. The following holds for the leading order term 
\begin{align}
|-\int_{C^{t_{1}}_{q}}\frac{t_{q}\varphi^{2}}{Nt^{2}(t-t_{q})}\mu_{\hat{g}}|_{C^{t_{1}}_{q}}|\lesssim \sup_{t\in[t_{1},t_{q})}\nonumber(||t|-|t_{q}||\int_{\mathbb{S}^{2}}\varphi^{2}\mu_{\hat{g}}|_{\mathbb{S}^{2}})\frac{|t_{1}-t_{q}|}{|t_{1}|}\\\nonumber 
\leq \sup_{t\in[t_{1},t_{q})}\nonumber(||t|-|t_{q}||\int_{\mathbb{S}^{2}}\varphi^{2}\mu_{\hat{g}}|_{\mathbb{S}^{2}}).
\end{align}
Now writing $\mu_{\hat{g}}|_{\mathbb{S}^{2}_{|t_{1}|}}=\sqrt{-\det(\hat{g}(|t_{1}|,\theta,\phi))}d\theta\wedge d\phi$, in view of the inequality (\ref{eq:boundary2}) and lemma (3), $\int_{\mathbb{S}^{2}_{|t_{1}|}}\frac{\varphi^{2}}{x_{0}}\mu_{\hat{g}}|_{\mathbb{S}^{2}_{t_{1}}}=zo(t_{1})$ and $\sup_{t\in[t_{1},t_{q})}\nonumber(||t|-|t_{q}||\int_{\mathbb{S}^{2}}\varphi^{2}\mu_{\hat{g}}|_{\mathbb{S}^{2}})=zo(t_{1})$. Here we have used the lemma ($3$) as follows: $\int_{S_{t_{1}}}\varphi^{6}\mu_{\hat{g}}|_{S_{t_{1}}}=zo(t_{1})$ implies $\int_{S_{t_{2}}}\varphi^{6}\mu_{\hat{g}}|_{S_{t_{2}}}=zo(t_{2})$ for $|t_{2}|<|t_{1}|$ (recall $zo(t)$ denotes positive continuous functions that vanish as $t\to0$). In addition, since for a measurable function $f$, $\int_{B}|f|\leq \int_{A}|f|$ for a measurable $B\subseteq A$, inequality (\ref{eq:boundary2}) yields $\sup_{t\in[t_{1},t_{q})}\nonumber(||t|-|t_{q}||\int_{\mathbb{S}^{2}}\varphi^{2}\mu_{\hat{g}}|_{\mathbb{S}^{2}})\lesssim \left(\int_{B_{t}}\varphi^{6}\mu_{\hat{g}}|_{B_{t}}\right)^{1/4}E^{1/4}\leq\left(\int_{S_{t}}\varphi^{6}\mu_{\hat{g}}|_{S_{t}}\right)^{1/4}E^{1/4} =zo(t)$ for $|t_{q}|<|t|\leq|t_{1}|$ and $B_{t}\subseteq S_{t}$ (see figure \ref{fig:fig3}). Therefore $\mathcal{I}_{a}$ satisfies the estimate
\begin{align}
\mathcal{I}_{a}=\int_{C^{t_{1}}_{q}}(\frac{\varphi}{x_{0}}l(\varphi)\nonumber+l_{0}\frac{\varphi^{2}}{2x^{2}_{0}})\mu_{\hat{g}}|_{C^{t_{1}}_{q}}\lesssim zo(t_{1})+zo(t),~~0<|t_{q}|<|t|\leq|t_{1}|.
\end{align}
Now we choose $|t_{1}|>0$ by making $\mathcal{I}_{a}$ less than an arbitrarily small $\delta>0$ i.e., $\mathcal{I}_{a}(t_{1})<\delta$. Utilizing the previous estimates, we may write the following 
\begin{align}
\int_{C^{t_{1}}_{q}}T(R,l)\mu_{\hat{g}}|_{C^{t_{1}}_{q}}+\int_{D^{t_{1}}_{p}-J^{t_{1}}_{q}}\left(-\frac{1}{2x_{0}}\nabla_{0}\varphi\nabla^{0}\varphi+\frac{1}{2x_{0}}\nabla_{i}\varphi\nabla^{i}\varphi\nonumber-\frac{x^{i}}{x^{2}_{0}}\nabla_{0}\varphi\nabla_{i}\varphi\right.\\\nonumber
\left.+\frac{1}{2x_{0}}\varphi^{6}-\frac{\varphi^{2}g_{00}}{x^{3}_{0}}+\mathcal{L}_{\mu\nu}T^{\mu\nu}\right)\mu_{\hat{g}}\\\nonumber 
\lesssim \delta.
\end{align}
Now we note an important fact that the extra term involving the bulk integral on the left hand side of the previous equation is positive definite modulo lower order harmless terms, that is 
\begin{align}
\int_{D^{t_{1}}_{p}-J^{t_{1}}_{q}}\left(-\frac{1}{2x_{0}}\nabla_{0}\varphi\nabla^{0}\varphi+\frac{1}{2x_{0}}\nabla_{i}\varphi\nabla^{i}\varphi\nonumber-\frac{x^{i}}{x^{2}_{0}}\nabla_{0}\varphi\nabla_{i}\varphi\right.\\\nonumber
\left.+\frac{1}{2x_{0}}\varphi^{6}-\frac{2\varphi^{2}g_{00}}{x^{3}_{0}}\right)\mu_{\hat{g}}>0,
\end{align}
where note that $g_{00},g^{00}<0$.
This simply follows from a straightforward calculation and using $x_{0}=-x^{0}=-t>0$ within $D^{t_{1}}_{p}-J^{t_{1}}_{q}$ lying in the causal past of $p$. Additionally note that $\int_{D^{t_{1}}_{p}-J^{t_{1}}_{q}}\mathcal{L}^{\mu\nu}T_{\mu\nu}\mu_{\hat{g}}\lesssim |t_{1}|^{2}$. Note an important fact that since $R$ and $l$ are past directed time-like and null vectors, $T(R,l)>0$.    
Therefore we have 
\begin{align}
\label{eq:estimatenew1}
\int_{C^{t_{1}}_{q}}T(R,l)\mu_{\hat{g}}|_{C^{t_{1}}_{q}}\lesssim \delta.
\end{align}
$T(R,l)$ contains terms involving $|l(\varphi)|^{2}, \varphi^{6}$ and additional positive terms. In order to obtain the  additional estimates which will finish the proof, we first need to estimate $\int_{C^{t_{1}}_{q}}|l(\varphi)|^{2}$ and $\int_{C^{t_{1}}_{q}}\varphi^{6}$. Even though $\int_{C^{t_{1}}_{q}}T(R,l)$ is positive definite, the individual terms which we want to estimate may contain negative (or small) coefficients. Therefore, we need to proceed case by case. Firstly, we explicitly evaluate $T(R,l)$ on $C^{t_{1}}_{q}$ using the expansions of $\partial_{t}$ and $\partial_{r}$ introduced in the previous section (\ref{eq:expansion1}). Noting that
\begin{align}
\hat{g}(x,l)=\hat{g}(x^{\mu}\partial_{\mu},l)=\hat{g}(t\partial_{t}+x^{i}\partial_{i},l)\\\nonumber 
=\hat{g}\left(t(al+b\bar{l}+c\lambda_{1}+d\lambda_{2})+\sqrt{\delta_{ij}x^{i}x^{j}}(a^{'}l+b^{'}\bar{l}+c^{'}\lambda_{1}+d^{'}\lambda_{2}),l\right)\\\nonumber 
=\frac{1}{2}(b^{'}\sqrt{\delta_{ij}x^{i}x^{j}}+bt),
\end{align}
we have 
\begin{align}
T(R,l)|=\frac{1}{x_{0}}\hat{g}(l,\partial\varphi) x^{\nu}\nabla_{\nu}\varphi-\frac{1}{2}\hat{g}(\partial\varphi,\partial\varphi)\nonumber\frac{1}{x_{0}}\hat{g}(x,l)-\hat{g}(x,l)\frac{\varphi^{6}}{6}
\end{align}
Explicit computation term by term yields 
\begin{align}
l(\varphi)x^{\mu}\partial_{\mu}\varphi=l(\varphi)(t\partial_{t}\varphi+x^{i}\partial_{i}\varphi)\\
=l(\varphi)\left(t(al(\varphi)+b\bar{l}(\varphi)+c\lambda_{1}(\varphi)+d\lambda_{2}(\varphi))\nonumber+\sqrt{\delta_{ij}x^{i}x^{j}}(a^{'}l(\varphi)\right.\\\nonumber \left.+b^{'}\bar{l}(\varphi)+c^{'}\lambda_{1}(\varphi)+d^{'}\lambda_{2}(\varphi))\right)\\\nonumber 
=(at+a^{'}\sqrt{\delta_{ij}x^{i}x^{j}})|l(\varphi)|^{2}+(bt+b^{'}\sqrt{\delta_{ij}x^{i}x^{j}})l(\varphi)\bar{l}(\varphi)\\\nonumber +(ct+c^{'}\sqrt{\delta_{ij}x^{i}x^{j}})l(\varphi)\lambda_{1}(\varphi)+(dt+d^{'}\sqrt{\delta_{ij}x^{i}x^{j}})l(\varphi)\lambda_{2}(\varphi),
\end{align}
and 
\begin{align}
\hat{g}(\partial\varphi,\partial\varphi)=4\hat{g}(\partial \varphi,l)\hat{g}(\partial\varphi,\bar{l})+|\hat{g}(\partial\varphi,\lambda_{i})|^{2}\\\nonumber
=4l(\varphi)\bar{l}(\varphi)+|\hat{g}(\partial\varphi,\lambda_{i})|^{2}.
\end{align}
Therefore $T(R,l)$ becomes 
\begin{align}
T(R,l)\\\nonumber
=\frac{1}{x_{0}}\left((at+a^{'}\sqrt{\delta_{ij}x^{i}x^{j}})|l(\varphi)|^{2}+(bt+b^{'}\sqrt{\delta_{ij}x^{i}x^{j}})l(\varphi)\bar{l}(\varphi)\nonumber+(ct+c^{'}\sqrt{\delta_{ij}x^{i}x^{j}})\right.\\\nonumber 
\left.
l(\varphi)\lambda_{1}(\varphi)+(dt+d^{'}\sqrt{\delta_{ij}x^{i}x^{j}})l(\varphi)\lambda_{2}(\varphi)\right)-\frac{(bt+b^{'}\sqrt{\delta_{ij}x^{i}x^{j}})}{4x_{0}}\left(4l(\varphi)\bar{l}(\varphi)+|\hat{g}(\partial\varphi,\lambda_{i})|^{2}\right)\\\nonumber 
-\frac{bt+b^{'}\sqrt{\delta_{ij}x^{i}x^{j}}}{2x_{0}}\frac{\varphi^{6}}{6}\\\nonumber 
=\frac{(at+a^{'}\sqrt{\delta_{ij}x^{i}x^{j}})}{x_{0}}|l(\varphi)|^{2}-\frac{bt+b^{'}\sqrt{\delta_{ij}x^{i}x^{j}}}{4}(|\lambda_{1}(\varphi)|^{2}+|\lambda_{2}(\varphi)|^{2})-\frac{bt+b^{'}\sqrt{\delta_{ij}x^{i}x^{j}}}{2x_{0}}\frac{\varphi^{6}}{6}\\\nonumber 
+\frac{(ct+c^{'}\sqrt{\delta_{ij}x^{i}x^{j}})l(\varphi)\lambda_{1}(\varphi)}{x_{0}}+\frac{(dt+d^{'}\sqrt{\delta_{ij}x^{i}x^{j}})l(\varphi)\lambda_{2}(\varphi)}{x_{0}}.
\end{align}
The estimate (\ref{eq:estimatenew1}) may now be written explicitly as follows
\begin{align}
\label{eq:estimatenew2}
\int_{C^{t_{1}}_{q}}\left(\frac{(at+a^{'}\sqrt{\delta_{ij}x^{i}x^{j}})}{x_{0}}|l(\varphi)|^{2}-\frac{bt+b^{'}\sqrt{\delta_{ij}x^{i}x^{j}}}{4x_{0}}(|\lambda_{1}(\varphi)|^{2}\nonumber+|\lambda_{2}(\varphi)|^{2})\right.\\ 
\left.-\frac{bt+b^{'}\sqrt{\delta_{ij}x^{i}x^{j}}}{2x_{0}}\frac{\varphi^{6}}{6}+\frac{(ct+c^{'}\sqrt{\delta_{ij}x^{i}x^{j}})l(\varphi)\lambda_{1}(\varphi)}{x_{0}}\right.\\\nonumber
\left.
+\frac{(dt+d^{'}\sqrt{\delta_{ij}x^{i}x^{j}})l(\varphi)\lambda_{2}(\varphi)}{x_{0}}\right)\mu_{\hat{g}}|_{C^{t_{1}}_{q}}\lesssim \delta.
\end{align}
Now if we use the estimates (\ref{eq:expanse1}-\ref{eq:expanse2}), then we see that 
\begin{align}
\int_{C^{t_{1}}_{q}}\left(\frac{(ct+c^{'}\sqrt{\delta_{ij}x^{i}x^{j}})l(\varphi)\lambda_{1}(\varphi)}{x_{0}}\nonumber+\frac{(dt+d^{'}\sqrt{\delta_{ij}x^{i}x^{j}})l(\varphi)\lambda_{2}(\varphi)}{x_{0}}\right)\mu_{\hat{g}}|_{C^{t_{1}}_{q}}\lesssim |t_{1}|^{2}\\\nonumber 
\lesssim \delta^{2},
\end{align}
where the involved constants depend on the initial energy. Therefore the estimate (\ref{eq:estimatenew2}) reduces to 
\begin{align}
\int_{C^{t_{1}}_{q}}\left(\frac{(-Nt+\sqrt{g_{ij}x^{i}x^{j}})}{x_{0}}|l(\varphi)|^{2}-\frac{Nt+\sqrt{g_{ij}x^{i}x^{j}}}{4x_{0}}(|\lambda_{1}(\varphi)|^{2}\nonumber+|\lambda_{2}(\varphi)|^{2})\right.\\\nonumber 
\left.-\frac{Nt+\sqrt{g_{ij}x^{i}x^{j}}}{2x_{0}}\frac{\varphi^{6}}{6}\right)\mu_{\hat{g}}|_{C^{t_{1}}_{q}}\lesssim \delta
\end{align}
i.e., 
\begin{align}
\int_{C^{t_{1}}_{q}}\left((N-\frac{\sqrt{g_{ij}x^{i}x^{j}}}{t})|l(\varphi)|^{2}+(N+\frac{\sqrt{g_{ij}x^{i}x^{j}}}{t})(|\lambda_{1}(\varphi)|^{2}\nonumber+|\lambda_{2}(\varphi)|^{2})\right.\\\nonumber 
\left.+(N+\frac{\sqrt{g_{ij}x^{i}x^{j}}}{t})\frac{\varphi^{6}}{6}\right)\mu_{\hat{g}}|_{C^{t_{1}}_{q}}\lesssim \delta.
\end{align}
Now in the past light cone of $p$, we always have $t^{2}>r^{2}$ (when expressed in geodesic normal coordinate variables and using $\hat{g}_{\mu\nu}x^{\nu}=\eta_{\mu\nu}x^{\nu}$ in the normal neighbourhood).
Expressed in terms of the full metric $\hat{g}$, it becomes 
\begin{align}
(-N^{2}+|Y|^{2})t^{2}+g_{ij}x^{i}x^{j}+2g_{ij}Y^{i}x^{j}t\leq 0,
\end{align}
that is 
\begin{align}
N^{2}t^{2}-g_{ij}x^{i}x^{j}\geq |Y|^{2}t^{2}+2g_{ij}Y^{i}x^{j}t
\end{align}
Now consider the first case $N^{2}t^{2}-g_{ij}x^{i}x^{j}<0$. Then $|Y|^{2}t^{2}+2g_{ij}Y^{i}x^{i}t\leq0$, and for $|t|<|t_{1}|$, choosing sufficiently small $|t_{1}|$ and since $|Y|^{2}t^{2}\lesssim |t|^{6}$ and $|g_{ij}Y^{i}x^{j}t|\lesssim |t|^{4}$,  
\begin{align}
\frac{\sqrt{g_{ij}x^{i}x^{j}}}{|t|}=N+\delta^{2}.
\end{align}
Following this fact, the last two terms of the integral are of size $\delta^{2}$ i.e.,
\begin{align}
-\int_{C^{t_{1}}_{q}}(N+\frac{\sqrt{g_{ij}x^{i}x^{j}}}{t})(|\lambda_{1}(\varphi)|^{2}\nonumber+|\lambda_{2}(\varphi)|^{2}+\frac{\varphi^{6}}{6})\mu_{\hat{g}}|_{C^{t_{1}}_{q}}\lesssim \delta^{2}
\end{align}
and therefore 
\begin{align}
\int_{C^{t_{1}}_{q}}(N-\frac{\sqrt{g_{ij}x^{i}x^{j}}}{t})|l(\varphi)|^{2}\mu_{\hat{g}}|_{C^{t_{1}}_{q}}\lesssim \delta
\end{align}
since $N-\frac{\sqrt{g_{ij}x^{i}x^{j}}}{t}>0$ always due to $t<0$. Moreover following (\ref{eq:estimates_normal}) and $t<0$, $0<N-\frac{\sqrt{g_{ij}x^{i}x^{j}}}{t}=O(1)$ and therefore $\int_{C^{t_{1}}_{q}}|l(\varphi)|^{2}\mu_{\hat{g}}|_{C^{t_{1}}_{q}}\lesssim\delta$. Now if $|Nt|>\sqrt{g_{ij}x^{i}x^{j}}$, then both the terms $N-\frac{\sqrt{g_{ij}x^{i}x^{j}}}{t}$ and $N+\frac{\sqrt{g_{ij}x^{i}x^{j}}}{t}$ are positive.  
Now noting $t<0$, we have two different cases 
\begin{align}
A. \frac{\sqrt{g_{ij}x^{i}x^{j}}}{|t|}<<N,~~~~B.~N>\frac{\sqrt{g_{ij}x^{i}x^{j}}}{|t|}>\beta
\end{align}
In the case $A$, we have the following 
\begin{align}
N-\frac{\sqrt{g_{ij}x^{i}x^{j}}}{t}>c_{1}=O(1)=>\int_{C^{t_{1}}_{q}}|l(\varphi)|^{2}\mu_{\hat{g}}|_{C^{t_{1}}_{q}}\lesssim \delta
\end{align}
and 
\begin{align}
N+\frac{\sqrt{g_{ij}x^{i}x^{j}}}{t}>c_{2}=O(1)=>\int_{C^{t_{1}}_{q}}\varphi^{6}\mu_{\hat{g}}|_{C^{t_{1}}_{q}}\lesssim \delta,
\end{align}
where $c_{1},c_{2}>0$. Now in the case $B$ we only have 
\begin{align}
N-\frac{\sqrt{g_{ij}x^{i}x^{j}}}{t}>C=O(1)=>\int_{C^{t_{1}}_{q}}|l(\varphi)|^{2}\mu_{\hat{g}}|_{C^{t_{1}}_{q}}\lesssim \delta.
\end{align}
 We will use this important property in the final analysis. 
Results obtained so far yields the following lemma\\
\textbf{Lemma 5:} \textit{Let $p\in M$ be such that in local coordinates $x(p)=0$ and let $\mathcal{G}$ be its geodesic normal neighbourhood and $C^{t_{1}}_{p}\subset \mathcal{G}$ be its past light cone extending up to the constant time hypersurface $t_{1}$ and $\delta>0$ be sufficiently small. Further assume that $|t_{1}|$ is sufficiently small. If $q$ is an interior point of the causal past $D^{t_{1}}_{p}$ of $p$ and $C^{t_{1}}_{q}$ its  past light cone extending up to $S_{t_{1}}$, then the following estimates hold for the two corresponding diffeomorphism invariant entities 
\begin{align}
\int_{C^{t_{1}}_{q}}|l(\varphi)|^{2}\mu_{\hat{g}}|_{C^{t_{1}}_{q}}\lesssim \delta, \int_{C^{t_{1}}_{q}}\varphi^{6}\mu_{\hat{g}}|_{C^{t_{1}}_{q}}\lesssim \delta
\end{align}
or 
\begin{align}
\int_{C^{t_{1}}_{q}}|l(\varphi)|^{2}\mu_{\hat{g}}|_{C^{t_{1}}_{q}}\lesssim \delta.
\end{align}
}

Now notice an important fact. The entities $\int_{C^{t_{1}}_{q}}|l(\varphi)|^{2}\mu_{\hat{g}}|_{C^{t_{1}}_{q}}$ and $\int_{C^{t_{1}}_{q}}\varphi^{6}\mu_{\hat{g}}|_{C^{t_{1}}_{q}}$ are diffeomorphism invariant. As long as the point $q$ lies within $D^{t_{1}}_{p}$ (up to the $t_{1}=constant$ hypersurface i.e., $\Sigma_{t_{1}}$ of course), these two diffeomorphism invariant integrals will remain small enough given that $|t_{1}|$ is chosen sufficiently small. One may now make a coordinate transformation by taking $q$  to be the centre of the normal coordinate system (see figure \ref{fig:fig3}). However, due to the diffeomorphism invariance property of these two integrals, they remain small enough.
This lemma together with the foregoing representation formula (integral equation to be precise) will yield the desired $L^{\infty}$ estimate. Invoking the integral equation (\ref{eq:repn}) from theorem 1  we have at $q$ ($\equiv x$ in local coordinates) 
\begin{align}
\varphi(x)=\frac{1}{2\pi}\int_{C^{-}_{q}}U(x,y)\varphi^{5}(y)\mu_{\Gamma}(y)+\frac{1}{2\pi}\int_{C^{-}_{q}}\hat\Box_{y} U(x,y)\varphi(y)\mu_{\Gamma}(y)\\\nonumber +\frac{1}{2\pi}\int_{\sigma_{q}}\left(2U(x,y)<\nabla_{y}\Gamma(x,y),\nabla_{y} \varphi(y)>+U(x,y)\Theta(y) \varphi(y)\right)d\sigma_{q}(y).
\end{align}
Now let us denote $\sup_{x\in (D^{t_{1}}_{p}\cup S_{t_{1}})-D^{t_{2}}_{p}}|\varphi(x)|$ by $\mathcal{M}(t_{2})$ ($|t_{2}|<|t_{1}|$). Assume $|\varphi|$ attains its maximum at $q\in (D^{t_{1}}_{p}\cup S_{t_{1}})-D^{t_{2}}_{p}$ and $q\to p$ as $t_{2}\to 0$. Noting 
$\lim_{x^{'}\to x}U(x,x^{'})=1$ and $\sup_{x}U(x,x^{'})\lesssim 1$, we may split the integral over $C^{-}_{q}$ into two parts: one on $C^{t_{1}}_{q}$ (i.e., the portion of the cone $C^{-}_{q}$ that lies above the hypersurface $t=t_{1}$) and the other one $C^{-}_{q}-C^{t_{1}}_{q}$ and write the following by taking supremum   
\begin{align}
\mathcal{M}(t_{2}) \leq C_{1}\mathcal{M}(t_{2})\int_{C^{t_{1}}_{q}}|\varphi(x^{'})|^{4}\frac{1}{u}\mu_{\hat{g}}|_{C^{t_{1}}_{q}}+\frac{1}{2\pi}\int_{C^{t_{1}}_{q}}|\hat\Box_{y}\nonumber U(x,y)|\frac{|\varphi(y)|}{|u|}\mu_{\hat{g}}|_{C^{t_{1}}_{q}}\\\nonumber 
 +C_{2}(t_{1}),
\end{align}
where the constant $C_{2}(t_{1})$ depends on the initial energy. We choose the hypersurface $\Sigma_{t_{1}}$ ($S_{t_{1}}=\Sigma_{t_{1}}\cap D^{t_{1}}_{p}$) earlier in such a way that the lemma 5 holds. Now unlike in flat spacetime, we have an additional term involving the covariant spacetime Laplacian acting on the bi-scalar $U$. We will have to show that this Huygens violating second term involving $|\hat\Box_{y} U(x,y)|$ contributes to a constant depending on the spacetime curvature and the initial data. Indeed we will perform an explicit computation to show that $|\hat\Box_{y} U(x,y)|\lesssim 1$ assuming $|Riem(\hat{g})|\lesssim 1$. Here we have executed the computation in geodesic normal coordinates based at $x=0$ (i.e., at $q$). Then, $U(0,x)=U(x)$ is given as follows
\begin{align}
U(x)=\frac{|\hat{g}(0)|^{1/4}}{|\hat{g}(x)|^{1/4}}=\frac{\mu_{\hat{g}}^{1/2}(0)}{\sqrt{\mu_{\hat{g}}(x)}},
\end{align}
the Laplacian of which is computed as
\begin{align}
\nabla^{\alpha}\nabla_{\alpha}U(x)=\frac{\mu_{\hat{g}}^{1/2}(0)}{\mu_{\hat{g}}(x)}\partial_{\alpha}(\mu_{\hat{g}}(x)\hat{g}^{\alpha\beta}\partial_{\beta}\mu^{-1/2}_{\hat{g}}(x))\\\nonumber 
%=-\frac{\mu_{\hat{g}}^{1/2}(0)}{4\mu_{\hat{g}}(x)}\partial_{\alpha}(\mu^{1/2}_{\hat{g}}(x)\hat{g}^{\alpha\beta}\hat{g}^{\mu\nu}\partial_{\beta}\hat{g}_{\mu\nu})\\\nonumber
=-\frac{\mu_{\hat{g}}^{1/2}(0)}{4\mu^{1/2}_{\hat{g}}(x)}\hat{g}^{\alpha\beta}\hat{g}^{\mu\nu}\partial_{\alpha}\partial_{\beta}\hat{g}_{\mu\nu}-\frac{\mu_{\hat{g}}^{1/2}(0)}{16\mu^{1/2}_{\hat{g}}(x)}\hat{g}^{\alpha\beta}\hat{g}^{\mu\nu}\hat{g}^{ab}\partial_{\alpha}\hat{g}_{ab}\partial_{\beta}\hat{g}_{\mu\nu}\\\nonumber 
-\frac{\mu_{\hat{g}}^{1/2}(0)}{4\mu^{1/2}_{\hat{g}}(x)}\partial_{\alpha}\hat{g}^{\alpha\beta}\hat{g}^{\mu\nu}\partial_{\beta}\hat{g}_{\mu\nu}-\frac{\mu_{\hat{g}}^{1/2}(0)}{4\mu^{1/2}_{\hat{g}}(x)}\hat{g}^{\alpha\beta}\partial_{\alpha}\hat{g}^{\mu\nu}\partial_{\beta}\hat{g}_{\mu\nu}.
\end{align}
Now the most dangerous point is the vertex of the cone $C^{t_{1}}_{q}$ i.e., $x=0$ and we want to show that at the vertex, this entity in fact remains bounded by an $O(1)$ term under the assumption of global hyperbolicity. Since the centre of the normal coordinate system is inertial, the first derivative of the metric vanishes there. But the second derivative does not vanish in general since it encodes the curvature information. Dropping the first derivative of the metric yields 
\begin{align}
\nabla^{\alpha}\nabla_{\alpha}U(x)|_{x=0}=-\frac{1}{4}\hat{g}^{\mu\nu}\hat{g}^{\alpha\beta}\partial_{\alpha}\partial_{\beta}\hat{g}_{\mu\nu}.
\end{align}
Now at the center of the geodesic normal coordinate system, the following must hold 
\begin{align}
\partial_{c}\partial_{d}\hat{g}_{ab}=\partial_{a}\partial_{b}\hat{g}_{cd},\\
\partial_{c}\partial_{d}\hat{g}_{ab}+\partial_{d}\partial_{b}\hat{g}_{ac}+\partial_{b}\partial_{c}\hat{g}_{ad}=0,
\end{align}
which yields 
\begin{align}
R_{\mu\nu\alpha\beta}=\partial_{\nu}\partial_{\alpha}\hat{g}_{\mu\beta}-\partial_{\beta}\partial_{\nu}\hat{g}_{\mu\alpha}.
\end{align}
Notice here that the anti-symmetry of $R_{\mu\nu\alpha\beta}$ is not apparent in this expression. However, this antisymmetry holds as a consequence of $\partial_{c}\partial_{d}\hat{g}_{ab}=\partial_{a}\partial_{b}\hat{g}_{cd}$ at the origin of the normal coordinates. We further obtain
\begin{align}
\partial_{\beta}\partial_{\nu}\hat{g}_{\mu\alpha}=-\frac{1}{3}(R_{\mu\nu\alpha\beta}+R_{\alpha\nu\mu\beta}),
\end{align}
that is 
\begin{align}
\hat{g}^{\mu\alpha}\hat{g}^{\beta\nu}\partial_{\beta}\partial_{\nu}\hat{g}_{\mu\alpha}=-\frac{2}{3}R(\hat{g})
\end{align}
only at the center of the normal coordinate system. The expression of the action of the co-variant Laplacian on the bi-scalar $U(x)$ becomes  
\begin{align}
\nabla^{\alpha}\nabla_{\alpha}U(0,x)|_{x=0}=\frac{1}{6}R(\hat{g}).
\end{align}
Now we are considering the metric to be a background field (i.e., no coupling with the scalar field). This leads to the fact that the scalar curvature $|R(\hat{g})|=O(1)$. Since the origin is the only possible blow up point for $\nabla^{\mu}\nabla_{\mu}U(0,x)$, we may safely conclude that the following holds \begin{align}
\sup_{x\in D^{-}_{q}}|\nabla^{\alpha}\nabla_{\alpha}U(0,x)|\lesssim 1.
\end{align}    

 Now applying Cauchy-Schwartz on the first term to the left, we obtain
\begin{align}
\mathcal{M}(t_{2})\leq C_{1}\mathcal{M}(t_{2})(\int_{C^{t_{1}}_{q}}\varphi^{6}\mu_{\hat{g}}|_{C^{t_{1}}_{q}})^{1/2}(\int_{C^{t_{1}}_{q}}\frac{\varphi^{2}}{u^{2}}\mu_{\hat{g}}|_{C^{t_{1}}_{q}})^{1/2}\nonumber+C^{'}_{2}\left(\int_{C^{t_{1}}_{q}}\frac{\varphi^{2}}{u^{2}}\mu_{\hat{g}}|_{C^{t_{1}}_{q}}\right)^{1/2}\\\nonumber+C_{3}(t_{1})\\\nonumber 
\leq C_{1}\mathcal{M}(t_{2})\left(\int_{C^{t_{1}}_{q}}\varphi^{6}\mu_{\hat{g}}|_{C^{t_{1}}_{q}}\right)^{1/2} \left(\int_{C^{t_{1}}_{q}}|l(\varphi)|^{2}\mu_{\hat{g}}|_{C^{t_{1}}_{q}}\nonumber+u^{2}_{1}\left(\int_{C^{t_{1}}_{q}}\varphi^{6}\mu_{\hat{g}}|_{C^{t_{1}}_{q}}\right)^{1/3}\right.\\
\left.+\left(\int_{S_{t_{1}}}\varphi^{6}\mu_{\hat{g}}|_{S_{t_{1}}}\right)^{1/4}\left((\int_{S_{t_{1}}}g(\partial\varphi,\partial\varphi)\mu_{\hat{g}}|_{S_{t_{1}}})^{1/4}\nonumber+(\int_{S_{t_{1}}}\varphi^{6}\mu_{\hat{g}}|_{S_{t_{1}}})^{1/12}\right)\right)^{1/2}\\\nonumber 
+C^{'}_{2}\left(\int_{C^{t_{1}}_{q}}|l(\varphi)|^{2}\mu_{\hat{g}}|_{C^{t_{1}}_{q}}\nonumber+u^{2}_{1}\left(\int_{C^{t_{1}}_{q}}\varphi^{6}\mu_{\hat{g}}|_{C^{t_{1}}_{q}}\right)^{1/3}+\left(\int_{S_{t_{1}}}\varphi^{6}\mu_{\hat{g}}|_{S_{t_{1}}}\right)^{1/4}\right.\\
\left.\left((\int_{S_{t_{1}}}g(\partial\varphi,\partial\varphi)\mu_{\hat{g}}|_{S_{t_{1}}})^{1/4}\nonumber+(\int_{S_{t_{1}}}\varphi^{6}\mu_{\hat{g}}|_{S_{t_{1}}})^{1/12}\right)\right)^{1/2}\\\nonumber
+C_{3}(t_{1}).
\end{align}
Now note that $\left(\int_{S_{t_{1}}}\varphi^{6}\mu_{\hat{g}}|_{S_{t_{1}}}\right)=zo(t_{1})$ from lemma $3$. In addition, we also make the trivial observation $\left(\int_{B_{t_{1}}}\varphi^{6}\mu_{\hat{g}}|_{B_{t_{1}}}\right)\leq \left(\int_{S_{t_{1}}}\varphi^{6}\mu_{\hat{g}}|_{S_{t_{1}}}\right)$. Therefore, since $|u_{1}|$ is sufficiently small, we may make $u^{2}_{1}\left(\int_{C^{t_{1}}_{q}}\varphi^{6}\mu_{\hat{g}}|_{C^{t_{1}}_{q}}\right)^{1/3}$ and $\left(\int_{S_{t_{1}}}\varphi^{6}\mu_{\hat{g}}|_{S_{t_{1}}}\right)^{1/4}$ small (notice that $\int_{C^{t_{1}}_{q}}\varphi^{6}\mu_{\hat{g}}|_{C^{t_{1}}_{q}}$ is bounded by energy), that is 
\begin{align}
u^{2}_{1}\left(\int_{C^{t_{1}}_{q}}\varphi^{6}\mu_{\hat{g}}|_{C^{t_{1}}_{q}}\right)^{1/3}\lesssim \delta^{2}, \left(\int_{S_{t_{1}}}\varphi^{6}\mu_{\hat{g}}|_{S_{t_{1}}}\right)^{1/4}\lesssim \delta.
\end{align}
Note that we still denote the initial hypersurface to be a $S_{t_{1}}$ even though we are working in the normal coordinate system based at $q$ now. This is done because the integral $\int_{S_{t_{1}}}\varphi^{6}\mu_{\hat{g}}|_{S_{t_{1}}}$ is diffeomorphism invariant. All of these estimates work due to the fact that the involved integrals which we want to be sufficiently small are diffeomorphism invariant.
In addition, from lemma $5$, we also have that the diffeomorphism invariant integral $\int_{C^{t_{1}}_{q}}|l(\varphi)|^{2}\mu_{\hat{g}}|_{C^{t_{1}}_{q}}$ satisfies
\begin{align}
\int_{C^{t_{1}}_{q}}|l(\varphi)|^{2}\mu_{\hat{g}}|_{C^{t_{1}}_{q}}\lesssim \delta
\end{align}
and therefore, we obtain the following 
\begin{align}
\mathcal{M}(t_{2})\leq C_{1}\mathcal{M}(t_{2})\delta+C_{2}(t_{1})
\end{align}
implying 
\begin{align}
\mathcal{M}(t_{2})\leq C_{2}(t_{1})<\infty~~~~i.e., \sup_{x\in (D^{t_{1}}_{p}\cup S_{t_{1}})-D^{t_{2}}_{p}}|\varphi(x)|\leq C_{2}(t_{1})
\end{align}
for sufficiently small $\delta$ and where the constant $C$ depends on the energy and the background geometry. Once we bound $\sup_{x\in (C^{t_{1}}_{p}\cup \sigma_{t_{1}})-C^{t_{2}}_{p}}|\varphi(x)| (\leq \sup_{x\in (D^{t_{1}}_{p}\cup S_{t_{1}})-D^{t_{2}}_{p}}|\varphi(x)|$) in terms of energy, $|\varphi(p)|$ is automatically bounded in terms of energy through the light cone formula for $\varphi(p)$. Since the energy does not blow up in finite time, this spacetime $L^{\infty}$ norm does not blow up in finite time either. One important point to note here is that $\varphi$ is a spacetime scalar and therefore the definition of the point-wise norm is unambiguous. For tensorial entities (e.g., in case of gravity or Yang-Mills theory), one needs to construct a gauge invariant point-wise norm. We finally obtain the crucial theorem we desire \\
\textbf{Theorem 2:} \textit{Let $M$ be a globally hyperbolic spacetime equipped with the Lorentzian metric $\hat{g}$ (\ref{eq:metric}) such that the point-wise norm of the Riemann curvature associated with $\hat{g}$ and its derivative are uniformly bounded from above. Then a classical solution of the semi-linear wave equation $\nabla^{\mu}\nabla_{\mu}\varphi=\alpha\varphi^{5}$, $\alpha>0$ (\ref{eq:main}) remains bounded point-wise on a globally hyperbolic background spacetime i.e., 
\begin{align}
\sup_{x\in A_{p}}|\varphi(x)|\leq C,
\end{align}
where $C$ depends on the $\dot{H}^{1}\times L^{2}$ norm of the initial data $(\varphi(0),m(0))$ and $A_{p}\subset M$ is the causal past of any point $p\in M$ including $p$ and extending up to the initial Cauchy hypersurface. Here $m=\frac{1}{N}(\partial_{t}-Y^{i}\partial_{i})\varphi$ is the momentum conjugate to $\varphi$ and $N$ and $Y$ are the usual lapse function and shift vector field of $\hat{g}$, respectively. 
}

\section{Sketch of the proof of global existence}
In this section, we give a rough sketch (for the sake of completeness) of the proof of global existence. We only provide a rough sketch since once the spacetime $L^{\infty}$ bound is obtained, the proof of global existence is a routine procedure. We will assume $\varphi\in \mathcal{S}(M)$ i.e., in the Schwartz class and appeal to an approximation argument (to get rid of the boundary terms while performing integration by parts over the entire space slice; such an argument of approximation is standard). We will use the boundedness of the energy in conjunction with an a priori bound on the spacetime point-wise norm of $\varphi$. First we sketch a proof of local existence of a a solution of the semilinear wave equation (\ref{eq:main}) in $C([0,t^{*}];H^{2}\times H^{1})$ where $t^{*}$ depends on the $H^{2}\times H^{1}$ norm of $(\varphi,m)$.  Roughly speaking, we will construct a sequence of approximate solutions $\{\varphi_{k},m_{k}\}_{k=1}^{\infty}$ and show using energy arguments that this sequence converges to a limit in $C([0,t^{*}];H^{2}\times H^{1})$, which solves the evolution equations. The obvious problem is that the bounded closed balls are not compact in infinite dimensions in general and therefore we need to explicitly work out the convergence. Once we obtain a local existence in $H^{2}$ norm (of $\varphi$), we need to show that this norm can not blow up in finite time in order for the global existence result to hold. In this stage, we will make use of the spacetime point-wise bound of the wavefield $\varphi$. We will follow the method developed by \cite{elliptichyperbolic} for proving the local existence theorem for a class of elliptic hyperbolic systems. We will not provide every detail but rather simply sketch the existence of a solution. The remaining procedure to establish continuity, uniqueness, and Cauchy stability is standard. Let us consider the equation of motion (\ref{eq:eom}) in the following form
\begin{align}
\partial_{t}\varphi-L_{Y}\varphi-Nm=0\\
\partial_{t}m-\nabla^{i}N\nabla_{i}\varphi-Ng^{ij}\nabla_{i}\nabla_{j}\varphi-L_{Y}m-m tr_{g}k=N\varphi^{5}.
\end{align}
Now write these as the following differential equations 
\begin{align}
\mathcal{L}_{N,Y,g}\mathcal{V}=\mathcal{F}[\varphi,m],
\end{align}
where the differential operator $\mathcal{L}$, the unknown $\mathcal{V}$, and the nonlinearity $\mathcal{F}[\varphi,m]$ read
\begin{align}
\mathcal{L}_{N,Y,g}\mathcal{V}=\left[\begin{array}{c}
\partial_{t}\varphi-L_{Y}\varphi-Nm\\
\partial_{t}m-\nabla^{i}N\nabla_{i}\varphi-Ng^{ij}\nabla_{i}\nabla_{j}\varphi-L_{Y}m-m tr_{g}k
\end{array}
\right],
\end{align}
\begin{align}
\mathcal{V}=\left[\begin{array}{c}
\varphi\\
m
\end{array}
\right], 
\mathcal{F}[\varphi,m]=\left[\begin{array}{c}
0\\
N\varphi^{5}
\end{array}
\right].
\end{align}
We have shown earlier that the energy defined naturally through the stress-energy tensor remains bounded on a globally hyperbolic background. However, here we will proceed in a different way where the boundedness of energy is not apparent. Let us now define an ad hoc energy associated with this system as follows
\begin{align}
\mathcal{E}_{1}[\mathcal{V}]=\int_{\Sigma_{t}}\left(\frac{1}{2}\varphi^{2}+\frac{1}{2}|\nabla\varphi|^{2}_{g}+\frac{1}{2}m^{2}\right)N\mu_{g},
\end{align}
where $\Sigma_{t}$ is a $t=$ constant hypersurface.
By construction we have $||\varphi||^{2}_{H^{1}}+||m||^{2}_{L^{2}}\approx \mathcal{E}_{1}$ and in the view of $\dot{H}^{1}(\Sigma_{t})\hookrightarrow L^{2}(\Sigma_{t})$, $||\varphi||^{2}_{\dot{H}^{1}}+||m||^{2}_{L^{2}}\approx \mathcal{E}_{1}$. A simple calculation using the evolution equations yields the following energy inequality 
\begin{align}
\partial_{t}\sqrt{\mathcal{E}_{1}(t)} \leq C(||k||_{L^{\infty}},||L_{Y}g||_{L^{\infty}},||N||_{L^{\infty}},\nonumber||\nabla N||_{L^{\infty}})(\sqrt{\mathcal{E}_{1}(t)}\\\nonumber
 +||\mathcal{F}[\varphi,m]||_{H^{1}\times L^{2}}).
\end{align}
Since we are on a globally hyperbolic background, set $C(||k||_{L^{\infty}},||L_{Y}g||_{L^{\infty}},||N||_{L^{\infty}},\\||\nabla N||_{L^{\infty}})<\infty$. Integration of the previous differential inequality yields
\begin{align}
\sqrt{\mathcal{E}_{1}(t)}-\sqrt{\mathcal{E}_{1}(0)} \leq C\int_{0}^{t}\sqrt{\mathcal{E}_{1}(t^{'})}dt^{'}+C\int_{0}^{t}||\mathcal{F}[\varphi,m]||_{H^{1}\times L^{2}}dt^{'}.
\end{align}
An application of Gr\"onwall's inequality yields
\begin{align}
\sqrt{\mathcal{E}_{1}(t)}\leq (\sqrt{\mathcal{E}_{1}(0)}+C||\mathcal{F}[\varphi,m]||_{L^{1}([0,t];H^{1}\times L^{2})})e^{Ct}.
\end{align}
Now notice $||\mathcal{F}[\varphi,m]||_{L^{1}([0,t];H^{1}\times L^{2})}$ is controlled by $t||\varphi||^{4}_{L^{\infty}([0,t];L^{\infty})}\\||[\varphi,m]||_{L^{\infty}([0,t];H^{1}\times L^{2})}$ and subsequently by $t||\varphi||^{4}_{L^{\infty}([0,t];H^{s})}\\||[\varphi,m]||_{L^{\infty}([0,t];H^{1}\times L^{2})}$ for $s>\frac{3}{2}$ due to Sobolev Embedding. Therefore, we need to control the $H^{s}\times H^{s-1}$ norm of $[\varphi,m]$ or for $n=3$, the $H^{2}\times H^{1}$ norm of $[\varphi,m]$ would be sufficient. We define the squared $H^{2}\times H^{1}$ norm of $[\varphi,m]$ as follows 
\begin{align}
\mathcal{E}_{2}:=\frac{1}{2}\int_{M_{t}}\left(\varphi^{2}+|\nabla\varphi|^{2}_{g}+|\nabla^{2}\varphi|^{2}_{g}+m^{2}+|\nabla m|^{2}_{g}\right)N\mu_{g},
\end{align}
where $|\nabla^{2}\varphi|^{2}_{g}$ is defined as $|\nabla^{2}\varphi|^{2}_{g}:=g^{IK}g^{JL}\nabla_{I}\nabla_{J}\varphi\nabla_{K}\nabla_{L}\varphi$. By an exact similar calculation, we obtain \begin{align}
\sqrt{\mathcal{E}_{2}(t)}\leq (\sqrt{\mathcal{E}_{2}(0)}+C||\mathcal{F}[\varphi,m]||_{L^{1}([0,t];H^{2}\times H^{1})})e^{Ct}.
\end{align}
Now notice that $||\mathcal{F}[\varphi,m]||_{L^{1}([0,t];H^{2}\times H^{1})}$ is dominated by $t||\varphi||^{4}_{L^{\infty}([0,t];L^{\infty})}\\||[\varphi,m]||_{L^{\infty}([0,t];H^{2}\times H^{1})}$ and subsequently by $t||\varphi||^{4}_{L^{\infty}([0,t];H^{2})}\\||[\varphi,m]||_{L^{\infty}([0,t];H^{2}\times H^{1})}$ therefore closing the argument.

Now we will sketch a proof of the local existence theorem by an iteration argument. Recall that we have the data on the initial hypersurface $t=0$ and it is given by $\mathcal{V}^{0}:=\mathcal{V}(0)=[\varphi(0),m(0)]^{T}$. Let us construct a sequence $\{\mathcal{V}^{0}_{k}\}_{k=1}^{\infty}\in C^{\infty}_{0}\cap B_{R}(\mathcal{V}^{0})$ by applying an approximation to the identity on $\mathcal{V}^{0}$ such that $\lim_{k\to\infty}\mathcal{V}^{0}_{k}=\mathcal{V}^{0}$. Here $C^{\infty}_{0}$ is the space of compactly supported smooth functions and $B_{R}(\mathcal{V}^{0})$ is a ball of radius $R$ in $H^{2}\times H^{1}$ centered at $\mathcal{V}^{0}$. Now we will construct a sequence of approximate solutions $\{\mathcal{V}_{k}\}_{k=1}^{\infty}\subset C([0,t^{*}];B_{R}(\mathcal{V}^{0}))$ for a suitable $t^{*}<1$ (we choose $t^{*}<1$ so that the involved constants do not depend on time) with initial data for $\mathcal{V}_{k}$ given by $\mathcal{V}^{0}_{k}$. Let us simply write $\mathcal{L}$ for $\mathcal{L}_{N,Y,g}$ and $\mathcal{H}^{s}$ for $H^{s}(\Sigma_{t})\times H^{s-1}(\Sigma_{t})$. We determine the sequence $\{\mathcal{V}_{k}\}_{k=1}^{\infty}$ through solving the following set of linear hyperbolic PDEs
\begin{align}
\mathcal{L}\mathcal{V}_{k+1}=\mathcal{F}_{k},~~~~\mathcal{V}_{k+1}(0)=\mathcal{V}^{0}_{k+1},
\end{align}
where $\mathcal{F}_{k}:=\mathcal{F}[\mathcal{V}_{k}]=\mathcal{F}[\varphi_{k},m_{k}]$ for $k>1$, $\mathcal{V}_{1}:=\mathcal{V}^{0}_{1}$, and set $\mathcal{F}_{1}=0$. Now existence of solutions of linear hyperbolic PDE with smooth coefficients given initial conditions is well established. Now we construct the initial sequence $\{\mathcal{V}^{0}_{k}\}_{k=1}^{\infty}$ such that 
\begin{align}
\mathcal{V}^{0}_{k}\in B_{R/4}(\mathcal{V}^{0})~\forall k\geq 1,~~~~C||\mathcal{V}^{0}_{k}-\mathcal{V}^{0}_{k^{'}}||_{\mathcal{H}^{2}}\leq \frac{R}{4}~\forall k,k^{'}\geq 1.
\end{align}
We will prove that the sequence $\{\mathcal{V}_{k}\}_{k=1}^{\infty}$ converges to $\mathcal{V}$ in $\mathcal{H}^{2}$ and the limit $\mathcal{V}$ solves the equation $\mathcal{L}\mathcal{V}=\mathcal{F}[\mathcal{V}]$.\\
We accomplish this in three steps. We first show that $\{\mathcal{V}_{k}\}_{k=1}^{\infty} \subset L^{\infty}([0,t^{*}];\\B_{R}(\mathcal{V}^{0}))$ for a suitably chosen time $t^{*}<1$. This is equivalent to proving that there exists a time $t^{*}<1$ such that if $\mathcal{V}_{k}\in L^{\infty}([0,t^{*}];B_{R}(\mathcal{V}^{0}))$ then $\mathcal{V}_{k+1}\in L^{\infty}([0,t^{*}];B_{R}(\mathcal{V}^{0}))$. This simply follows from the difference equation and the energy inequality, that is,
\begin{align}
\mathcal{L}(\mathcal{V}_{k+1}-\mathcal{V}_{1})=\mathcal{F}_{k}-\mathcal{L}\mathcal{V}_{1}
\end{align}
implying 
\begin{align}
||\mathcal{V}_{k+1}-\mathcal{V}_{1}||_{L^{\infty}([0,t^{*}];H^{2}\times H^{1})}\\
\leq C(||\mathcal{V}^{0}_{k+1}-\mathcal{V}^{0}_{1}||_{L^{\infty}([0,t^{*}];H^{2}\times H^{1})}\nonumber+||\mathcal{F}_{k}||_{L^{1}([0,t^{*}];H^{2}\times H^{1})}\\\nonumber 
+||\mathcal{L}\mathcal{V}_{1}||_{L^{1}([0,t^{*}];H^{2}\times H^{1})})\\\nonumber 
\leq C(||\mathcal{V}^{0}_{k+1}-\mathcal{V}^{0}_{1}||_{L^{\infty}([0,t^{*}];H^{2}\times H^{1})}+t||\mathcal{F}_{k}||_{L^{\infty}([0,t^{*}];H^{2}\times H^{1})}+t||\mathcal{L}\mathcal{V}_{1}||_{L^{1}([0,t^{*}];H^{2}\times H^{1})})\\\nonumber 
\leq C(||\mathcal{V}^{0}_{k+1}-\mathcal{V}^{0}_{1}||_{L^{\infty}([0,t^{*}];H^{2}\times H^{1})}+||\mathcal{F}_{k}||_{L^{\infty}([0,t^{*}];H^{2}\times H^{1})}++||\mathcal{L}\mathcal{V}_{1}||_{L^{1}([0,t^{*}];H^{2}\times H^{1})})
\end{align}
and therefore there exists a suitable $t^{*}<1$ such that if $\mathcal{V}_{k}\in L^{\infty}([0,t^{*}];B_{R}(\mathcal{V}^{0}))$, then $\mathcal{V}_{k+1}-\mathcal{V}_{1}\in L^{\infty}([0,t^{*}];B_{R/2}(\mathcal{V}^{0}))$. Since $\mathcal{V}_{1}=\mathcal{V}^{0}_{1}\in B_{R/4}(\mathcal{V}^{0})$ by construction, we obtain $\mathcal{V}_{k+1}\in L^{\infty}([0,t^{*}];B_{R}(\mathcal{V}^{0}))$.

Secondly, we show that the sequence $\{\mathcal{V}_{k}\}_{k=1}^{\infty}$ converges in $L^{\infty}([0,t^{*}];H^{1}\times L^{2})$ for a suitable $t^{*}<1$. We observe  
\begin{align}
\mathcal{L}(\mathcal{V}_{k+1}-\mathcal{V}_{k^{'}+1})=\mathcal{F}_{k}-\mathcal{F}_{k^{'}}
\end{align}
which through the energy inequality yields 
\begin{align}
||\mathcal{V}_{k+1}-\mathcal{V}_{k^{'}+1}||_{L^{\infty}([0,t^{*}];\mathcal{H}^{1})}\nonumber \leq C(||\mathcal{V}^{0}_{k+1}-\mathcal{V}^{0}_{k^{'}+1}||_{\mathcal{H}^{1}}+||\mathcal{F}_{k}-\mathcal{F}_{k^{'}}||_{L^{1}([0,t^{*}];\mathcal{H}^{1})}).
\end{align}
Now let us evaluate the following 
\begin{align}
||\mathcal{F}_{k}-\mathcal{F}_{k^{'}}||_{L^{1}([0,t^{*}];\mathcal{H}^{1})}=\int_{0}^{t^{*}}||\mathcal{F}_{k}-\mathcal{F}_{k^{'}}||_{\mathcal{H}^{1}}dt^{'}\nonumber=\int_{0}^{t^{*}}||N(\varphi^{5}_{k}-\varphi^{5}_{k^{'}})||_{L^{2}}dt^{'}\\\nonumber
\leq C||\varphi_{k}||^{4}_{L^{\infty}([0,t^{*}];L^{\infty})}||\varphi_{k}-\varphi_{k^{'}}||_{L^{\infty}([0,t^{*}];L^{2})}\\\nonumber 
\leq C||\varphi_{k}||^{4}_{L^{\infty}([0,t^{*}];H^{2})}||\mathcal{V}_{k}-\mathcal{V}_{k^{'}}||_{L^{\infty}([0,t^{*}];\mathcal{H}^{1})}
\end{align}
due to Sobolev embedding $H^{2}(M)\hookrightarrow L^{\infty}(M)$. This yields using the boundedness of $||\varphi_{k}||_{L^{\infty}([0,t^{*}];H^{2})}$ sketched in the previous step and a $t^{*}<1$ depending on the $H^{2}$ norm of $\varphi$
\begin{align}
||\mathcal{V}_{k+1}-\mathcal{V}_{k^{'}+1}||_{L^{\infty}([0,t^{*}];\mathcal{H}^{1})}\\\nonumber
\leq C(R)||\mathcal{V}^{0}_{k+1}-\mathcal{V}^{0}_{k^{'}+1}||_{\mathcal{H}^{1}}+\frac{1}{2}||\mathcal{V}_{k}-\mathcal{V}_{k^{'}}||_{L^{\infty}([0,t^{*}];\mathcal{H}^{1})}.
\end{align}
Now after passing to a suitable sub-sequence, we may write $C(R)\sum_{k=1}^{\infty}||\mathcal{V}^{0}_{k+1}-\mathcal{V}^{0}_{k}||_{\mathcal{H}^{1}}<\frac{R}{2}$ since the constructed initial sequence $\{\mathcal{V}^{0}_{k}\}_{k=1}^{\infty}$ is Cauchy in $\mathcal{H}^{1}$. The previous inequality may be written after passing to a suitable sub-sequence as 
\begin{align}
||\mathcal{V}_{k+1}-\mathcal{V}_{k}||_{\mathcal{H}^{1}}< C(R)||\mathcal{V}^{0}_{k+1}-\mathcal{V}^{0}_{k}||_{\mathcal{H}^{1}}+\frac{1}{2}||\mathcal{V}_{k}-\mathcal{V}_{k-1}||_{\mathcal{H}^{1}}~~~~\forall k\geq 2
\end{align}
which yields 
\begin{align}
\sum_{k=3}^{\infty}||\mathcal{V}_{k}\nonumber-\mathcal{V}_{k-1}||_{L^{\infty}([0,t^{*}];\mathcal{H}^{1})}\\\nonumber
< 2C(R)\sum_{k=2}^{\infty}||\mathcal{V}^{0}_{k}-\mathcal{V}^{0}_{k-1}||_{\mathcal{H}^{1}}\nonumber+||\mathcal{V}_{2}-\mathcal{V}_{1}||_{L^{\infty}([0,t^{*}];\mathcal{H}^{1})}\\\nonumber
<R+||\mathcal{V}_{2}-\mathcal{V}_{1}||_{L^{\infty}([0,t^{*}];\mathcal{H}^{1})}.
\end{align}
Therefore $\{\mathcal{V}_{k}\}_{k=1}^{\infty}$ converges to $\mathcal{V}$ in $L^{\infty}([0,t^{*}];\mathcal{H}^{1})$ after passing to a suitable sub-sequence. Now, we want to show that $\mathcal{V}\in L^{\infty}([0,t^{*}];H^{2}\times H^{1})$ and it solves the evolution equation. By construction we have $\mathcal{V}(0,.)=\mathcal{V}^{0}$ and therefore we only need to show $\mathcal{L}\mathcal{V}=\mathcal{F}[\mathcal{V}]$. Notice the following 
\begin{align}
\mathcal{L}\mathcal{V}-\mathcal{F}[\mathcal{V}]=\mathcal{L}(\mathcal{V}-\mathcal{V}_{k})-(\mathcal{F}[\mathcal{V}]-\mathcal{F}[\mathcal{V}_{k}]).
\end{align}
We have already shown that $\{\mathcal{V}_{k}\}_{k=1}^{\infty}$ is Cauchy in $L^{\infty}([0,t^{*}];\mathcal{H}^{1})$ and therefore $(\mathcal{F}[\mathcal{V}]-\mathcal{F}[\mathcal{V}_{k}])$ converges in $L^{\infty}([0,t^{*}];\mathcal{H}^{1})$. Since the background geometry is assumed to be sufficiently regular (i.e., $||k||_{L^{\infty}(M)},||\nabla N||_{L^{\infty}(M)},\\ ||Y||_{L^{\infty}(M)}, ||\nabla Y||_{L^{\infty}(M)}<C$ and in addition curvature and certain of its derivatives are also point-wise bounded), $\{\mathcal{V}_{k}\}_{k=1}^{\infty}$ is Cauchy in $L^{\infty}([0,t^{*}];L^{2}\times H^{-1})$ and therefore $\mathcal{L}(\mathcal{V}-\mathcal{V}_{m})$ approaches 0 in $L^{\infty}([0,t^{*}];L^{2}\times H^{-1})$. However, since the left hand side is independent of $k$, we have 
\begin{align}
L\mathcal{V}=\mathcal{F}[\mathcal{V}],
\end{align}
that is, $\mathcal{V}=(\varphi,m)$ with finite energy solves the wave equation. Lastly, we argue that $\mathcal{V}=(\varphi,m)\in L^{\infty}([0,t^{*}];H^{2}\times H^{1})$. Let $C$ be a constant such that $||\mathcal{V}_{k}||_{L^{\infty}([0,t^{*}];H^{2}\times H^{1})}\leq C~\forall k$ using the first step. Now $\mathcal{V}=(\varphi,m)$ is the limit of $\mathcal{V}_{k}$ in $L^{\infty}([0,t];H^{1}\times L^{2})$ and therefore from uniform boundedness of $||\mathcal{V}_{k}||_{L^{\infty}([0,t^{*}];H^{2}\times H^{1})}$, we have $\mathcal{V}=(\varphi,m)\in L^{\infty}([0,t^{*}];H^{2}\times H^{1})$.

Uniqueness of the solution follows trivially from the energy inequality. Continuity and Cauchy stability of the solutions may be obtained in a standard way (presented in \cite{elliptichyperbolic} in detail). 
 This concludes the sketch of the proof of establishing the existence a solution $(\varphi,m)$ of (\ref{eq:lagrange}) in $C([0,t^{*}];H^{2}\times H^{1})$ which yields either $t^{*}=\infty$ or that the $H^{2}\times H^{1}$ norm of $[\varphi,m]$ blows up as $t\to t^{*}$. Therefore to prove global existence, we only need to show the boundedness of the $H^{2}\times H^{1}$ norm of $[\varphi,m]$. We go back to the energy inequalities 
 \begin{align}
 \partial_{t}\sqrt{\mathcal{E}_{1}(t)} \leq C(t)(\sqrt{\mathcal{E}_{1}(t)}+||\mathcal{F}[\varphi,m]||_{H^{1}\times L^{2}})\\\nonumber 
  \leq C(t)(1+||\varphi(t)||^{4}_{L^{\infty}})\sqrt{\mathcal{E}_{1}(t)},\\\nonumber 
 \partial_{t}\sqrt{\mathcal{E}_{2}(t)} \leq C(t)(\sqrt{\mathcal{E}_{2}(t)}+||\mathcal{F}[\varphi,m]||_{H^{2}\times H^{1}}),\\ 
  \leq C(t)(1+||\varphi(t)||^{4}_{L^{\infty}})\sqrt{\mathcal{E}_{2}(t)}
 \end{align}
 where $C(t)$ depends on the background geometry. Given boundedness of $||\varphi(t)||^{4}_{L^{\infty}}$, we observe that $\mathcal{E}_{1}(t)$ and $\mathcal{E}_{2}(t)$ can not blow up in finite time. Therefore $[\varphi,m]_{H^{1}\times L^{2}}$ $[\varphi,m]_{H^{2}\times H^{1}}$ can not blow up in finite time. Therefore $t^{*}=\infty$. This concludes the sketch of the proof of global existence.   

\section{Concluding Remarks} 
Here we have established a global existence result for the semi-linear wave equation with critical nonlinearity. The most important (and difficult) part of the result is the proof of a spacetime $L^{\infty}$ bound on the solution. Once such a bound is obtained, the rest is standard procedure. Due to the critical nature of the non-linearity, one roughly has a balance between the energy dispersion by the derivative term and the energy concentration by the non-linearity. These border-line cases are generally difficult to deal with since obtaining a point-wise bound on the solution and thereby establishing that the dispersive effect is slightly dominant is not obvious. In order to accomplish such a point-wise bound, employment of the integral equation (theorem 1) has proven to be crucial. In addition to the light cone integrals, the the so called `approximate' Killing and conformal Killing fields played an important role. In the case of Minkowski space this result has existed since the classical work of Grillakis \cite{grillakis1990regularity}. In curved spacetimes, however, the challenge is to derive an integral equation for the solution of the wave equation (\ref{eq:eom}). Such an integral equation may easily be derived for Minkowski space and was carried out in \cite{grillakis1990regularity}. In a curved spacetime, as we have seen in the current article, we required some additional machinery. Such an integral equation in the context of proving global existence for hyperbolic equations is however not so uncommon. In the classical paper, \cite{eardley1982global2} used a similar integral equation satisfied by the Yang-Mills field (curvature of the associated gauge bundle) propagating on the Minkowski spacetime to derive a point-wise (spacetime) bound of the same. This indeed yielded the global existence result. Motivated by the use of this integral equation associated with hyperbolic equations, Moncrief derived an integral equation for the spacetime curvature $2-$form \cite{moncrief2005integral}. This equation has a number of similarities with that of the Yang-Mill fields propagating on a curved spacetime. Following the results of \cite{moncrief2005integral}, \cite{klainerman2007kirchoff} derived a similar `approximate' integral equation for hyperbolic PDEs utilizing which \cite{ghanem2016global} proved a spacetime bound for the Yang-Mills curvature. Recently Moncrief and the current author are working on giving a new simplified proof of global existence of Yang-Mills fields on a globally hyperbolic background spacetimes utilizing the light cone integrals. The results seem to be extremely promising.

One of the main aspects of our study is that we needed point-wise control on the background geometry namely the point-wise norm of the strain tensor associated with the timelike vector field $n$. On the other hand, the breakdown criteria for the vacuum Einstein equation obtained by \cite{klainerman2010breakdown} was the blow-up of the point-wise norm of this strain tensor. In fact, the boundedness of the point-wise norm of this strain tensor controlled the point-wise behavior of the spacetime curvature. Now if we consider the full coupling of the scalar field (with critical nonlinearity) and gravity i.e., study the current problem in a setting where gravity is no longer a background field, the natural question arises whether the same breakdown criteria persists. On the other hand, a coupling with gravity may require additional criteria. One such result is already available. \cite{shao2011breakdown} studied the breakdown criteria for the non-vacuum Einstein equations including Maxwell and Klein Gordon fields as sources. The continuation criteria that was obtained required a point-wise bound of the strain tensor of $n$ in addition to the point-wise bound on the derivative of the Klein-Gordon field (for the Einstein-Klein Gordon system) or the Maxwell field (for the Einstein-Maxwell system). We hope to study the continuation criteria of Einsteinian spacetimes with a non-linear scalar field source term (having critical nonlinearity). Since we have Moncrief's integral equation for spacetime curvature at our disposal, we may simply couple the gravity with the nonlinear scalar field and obtain a system of coupled light cone integral equations for both the spacetime curvature and the scalar field and perform the subsequent analysis using suitably defined energies. Since we are ultimately interested in studying the gravity problem with large data, these studies are expected to shed new lights or provide new directions on the matter.  

Our result, while in its own right is an interesting mathematical result, is only a warm up exercise for the ultimate gravity problem as we mentioned previously. Of course one knows that obtaining a point-wise bound for the spacetime curvature in the fully general dynamical gravity problem (i.e., when gravity is no longer a background field) is unrealistic since there are explicit examples of singularity formation. There are known examples (explict solutions) where global existence is violated via the formation of black holes. These examples include Schwarzschild and Kerr spacetimes, where a true curvature singularity occurs within the event horizon of the black hole. Even in the absence of any matter source, pure gravity could `blow up' (through curvature concentration) i.e., gravitational singularities could prevent global existence or the global hyperbolicity of the spacetimes may simply be lost through the formation of Cauchy horizons (as in Taub-NUT spacetimes for example). However, there are hopes to prove global existence with arbitrarily large data in a number of cases (spacetimes of certain topological types with imposed symmetries). These include the so called $U(1)$ problem where the underlying manifold is $\mathbb{R}\times \mathcal{K}_{g}\times \mathbb{S}^{1}$, $\mathcal{K}_{g}$ being the closed Riemann surface with genus $g$, expanding spacetimes foliated by compact hyperbolic manifolds \cite{andersson2004future, andersson2011einstein, mondal2020attractors, mondal2020linear}, Milne spacetime \cite{andersson2004future} (perhaps with a positive cosmological constant to avoid black hole formation through curvature concentration) etc. Each of these spacetimes has a certain speciality. In the later two cases, the rapid expansion (accelerated with a positive cosmological constant) does not allow the curvature to concentrate at the level of small data (disperses the energy). One hope would be that this property persists when the limit on the size of the data is removed. In the $U(1)$ case, the full $3+1$ gravity problem may be reduced to $2+1$ gravity coupled to wave map fields \cite{choquet2003nonlinear, choquet2001future} with target being the hyperbolic plane. The global existence problem for such wave maps on a fixed (2 + 1 dimensional) Minkowski background has been solved \cite{krieger2012concentration, tao2008global}. Since, these wave map fields are essentially components of the full spacetime Riemann curvature tensor, if one may obtain point-wise estimates of the later through the light cone estimates, then the former would automatically be under control allowing one to `tame' the gravity. Since the light cone estimate technique is proven to work for a few rather non-trivial problems, it is fair to hope that under special circumstances, perhaps one may be able to control gravity. These issues where a direct application of light cone estimates (for suitable entities) becomes relevant are currently under intense investigation.

Apart from its importance towards a greater goal of tackling the gravity problem (and additional hyperbolic equations), this result of global existence for a critically nonlinear wave field is itself motivating. This is due to the fact that the global existence indicates that the scalar field with critical non-linearity indeed respects classical determinism. In that sense it serves as a `good' source while coupled to gravity. Even though this massless field with critical nonlinearity is not known to describe an interesting physical systems, it is certainly worth studying the global existence problem with coupling to gravity at least in a small data regime. Due to its energy critical nature, coupling to gravity may provide deep technical insights which may be helpful for the large data gravity problem itself. In addition, the techniques in this paper may be applied to critically nonlinear massive wave fields (i.e., the Klein-Gordon fields) after straightforward modifications of some of the calculations. 

\section*{Acknowledgements}
P.M would like to thank Prof. Vincent Moncrief for numerous useful discussions related to this project and for his help improving the manuscript. This work was supported by Yale University and CMSA at Harvard University.

%%  The bibliography

\end{document}